\documentclass[twocolumn]{aastex631}
\long\def\comment#1{}

\usepackage[hang,flushmargin]{footmisc} 

\usepackage[T1]{fontenc}
\usepackage{pifont}
\newcommand{\Plus}{\mathord{\text{\ding{58}}}}

\usepackage{makecell}
\usepackage{boldline}
\usepackage{tabularx}
\usepackage{booktabs}

\newcolumntype{Y}{>{\centering \arraybackslash}}

\newcolumntype{M}[1]{>{\centering\arraybackslash}p{#1\dimexpr\columnwidth}}

\makeatletter
\g@addto@macro\@floatboxreset\centering
\makeatother

\usepackage{amsmath}

\usepackage[normalem]{ulem}
\useunder{\uline}{\ul}{}

\shorttitle{Contact forces during the formation of planetesimals by gravitational collapse}
\shortauthors{Barnes et al.}
\graphicspath{{./}{figures/}}

\begin{document}

\title{Considering contact forces during the formation of planetesimals by gravitational collapse: mutual orbits, spin states, and shapes}

\correspondingauthor{Jackson T. Barnes}
\email{barne383@msu.edu}

\author[0000-0003-4041-0074]{Jackson T. Barnes}
\affiliation{Department of Earth and Environmental Sciences,
Michigan State University,
288 Farm Ln,
East Lansing, MI 48824, USA}

\author[0000-0001-5475-9379]{Stephen R. Schwartz}
\affiliation{Planetary Science Institute,
1700 East Fort Lowell, Suite 106,
Tucson, AZ 85719-2395}
\affiliation{Instituto de Física Aplicada a las Ciencias y la Tecnologías,
Universidad de Alicante, San Vicente del Raspeig,
E-03690 Alicante, Spain}

\author[0000-0002-4952-9007]{Seth A. Jacobson}
\affiliation{Department of Earth and Environmental Sciences,
Michigan State University,
288 Farm Ln,
East Lansing, MI 48824, USA}

\begin{abstract}
In this work, we apply a soft-sphere discrete element method (SSDEM) within the PKDGRAV \textit{N}-body integrator to investigate the formation of planetesimal systems through the gravitational collapse of clouds of super-particles.
Previously published numerical models have demonstrated that the gravitational collapse of pebble clouds is an efficient pathway to produce binary planetesimal systems.
However, such investigations were limited by their use of a perfect-merger and inflated-radii super-particle approach, which inhibits any analysis of planetesimal shapes and spin states, precludes the formation of the tightest binary orbits, and produces significantly under-dense planetesimals.
The SSDEM enables super-particles to rest upon each other through mutual surface penetration and by simulating contact physics. 
Super-particles do not need to be inflated and collisions are not treated as perfect mergers; we can thus track the evolution of planetesimal shapes, spins, and tight binary orbits.
We demonstrate that the SSDEM is an excellent method to model the collapse process, and is capable of producing many binary planetesimal systems from a single cloud.
The most-massive systems often exhibit low-inclination ($i~\lesssim~15^{\circ}$) and moderately eccentric ($e~\lesssim~0.40$) orbits within a tight range of semi-major axes normalized by their Hill radii ($a/R_{\text{Hill}}\sim 0.02$--$0.15$).
Conversely, less-massive planetesimal systems display a wider range of eccentricities and inclinations and maintain a wider range of $a/R_{\text{Hill}}$ ($0.15$ to $\geq$~0.50).
These results confirm the findings of previously published perfect-merging models while also producing novel results about planetesimal spin and shape properties.
Newly-formed planetesimals exhibit 10-hr rotation periods on average and can be characterized by a wide variety of shapes (spherical, oblate, top-shaped, flattened, egg-shaped, or prolate), with the most-massive planetesimals primarily forming as spheres and oblate-spheroids.
\end{abstract}


\section{Introduction}
\label{sec:introduction}
In order to understand the origins of planets, it is necessary to understand planetesimal formation. 
However, this key growth step that creates objects held together by gravity rather than material strength remains mysterious.
Unfortunately, it is not possible to directly observe planetesimal formation because it occurred approximately 4.6 Ga in the solar system, and it is also not possible to telescopically resolve the fine scales of planetesimal formation in distant protoplanetary disks. 
In the solar system, we are only capable of observing relict planetesimals (e.g., asteroids, comets, and Kuiper Belt objects) leftover from their formation. 
We must therefore also rely on laboratory experiments and numerical simulations in order to create a self-consistent model of planetesimal formation. 

The growth of dust aggregates ($\mu$m- to 100 $\mu$m-sizes) to pebbles (mm- to cm-sizes) through collisions between particle pairs has been replicated in both laboratory and theoretical experiments, yet growth beyond pebble sizes via pairwise collisions is not effective \citep{Guttler2010,Zsom2010,Windmark2012a}.
Condensation may also play a role in growing and establishing the size of pebbles particularly near condensation fronts in the protoplanetary disk \citep{Drazkowska2016,Schoonenberg2017,Ros2019}, but no proposed scenario results in growth to planetesimal sizes.
Laboratory experiments have shown that when pebbles collide at velocities relevant to the protoplanetary disk, they bounce or fragment instead of growing  \citep{Guttler2010,Zsom2010,Windmark2012a}.
Bouncing occurs during low velocity collisions because the mutual region of contact is too small and renders adhesion forces between them insufficient, and fragmentation occurs at higher velocity collisions when the stress of impact exceeds the internal strength of the particles \citep[for further details, see][]{Testi2014,Simon2022}.
While both bouncing and fragmentation lead to loss of relative kinetic energy, the surviving particles are very unlikely to encounter each other again to continue this loss process.
Therefore, the particles cannot achieve energies low enough so that the they can rest upon each other.
However, in a bound pebble cloud, dissipative collisions lead to lower relative velocities until collisions are gentle enough that they lead to coagulation and growth. 

As solid material in the protoplanetary disk grows to pebble-sizes, these solids move through the disk.
Pebbles settle near the mid-plane \citep{Nakagawa1986}, but they do not form a razor-thin disk like Saturn's rings and instead form a thicker disk determined by a balance between gravitational settling and vertically diffusive turbulence \citep{Youdin2007a}.
The thermal gradient through the disk created by the Sun creates a decreasing pressure gradient with increasing radial distance, and this radial pressure gradient partially supports the gas disk against the gravity of the Sun.
The orbital revolution of gas is therefore slowed relative to solid objects like pebbles, which orbit at an unsupported, Keplerian rate.
This velocity difference creates a headwind between pebbles and the gas disk, and the aerodynamic drag due to this headwind saps energy from the pebbles causing their orbits to shrink, spiraling them inward toward the Sun \citep{Weidenschilling1977b,Hayashi1981}.
While the global radial pressure gradient is negative, it is possible for various phenomena to create significant variations in the magnitude and even the sign of the pressure gradient in a local region of the disk, i.e., a pressure bump \citep[for a review of pressure structures that can be built in the disk, see][]{Johansen2015}.
Indeed, the pebbles themselves can affect the hydrodynamics of the gas disk and create a feedback loop.
From these interactions between pebbles and the gas disk, dynamic instabilities such as the streaming instability can occur. 

\subsection{Streaming instability and gravitational collapse}
\label{sec:collapse}
The streaming instability creates over-densities of pebbles within the protoplanetary disk, which undergo gravitational collapse to directly form planetesimals \citep{Youdin2005b, Youdin2007b, Johansen2007b}. 
As local clumps of dust in the protoplanetary disk drift radially inwards, they drag along surrounding gas due to momentum feedback from the dust onto the gas.
Subsequently, Coriolis accelerations modify the radial inflow of gas into an azimuthal flow and the gas then drags the dust clumps, resulting in a super-Keplerian dust velocity and thus, a deflection of the clumps radially outward. 
This results in a reduced radial drift and local dust pileup which increases the overall density of the dust clumps.
The entire sequence then repeats and eventually runs away, creating increasingly more-dense clumps \citep[for a detailed review of the hydrodynamics of the streaming instability, see][]{Squire2020}.

The streaming instability creates pebble clumps that continue to increase in mass density until they exceed the local Roche density \citep[$\rho_R = 9 \Omega_K^{2} / 4 \pi G$, where $\Omega_K$ is the heliocentric Keplerian orbital frequency and $\pi$ and $G$ are the familiar constants;][]{Johansen2007a}, and they are therefore no longer sheared by stellar tidal forces.
Within the clumps, pebbles bounce off of one another or fragment, but remain trapped within the clump's gravitational potential.
Thus, as the pebbles' kinetic energy is effectively dissipated, their collision velocities are reduced and self-gravity dominates. 
Pebbles therefore collide both gently and more frequently, and sticking and resting is feasible leading to growth and the formation of planetesimals \citep[for a review, see][]{Simon2022}.

As a pebble cloud collapses, excess angular momentum often prevents the formation of single planetesimals and instead results in the formation of binary systems \citep{Nesvorny2010,Robinson2020,Nesvorny2021}.
As the pebble cloud mass becomes concentrated within a much more compact radius, its rotation rate must also proportionally increase due to angular momentum conservation during the collapse process.
The collapsed cloud's rotation rate can exceed the rotational spin limit for a single planetesimal, and thus it must form a multi-component system.
\citet{Nesvorny2010} found that the majority of a collapsing cloud's mass accretes into a binary system with a wide orbit and equal-sized components.
\citet{Robinson2020} independently confirmed these results but also noted the formation of many smaller planetesimals including other multi-component bound systems out of the same collapsing cloud.
\citet{Nesvorny2021} explored the consequences of using more realistic initial conditions for the collapsing cloud taken directly from hydrodynamic shearing-box simulations of the pebble disk.
All of these prior works assumed perfect merging between colliding super-particles within their simulations \citep{Nesvorny2010,Nesvorny2021,Robinson2020}, so the role of contact physics is unclear and the spin and shape distribution of the formed planetesimals has not been examined.

\subsection{Mutual orbits, spins, and shapes of relict planetesimals}
\label{sec:relicts}
If planetesimal systems formed as a direct result of the gravitational collapse of pebble clouds, then the collapse process will determine their initial mutual orbits, spins, and shapes.
A direct hypothesis from gravitational collapse is that as energy is dissipated and angular momentum is conserved, rapidly spinning planetesimals with distended shapes are created. 
In addition, contact binaries can form if a collapsing cloud only slightly exceeds a critical angular momentum threshold, and equal-mass binaries on wide orbits form when the angular momentum is much greater than this threshold.
This hypothesis can be tested by comparing the outcomes of numerical simulations of gravitational collapse to relict planetesimals such as those in the Kuiper and Main belts.

The mutual orbit properties of binary relict planetesimal systems can provide useful constraints for the conditions required for their formation.
A significant fraction ($\geq$~30\%) of cold classical Kuiper Belt objects are equal-sized binaries \citep{Sheppard2004, Fraser2017, Thirouin2019b}.
They often maintain wide (10$^3$--10$^5$~km) mutual orbit separations \citep{Petit2004, Noll2020} that likely would not have survived implantation via Neptune's migration \citep{Nesvorny2019b, Volk2019, Nesvorny2022} or any significant gravitational interactions.
In particular, widely cold classical systems are often characterized by low-inclinations, whereas tighter orbiting systems exhibit a wide range of inclinations \citep{Parker2011, Grundy2019}.
Furthermore, they possess similarly-colored components indicating an in-situ formation in a region of uniform composition  \citep{Jacobson2007, Benecchi2009, Fraser2021}.
Thus, these components are likely to have formed together and been in orbit about each other since formation \citep{Nesvorny2010, Fraser2021, Nesvorny2022}.
The pristine nature of the cold classicals can serve as a means to understand potential conditions to form binary planetesimal orbits.
The orbits of other relict multi-component planetesimal systems---excluding those in the cold classicals---were likely significantly altered due to their more dramatic dynamical history.
However, interpreting their current state requires a knowledge of what the initial systems may have looked like, further motivating this work.

While prior work has focused on the mutual orbits of collapsed planetesimal systems, this work also assesses the spins and shapes of collapsed planetesimals and compares them to relict planetesimals.
The spins of large (D~$>$~40~km) Main Belt asteroids preserve a non-Maxwellian distribution, i.e., inconsistent with a collisional origin \citep[see Figure~2 in][]{Pravec2002}.
Specifically, the observed distribution has an excess relative to a Maxwellian of rapidly rotating large asteroids, and there is also an upward inflection in the average trend of asteroid rotation rates at diameters greater than 100~km (reproduced in the ensuing Figure~\ref{fig:diameter_v_spin}).
These population statistics have been interpreted as primordial since they are not in a collisional equilibrium, but the origin of this primordial distribution is unknown \citep{Salo1987}. 
Relatedly, direct observations of the rotational periods of cold classical Kuiper Belt objects reveal another useful constraint regarding relict planetesimal spins \citep[e.g.,][]{Thirouin2019b}, although these rotation rates might be subject to subsequent collisional and/or tidal evolution.
Assessing the role of subsequent processes again requires an estimate of the initial rotation state.
It therefore remains critical to fully understand whether primordial planetesimals were rapidly rotating following their formation or if angular momentum was efficiently removed by material ejected from the gravitationally collapsing cloud.

Moreover, the origins of planetesimal shapes remain poorly understood, yet these shapes are essential to interpreting their geophysics and evolutionary histories.
For example, the unexpected shape of the cold classical Kuiper Belt object, Arrokoth, may be the direct consequence of planetesimal formation \citep{Stern2019, Spencer2020} or subsequent evolution such as the in-spiral of a binary due to gas drag \citep{McKinnon2020}, gas drag assisted by Kozai-Lidov oscillations \citep{Lyra2021}, and/or sublimation shaping its flattened lobes \citep{Zhao2021}. 
More generally, there is not yet a consensus as to what the primordial shapes of planetesimals should be.
Lightcurve analyses of primordial Kuiper Belt objects indicate that bilobate and flattened shapes with uniform proportions may be common, but significant uncertainties remain \citep{Thirouin2019a,Showalter2021}. 
Likewise, occultation measurements of the targets of NASA's Lucy mission to the primordial Jupiter Trojans \citep{Levison2021} indicate a potential wide array of shapes, from prolate yet irregular ellipsoids \citep{Mottola2020} to more elongated shapes \citep{Buie2018,Mottola2023}.
Results from future Lucy flybys will diminish current uncertainties in the derived shape models.
Relatedly, future observations conducted at the Vera Rubin Observatory plan to significantly increase the observed population of outer solar system objects to approximately $30,000$, many of which will have measured shapes and spins \citep{Ivezic2019}.

\subsection{Recent models of gravitational collapse}
\citet{Nesvorny2010,Nesvorny2021} and \citet{Robinson2020} used \textit{N}-body models to simulate the gravitational collapse of rotating clouds of super-particles in order to reproduce the mutual orbits of binary planetesimal systems.
Super-particles are required to meet the primary challenge of modeling the gravitational collapse of a pebble cloud because a cloud initially contains an approximate sextillion ($10^{21}$) cm-sized pebbles \citep{Johansen2015}---an impossible number to ever directly simulate.
Super-particles in prior simulations represented sub-clouds of particles with masses equivalent to approximate 1~km-sized objects and inflated radii equivalent to 10--100~km-radius objects.
Excess angular momentum is imparted onto pebble clouds from vorticity leftover from the streaming instability \citep[e.g.,][]{Johansen2007b, Li2019a}.
For simplicity, \citet{Nesvorny2010,Nesvorny2021} and \citet{Robinson2020} initialized their clouds assuming solid-body rotation and exploring different angular rates.
They found a high ($\approx$~100\%) binary planetesimal formation rate with a majority of each cloud's initial mass ($\geq$~50\%) located in a near-equal mass binary system.
Moreover, their formed binary systems exhibit moderate eccentricities and low inclinations, as well as binary semi-major axes ranging from approximately 10$^3$--10$^5$~km.
These mutual orbits are consistent with objects found in the cold classical Kuiper Belt.

The perfect-merger and inflated-radii super-particle modeling approach successfully reproduces the mutual orbits of binary Kuiper Belt objects, but this method cannot reproduce important information about planetesimals' shapes and spin states.
For all impacts, colliding particles combine into a single new spherical particle conserving both mass and momentum. 
Particle contacts are effectively instantaneous as interparticle collisions last for an infinitesimally short duration \citep{Richardson2000, Richardson2011}.
In other words, merging collisions are assumed to be perfectly inelastic, and dynamical energy is efficiently removed from the collapsing cloud.
Thus, all information about shape and spin is lost since all formed planetesimals are spheres.
Moreover, in order to enhance the collision rate, super-particle surface areas were enhanced to mimic the surface areas of a large sub-cloud of pebbles.
As a consequence, the densities of these super-particles are exceptionally small ($\ll$~1~g~cm$^{-3}$) and unrealistic.
Inflating particles also precludes the formation of the tightest binary orbits, since any closely-orbiting planetesimals come into contact due to their inflated sizes.
The perfect-merger method therefore leaves critical questions unanswered, and limits a full investigation of the gravitational collapse process.

\section{Methods}\label{Methods}
The goals to model the mutual orbits, spin states, and shapes of formed planetesimals required a new approach to model gravitationally collapsing systems.
We therefore use the \textit{N}-body integrator PKDGRAV \citep{Stadel2001,Richardson2000} and its soft-sphere discrete element method \citep[SSDEM;][]{Schwartz2012} (ver. 08/03/2018) to model clouds of particles with realistic densities that collapse to form planetesimal systems.
PKDGRAV integrates the equations of motion of all particles in a system using a second-order leapfrog integrator, and it uses a \textit{k}-D tree to efficiently calculate the gravitational interactions between a large number of particles distributed across many processors \citep{Dikaiakos1996,Richardson2000,Stadel2001}.
In particular, the PKDGRAV SSDEM implementation has been designed to take into account short-distance particle interactions with contact forces as well as long-distance gravitational interactions \citep{Schwartz2012,Zhang2017}, which has been validated by direct comparison to laboratory experiments \citep[e.g.,][]{Schwartz2013}.
PKDGRAV has found success in many different use cases within the planetary sciences including planetesimal dynamics \citep{Richardson2000,Leinhardt2005,Leinhardt2009}, small body moon formation \citep{Walsh2008,Agrusa2024}, behaviors of granular systems relevant to small bodies \citep{Matsumura2014,Yu2014,Maurel2017}, collisions between small bodies \citep{Durda2004,Ballouz2014,Ballouz2015,Zhang2015}, and planetesimal formation \citep{Nesvorny2010, Nesvorny2020b, Nesvorny2021}.

\begin{figure*}[t!]
    \centering
    \includegraphics[width=\textwidth]{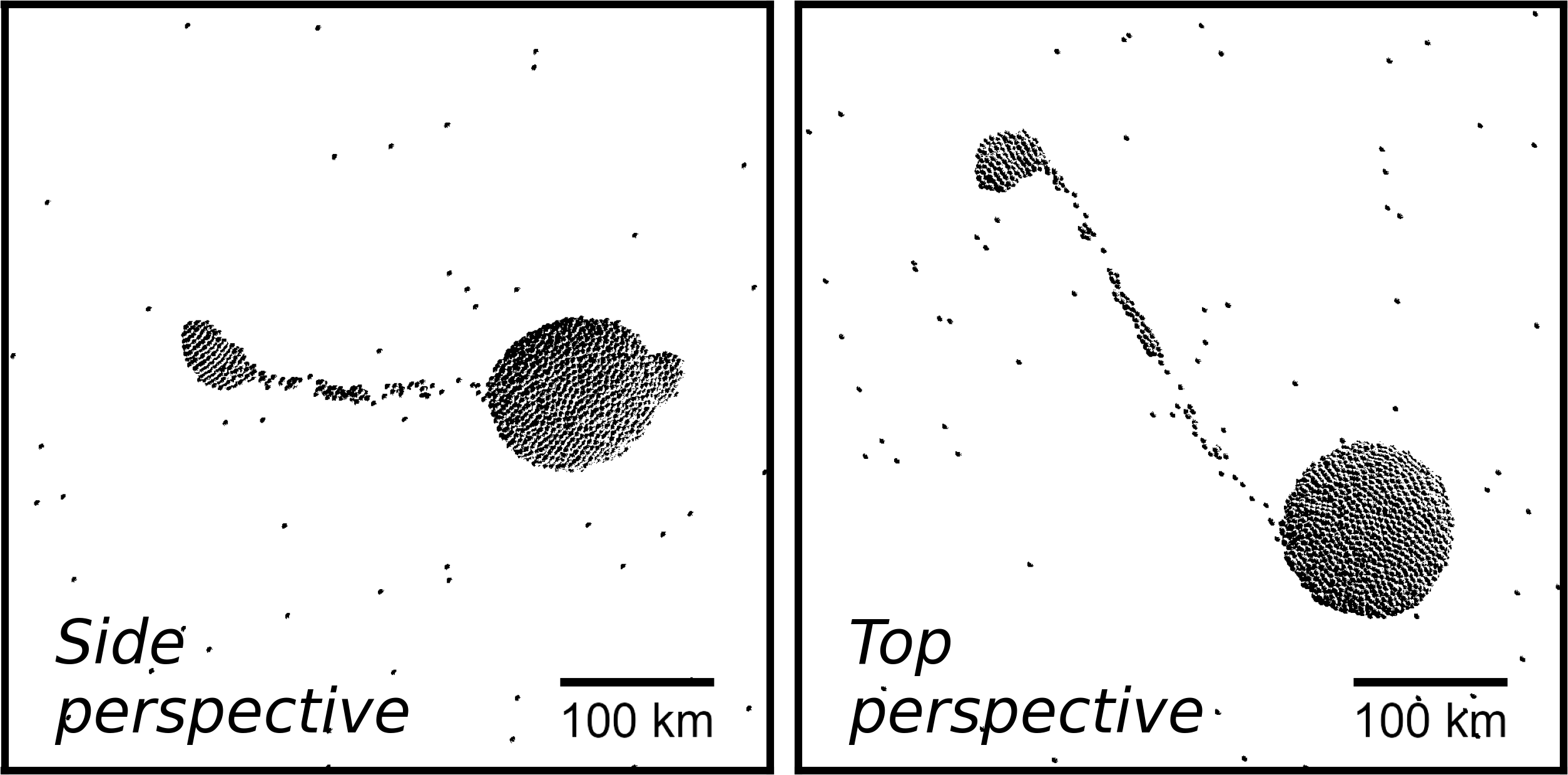}
    \caption{A snapshot of the aftermath of a collision between two planetesimals is shown from two views: along the equator and from above the spin pole of the most massive planetesimal.
    The soft-sphere discrete element method (SSDEM) can model accretion and decretion process, as evidenced by the post-impact debris tail and hit'n'run projectile. 
    In this case, a perfect merger method would have combined the bodies during the collision into a single spherical particle containing all the mass and momentum.}
    \label{fig:shapes_composite_2}
\end{figure*}
In the soft-sphere discrete element method (SSDEM) contact model \citep{Cundall1978}, particles interact with each other via mutual surface penetration. 
It computes the effects of particle collisions as well as simulates long-lasting contacts.
The latter capability allows for modeling planetesimals as aggregates of particles, in which particle interactions and behaviors are governed by mutual gravitational forces and contact forces that lead to a variety of potential shapes.
Because of the SSDEM, the simulation is able to track the motions of each planetesimal's constituent particles throughout a simulation, so that one can determine the planetesimal's resulting spin state. 
In addition, the densities of the accreted planetesimals are also realistic since they are composed of realistically dense super-particles. 
An SSDEM simulation can also form tightly orbiting binary systems, as their formation is not inhibited by under-dense, inflated super-particle radii.
Additionally, because SSDEM planetesimals are particle aggregates, it is possible that collisions with other particles or aggregates may lead to particle loss from the aggregate.
Thus, SSDEM models can capture both accretion and decretion, as shown in Figure~\ref{fig:shapes_composite_2}, a possibly important process that perfect merging models neglect.

Modeling the complete collapse process with the soft-sphere discrete element method (SSDEM) is computationally intensive.
The SSDEM requires small timesteps in order to properly resolve all particle overlaps and mutual contact forces.
In fact, the 4.7-s SSDEM timestep used in the present work is three orders of magnitude shorter than the 7.2-hr timestep used by the perfect merging method of \citet{Nesvorny2010}.
In order to compensate for the shorter timestep, we chose to simulate systems with a smaller gravitational collapse timescale.
The gravitational collapse (free-fall) timescale $\tau \propto \sqrt{1/G \varrho}$ is set by the overall particle cloud density $\varrho$ \citep{Binney2008}, where $G$ is Newton's gravitational constant.
The overall particle cloud density $\varrho \propto M_{\text{cloud}} / R_{\text{Hill}}^{3}$ is determined by the total mass of all particles in the cloud~$M_{\text{cloud}}$ divided by the volume of a significant fraction of the cloud's Hill sphere, where the Hill radius $R_{\text{Hill}} \propto a_\odot M_{\text{cloud}}^{1/3}$ \citep{Murray1999}.
Thus, the gravitational collapse timescale $\tau \propto a_\odot^{3/2}$ depends only on the heliocentric semi-major axis of the cloud's orbit~$a_\odot$.
Note that the collapse timescale does not depend on the cloud's mass, only its average orbital distance from the Sun.
Accordingly, the collapse timescale of a cloud on an orbit with a heliocentric semi-major axis near Earth's orbit $a_\odot = 1$ AU is over two orders of magnitude shorter than one on an orbit in the Kuiper Belt $a_\odot= 30$ AU.
Therefore, as opposed to \citet{Nesvorny2010} or \citet{Robinson2020} who focus on locations in the Kuiper Belt, we focus on planetesimal formation location closer to the Sun at 1 AU.

While the time between collisions is shorter at 1 AU than at 30 AU, the outcome of each collision and the properties of effected planetesimals do not depend on the time between collisions. 
Instead, these outcomes depend on the geometry and velocity of each collision.
Since collisions are determined by gravitational dynamics within the collapsing cloud, the typical collision velocity~$v_{\text{coll}}$ scales with the particle surface escape velocity $v_{\text{coll}} \propto v_{\text{esc}} = \sqrt{2 G m_p / r_p }$ where $m_p$ and $r_p$ are a characteristic planetesimal mass and radius at that time during the collapse.
While gravitational focusing increases the collisional cross-section of each planetesimal, this does not modify the expected likelihood of a given impact angle.
Thus, when considering two collapsing clouds at different heliocentric distances, similar-sized planetesimals in each cloud are undergoing similar collisions.
In other words, the choice of the cloud's heliocentric orbit does not have a significant effect on the accreted planetesimal properties.

\begin{table*}[t!]
\begin{tabularx}{0.314\textwidth}{cXcXcXcXcXcXcX}
\multicolumn{7}{@{}M{0.665}}{\textit{\textbf{Suite 1 --- Cloud Angular Velocity}}} \\
\toprule
\textit{$f$} & \textit{$\varepsilon_{n}$} & \textit{$\varepsilon_{t}$} & \textit{$\mu_{s}$} & \textit{$\mu_{r}$} & \textit{$\mu_{t}$} & \textit{Runs} \\ 
\midrule
0.2 & 0.5 & 1.0 & 0.0 & 0.0 & 0.0 & 5 \\
0.4 & 0.5 & 1.0 & 0.0 & 0.0 & 0.0 & 5 \\
0.6 & 0.5 & 1.0 & 0.0 & 0.0 & 0.0 & 5 \\
0.8 & 0.5 & 1.0 & 0.0 & 0.0 & 0.0 & 5 \\
1.0 & 0.5 & 1.0 & 0.0 & 0.0 & 0.0 & 5 \\
1.2 & 0.5 & 1.0 & 0.0 & 0.0 & 0.0 & 5 \\ 
\bottomrule
\\
\\
\end{tabularx}
\hspace{-0.25\columnwidth}
\begin{tabularx}{0.324\textwidth}{cXcXcXcXcXcXcX}
\multicolumn{7}{@{}M{0.686}}{\textit{\textbf{Suite 2 --- Particle Contact Physics}}} \\
\toprule
\textit{$f$} & \textit{$\varepsilon_{n}$} & \textit{$\varepsilon_{t}$} & \textit{$\mu_{s}$} & \textit{$\mu_{r}$} & \textit{$\mu_{t}$} & \textit{Runs} \\ 
\midrule
0.4 & 0.5 & 0.5 & 0.0 & 0.0 & 0.0 & 3 \\
0.4 & 0.5 & 0.5 & 0.5 & 0.5 & 0.5 & 3 \\
0.4 & 0.9 & 0.9 & 0.0 & 0.0 & 0.0 & 3 \\
0.4 & 0.9 & 0.9 & 0.5 & 0.5 & 0.5 & 3 \\
1.0 & 0.5 & 0.5 & 0.0 & 0.0 & 0.0 & 3 \\
1.0 & 0.5 & 0.5 & 0.5 & 0.5 & 0.5 & 3 \\
1.0 & 0.9 & 0.9 & 0.0 & 0.0 & 0.0 & 3 \\
1.0 & 0.9 & 0.9 & 0.5 & 0.5 & 0.5 & 3 \\ 
\bottomrule
\end{tabularx}
\caption{Parameters describing the soft-sphere discrete element method (SSDEM) N-body simulations conducted for this work.
Each simulation suite was defined by a choice of parameters: initial cloud rotation rate scale factor ($f$); coefficients of normal ($\varepsilon_{n}$) and tangential ($\varepsilon_{t}$) restitution; coefficients of static ($\mu_{s}$), rolling ($\mu_{r}$), and twisting ($\mu_{t}$) friction; and the total number of runs for each suite of simulations (\textit{Runs}).}
\label{table:suite_params}
\end{table*}

\subsection{Contact Physics}\label{section:contact_physics}
A key aspect of the soft-sphere discrete element method (SSDEM) is the inclusion of particle contact physics, which determine aggregate properties like shape and spin.
In the SSDEM, particle surfaces overlap and become subject to a host of frictional forces based on the history of that contact \citep{Schwartz2012,Zhang2017}.
Depending on the energies and the mechanical properties of the materials involved, particles may bounce, roll, or rest upon each other.
Two particles' mutual normal and tangential contact forces are modeled as spring and dashpot systems, which take as inputs the degree of overlap, incident velocities, and coefficients of dissipation like those of friction and restitution \citep{Cundall1978,Schwartz2012}.
Each of these coefficients governs the energy dissipation between the two colliding particles.
Since planetesimals will be simulated as aggregates of super-particles, the coefficients describing the interactions between those super-particles reflect an effective inter-particle contact physics rather than a contact physics based on physical model materials.

Restoring forces in the soft-sphere discrete element method (SSDEM) are modelled according to Hooke's law such that all super-particles are associated with a normal spring constant~$k_n$, which effectively sets the particle-particle interaction stiffness and the degree of overlap during mutual collisions.
\citet{Schwartz2012} used values of a spring constant~$k_n$ such that maximum particle overlaps remain under 1\%.
However, the spring constant~$k_n$ sets the most important requirement on each simulation's timestep, such that $t_{\text{step}} \propto \sqrt{1/k_n}$.
Due to the inherent computational intensity of the SSDEM simulations, we chose a spring constant $k_n \approx 10^{11}$~kg~s$^{-2}$, which means that we relaxed the criterion on the maximum particle overlap to approximately 20\%.
This chosen spring constant resulted in a simulation timestep of about 4.7~s, which enabled the completion of individual simulations within about 1~month of wall-clock time.

We explore the effects of particle contact physics during gravitational collapse by experimenting with different values for super-particles' coefficients of normal and tangential restitution as well as their coefficients of static, rolling, and twisting friction in the SSDEM. 
For convenience, we direct the reader to \citet{Schwartz2012} and \citet{Zhang2017} for the complete details on the formulations and full discussions of particle damping coefficients, coefficients of friction, coefficients of restitution, and associated restoring forces.
In the following, we provide only a necessary overview of each parameter as well as the range for which each parameter is varied, as shown in Table~\ref{table:suite_params}.
Notably, these parameters were chosen not to perfectly mimic the contact physics between individual grains of material, i.e., particles, but instead collections of grains, i.e., super-particles.
Therefore, we could not rely on experimentally-derived values for the contact physics parameters, and so it was necessary to explore a wide range of values to assess their possible effects on the gravitational collapse process.

The coefficients of normal and tangential restitution dictate the elasticity of a collision between two super-particles.
The normal coefficient of restitution~$\varepsilon_{n}$ is defined as the ratio of the relative normal velocity after and before a collision.
By varying the normal coefficient of restitution~$\varepsilon_{n}$ from 0.0 to 1.0, one can model a range of collisional outcomes from perfectly inelastic to perfectly elastic.
The tangential coefficient of restitution~$\varepsilon_{t}$ is defined as the ratio of the relative tangential velocity after and before a collision.
The tangential velocity is the velocity component projected onto the direction of tangential relative motion at the point of contact at the start of the overlap.
The tangential coefficient of restitution~$\varepsilon_{t}$ also varies from 0.0 to 1.0, low values represent a rough surface with high friction and 1.0 represents a perfectly smooth surface with no dissipation due to slippage.

We have conducted simulations which investigate two different values for the coefficients of normal and tangential restitution, which were varied together such that the coefficient of restitution $\varepsilon = \varepsilon_{n} = \varepsilon_{t} = $ 0.5 or 0.9.
Previous experiments of gravitational collapse have already thoroughly explored the consequences of perfectly inelastic collisions \citep[e.g.,][]{Nesvorny2010,Robinson2020}, so we explored the consequences of both inelastic ($\varepsilon_{n,t}=0.5$) and elastic ($\varepsilon_{n,t}=0.9$) collisions.
Values were varied together due to computational constraints.

The coefficient of static friction~$\mu_s$ determines whether slippage occurs between two particles due to tangential stress.
If static friction is not overcome, then the tangential force resulting from a collision is best approximated by the force of static friction.
If static friction is overcome, then the tangential force is simply equal to the particles' mutual tangential damping force.
Static friction is nonexistent when the coefficient of static friction~$\mu_s$ is 0.0, and it is proportional to the normal force when equal to 1.0, and it can take any value in between those values.

The coefficient of rolling friction~$\mu_r$ determines whether rotational energy is dissipated from the relative rotational motion between rolling particles, and it induces a torque on a particle due to this rotational motion at the point of particle contact.
Rolling friction is nonexistent when the coefficient of rolling friction is 0.0.
When the coefficient of friction is 1.0, the force of rolling friction is proportional to the induced torque created by the normal force which acts to reduce the particle's relative rotational velocity, and it is dependent on the length from the particles' centers to their mutual point of contact.
Thus, the coefficient of rolling friction~$\mu_r$ can take any value between 0.0 and 1.0.

The coefficient of twisting friction~$\mu_t$ determines the dissipation of energy associated with the relative rotation of the particles around a unit vector that is aligned with the vector between the centers of the two particles.
The induced torque on each particle due to twisting friction is determined by the relative spins of a particle and its neighbor along their axis of contact and the distance from their centers to the circumference of a circle describing their mutual surface intersection.
Twisting friction is nonexistent when the coefficient of twisting friction~$\mu_t$ is 0.0.
When the coefficient of twisting friction~$\mu_t$ is 1.0, it is proportional to the induced torque created by the normal force which acts to slow a particle's relative rotation about the normal axis, and it is dependent both on the spin vectors of a particle and its neighbor as well as the size of the particle overlap region.
Thus, the coefficient of twisting friction~$\mu_t$ can take any value between 0.0 and 1.0.

We have conducted simulations which investigate two different values for the coefficients of static, rolling, and twisting friction, which were varied together such that the coefficient of friction $\mu = \mu_{s} = \mu_{r} = \mu_{t} = $ 0.0 or 0.5.
We explored the consequences of no interparticle friction ($\mu_{s,r,t}=0.0$) and moderate friction ($\mu_{s,r,t}=0.5$).
While the simulations with moderate friction $\mu_{s,r,t}=0.5$ may be closer to reality, the simulations with no friction $\mu_{s,r,t}=0.0$ allowed an assessment of the impacts of our two choices coefficients of restitution $\varepsilon_{n,t}$ on planetesimal growth separate from friction.
Values of friction $\mu_{s,r,t}$ were varied together due to computational constraints.
 
\subsection{Initial Conditions}\label{InitCond}
The primary goal of this work is to understand the origins of planetesimals' mutual orbits, shapes, and spins, thus we needed to include contact physics. 
So that the soft-sphere discrete element method (SSDEM) outcomes can be directly compared with outcomes from the perfect-merger approach, we repeated the computational experiments from \citet{Nesvorny2010} and \citet{Robinson2020}.
The initial cloud mass was equivalent to a 100~km class planetesimal with a density typical of comets and planetesimals (1~g~cm$^{-3}$), and each cloud contained $N_p = 10^{5}$ super-particles.
The particles within each cloud had identical masses $M_p =  M_{\text{cloud}}/N_p \sim 4.19\times10^{13}$~kg and radii $r_p\sim2.15$~km, having assumed each super-particle is composed of a uniform density.
All particles were initially positioned within the spherical cloud, drawing their positions randomly from a distribution with uniform number density. 

Pebble clouds formed from the streaming instability exhibit rotation following their formation \citep[e.g.,][]{Johansen2007b, Li2019a}.
This rotation is often complex, and numerical models of the streaming instability cannot fully resolve the dynamics of the pebble clouds anyways.
Furthermore, existing numerical models of the streaming instability make assumptions about protoplanetary disk parameters that may not be accurate and may affect cloud rotational properties.
As such, the rotations of pebble clouds remains uncertain, and so for this work, we follow \citet{Nesvorny2010,Nesvorny2021} and \citet{Robinson2020} so that each super-particle cloud is initialized undergoing rotation as a solid-body described by a single initial angular velocity $\Omega_{\text{cloud}} = f \Omega_{\text{circ}}$, where $\Omega_{\text{circ}} = (GM_{\text{cloud}}/R_{\text{cloud}}^3)^{1/2}$ is the circular orbital angular frequency at the cloud’s edge, $R_{\text{cloud}}$ is the spherical radius of the cloud, and $f$ is the initial pebble cloud rotation rate scale factor.  
The radius of the cloud is $R_{\text{cloud}} = 0.6 R_{\text{Hill}}$, where $R_{\text{Hill}} = a_{\odot} (M_{\text{cloud}}/ 3 M_{\odot})^{1/3}$, $a_{\odot}$ and $M_{\text{cloud}}$ are the heliocentric semi-major axis and mass of the cloud, and $M_{\odot}$ is the mass of the Sun.

The first suite of simulations repeats a set of experiments by \citet{Nesvorny2010} and \citet{Robinson2020}, but with the SSDEM rather than the perfect merger model.
For a nominal set of contact physics coefficients ($\varepsilon_{n} = 0.5$, $\varepsilon_{t} = 1.0$, and $\mu = 0.0$), we vary the initial cloud angular velocity~$\Omega_{\text{cloud}}$ adjusting the scale factor $f \in \left[0.2, 1.2\right]$ in steps of~$0.2$.
Thus, while maintaining nominal particle contact physics parameters, we can isolate and effectively investigate the consequences of a range of cloud rotations from minimum- to maximum-circular rotation on planetesimal formation. 
All of the parameter choices for this first suite of simulations are shown in Table \ref{table:suite_params}.
We ran five simulations for each experiment in the first suite for a total of $30$ simulations.
The cloud in each simulation was independently generated.
Sections \ref{sec:growth} and \ref{sec:multiplicity} focus on results from these simulations only.

For the second suite of simulations, we varied the particle contact physics coefficients ($\varepsilon = 0.5$ or~$0.9$ and $\mu = 0.0$ or~$0.5$), while continuing to experiment with two different initial cloud angular velocity scale factors $f = 0.4$ or~$1.0$.
All of the parameter choices for this second suite of simulations are shown in Table \ref{table:suite_params}.
Simulations were run in triplicate for each experiment in the second suite for a total of 24 simulations.
As in the first suite, the cloud in each simulation was independently generated.
Sections \ref{sec:masses} through \ref{sec:shapes} include results from both the first and second suite of experiments, so throughout those sections we always indicate from which suite the results come.


\begin{figure*}[t!]
    \centering
    \includegraphics[width=\textwidth]{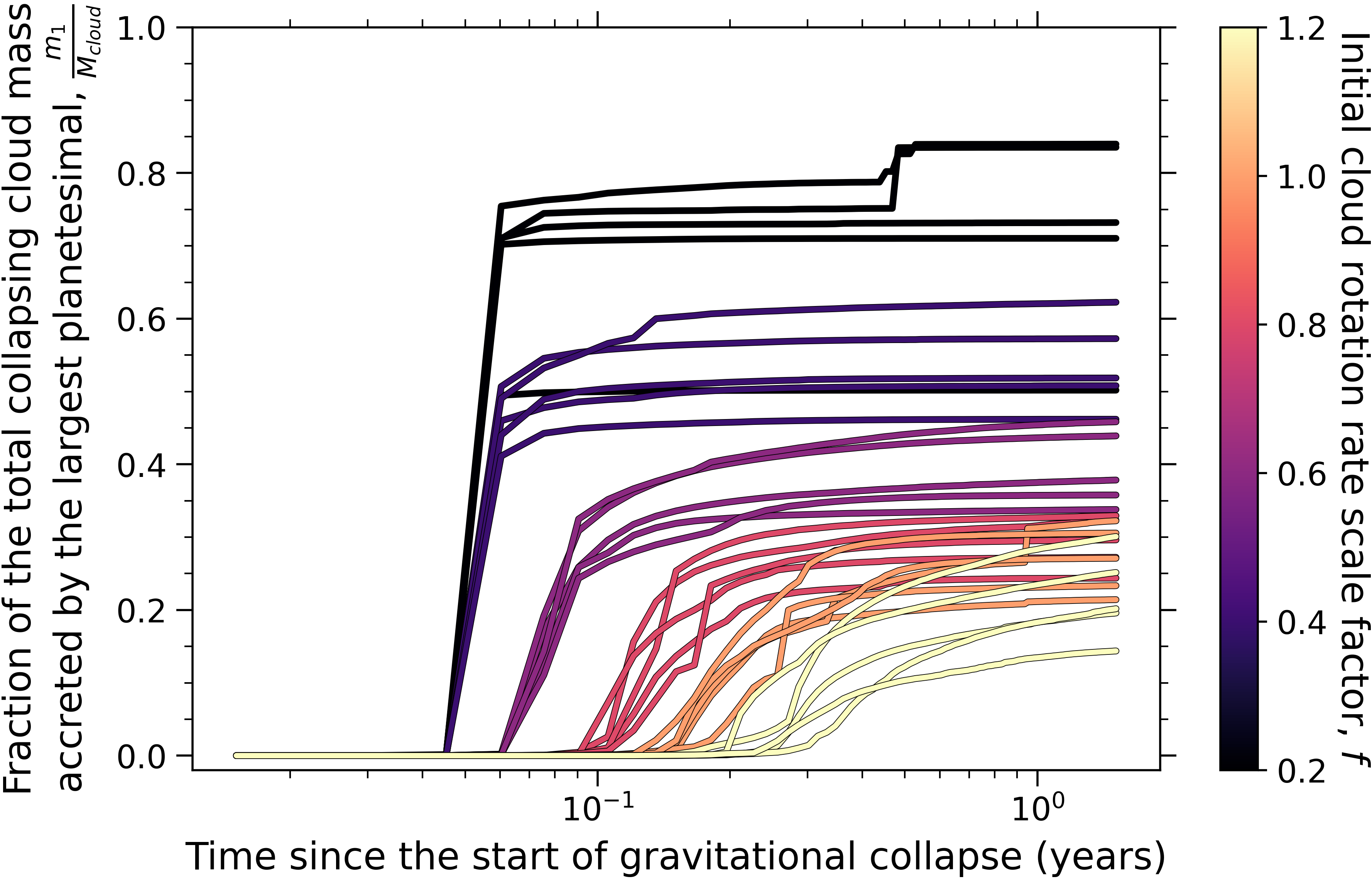}
    \caption{The fraction of the total collapsing cloud mass accreted by the largest planetesimal in each gravitationally collapsing cloud as a function of time since the start of each simulation.
    Lines are colored according to each simulation's initial cloud rotation rate scale factor~$f$.
    $30$ simulations are shown corresponding to $6$ initial cloud rotation rate scale factors~$f$ from $0.2$ to $1.2$ in steps of $0.2$ ($5$ simulations at each step). 
    See \textit{Suite 1 --- Cloud Angular Velocity} in Table~\ref{table:suite_params} for details.}
    \label{fig:planetesimalgrowth}
\end{figure*}

\section{Results}
In order to validate the usefulness of the soft-sphere discrete element method (SSDEM) for modeling planetesimal formation, we reproduce the experiments of \citet{Nesvorny2010} and \citet{Robinson2020}.
Importantly, we directly compare the results from the SSDEM to the perfect merging method for planetesimal mass ratios and accretion efficiencies as well as the mutual orbits of formed binary planetesimal systems. 
The results from both methods have strong similarities regarding planetesimal growth rates (see Section~\ref{sec:growth}) and the final masses and radii for binary planetesimal components (see Section~\ref{sec:masses}).
We also do a deeper dive into planetesimal multiplicity than prior analyses (see Section~\ref{sec:multiplicity}).
Furthermore, we also highlight key contrasts with prior works, such as finding that the SSDEM anticipates tighter binary mutual orbits (see Section~\ref{sec:orbits}), which are precluded by the inflated radii of the perfect merging methods \citep{Nesvorny2010,Robinson2020}.

We then investigate the shapes and spins of these formed planetesimals, which are the unique outcomes of the SSDEM approach.
First, we examine how rotation period (see Section~\ref{sec:spins}) varies with planetesimal size, and we compare these results to relict objects in the asteroid and Kuiper belts.
There are striking similarities between asteroids and SSDEM planetesimals with regard to how spin rate decreases for objects up to 100 km sizes.
However, beyond this 100-km threshold, asteroids and Kuiper Belt objects have rotation periods similar to SSDEM planetesimals, which indicates that their spins may be primordial.
Moreover, we investigate the relationships between initial cloud rotation state and rotation period, and we discover counterintuitively that slower rotating clouds create planetesimals with more rapid rotations and faster rotating clouds create planetesimals with slower rotations. 
Finally, we identify six distinct planetesimal shapes (spherical, oblate, prolate, flattened, top-shaped, and egg-shaped; see Section~\ref{sec:shapes}) and examine their relative frequencies across all simulations; spherical and oblate planetesimals are the most common and top-shaped and flattened planetesimals are rare.

\subsection{Planetesimal Growth}\label{sec:growth}
The SSDEM simulations reproduce the finding from \citet{Nesvorny2010} and \citet{Robinson2020} that planetesimal growth is both rapid and efficient. 
In Figure~\ref{fig:planetesimalgrowth}, we highlight the growth of the largest planetesimal in each simulation of a collapsing cloud from the suite of angular velocity simulations.
The largest planetesimal in each simulation (except for those from clouds with the highest initial angular velocity scale factor $f = 1.2$) completes its growth within approximately $1$~year at $a_\odot = 1$~AU, i.e., slightly longer than an orbital period.
Note that \citet{Nesvorny2010} and \citet{Robinson2020} modeled collapse at $a_\odot = 30$~AU and planetesimal growth in these prior works proceeds for approximately $100$~years.
Since the collapse timescale goes as $t_{coll} \propto a_\odot^{3/2}$ (see Section~\ref{Methods}) which is the same scaling as the orbital period, our work is entirely consistent with theirs, i.e., collapse occurs on the timescale of an orbital period.
While the majority of the planetesimal's mass is accreted within half of an heliocentric orbit, the most rapid planetesimal growth occurs during the first tenth of an orbit for initial angular velocity scale factors $f = 0.2$--$0.6$ and more slowly between $0.1$--$0.5$ of an orbit for $f = 0.8$--$1.2$. 

Planetesimal accretion efficiency, i.e., how much of the collapsing cloud is accreted by the largest planetesimal, scales inversely with the initial angular velocity of collapsing cloud.
The most efficient growth occurs for clouds which initially possess low angular velocity scale factors $f = 0.2$ or~$0.4$, and the least efficient growth occurs for clouds with the highest initial angular velocity scale factor $f = 1.2$.
The slower rate and reduced accretion efficiency of accretion at greater initial cloud angular velocities is due to the conservation of angular momentum providing support against cloud collapse.
In order for collapse to continue, some mass needs to be ejected from the main collapsing cloud system to remove angular momentum.
Relatedly, due to angular momentum support conserved from the rotation of the collapsing cloud, often some mass remains in orbit about the largest planetesimal preserving that angular momentum in that satellite's orbit.
Details regarding the formation of bound multi-component systems are discussed in detail in later sections.

\begin{table}[h!]
\noindent
\begin{center}
\begin{tabularx}{0.36\columnwidth}{YXYp{0.3\columnwidth}}
\multicolumn{2}{@{}M{0.36}}{\textit{\textbf{\shortstack{Suite 1 --- Cloud \\ Angular Velocity}}}} \\
\toprule
\textit{f} & \textit{N\textsubscript{p}} \\
\midrule
0.2 & $10\pm8$   \\    
0.4 & $31\pm6$   \\    
0.6 & $104\pm15$   \\    
0.8 & $163\pm10$   \\    
1.0 & $203\pm22$   \\    
1.2 & $202\pm30$   \\    
\bottomrule
\\
\\
\end{tabularx}
\hspace{-0.25\columnwidth}
\begin{tabularx}{0.47\columnwidth}{YXYXYXYp{0.25\columnwidth}}
\multicolumn{4}{@{}M{0.47}}{\textbf{\textit{\shortstack{Suite 2 --- Particle \\ Contact Physics}}}} \\
\toprule
\textit{f} & $\varepsilon$ & $\mu$ & \textit{N\textsubscript{p}} \\
\midrule 
0.4 & 0.5 & 0.0 & $26\pm7$    \\   
0.4 & 0.5 & 0.5 & $47\pm9$    \\   
0.4 & 0.9 & 0.0 & $8\pm4$     \\   
0.4 & 0.9 & 0.5 & $17\pm3$    \\   
1.0 & 0.5 & 0.0 & $199\pm30$  \\   
1.0 & 0.5 & 0.5 & $198\pm32$  \\   
1.0 & 0.9 & 0.0 & $130\pm11$  \\   
1.0 & 0.9 & 0.5 & $84\pm23$   \\   
\bottomrule
\end{tabularx}
\end{center}
\caption{The average number of individual planetesimals created from each collapsing cloud for each group of simulations in their associated suite (\textit{N\textsubscript{p}}).
Averages are shown with $3$-$\sigma$ standard errors.
Each simulation suite was defined by their initial cloud rotation rate scale factor~$f$ as well as particle coefficients of restitution~$\varepsilon$ and friction~$\mu$, see Table~\ref{table:suite_params} for details. 
For \textit{Suite 1 -- Cloud Angular Velocity}, values are determined from $5$~simulations.
For \textit{Suite 2 -- Particle Contact Physics}, values are determined from $3$~simulations.
Here we define a planetesimal as an object that has accreted $10$~or more particles.}
\label{table:planetesimal_stats}
\end{table}

\begin{figure*}[t!]
    \centering
    \includegraphics[width=\textwidth]{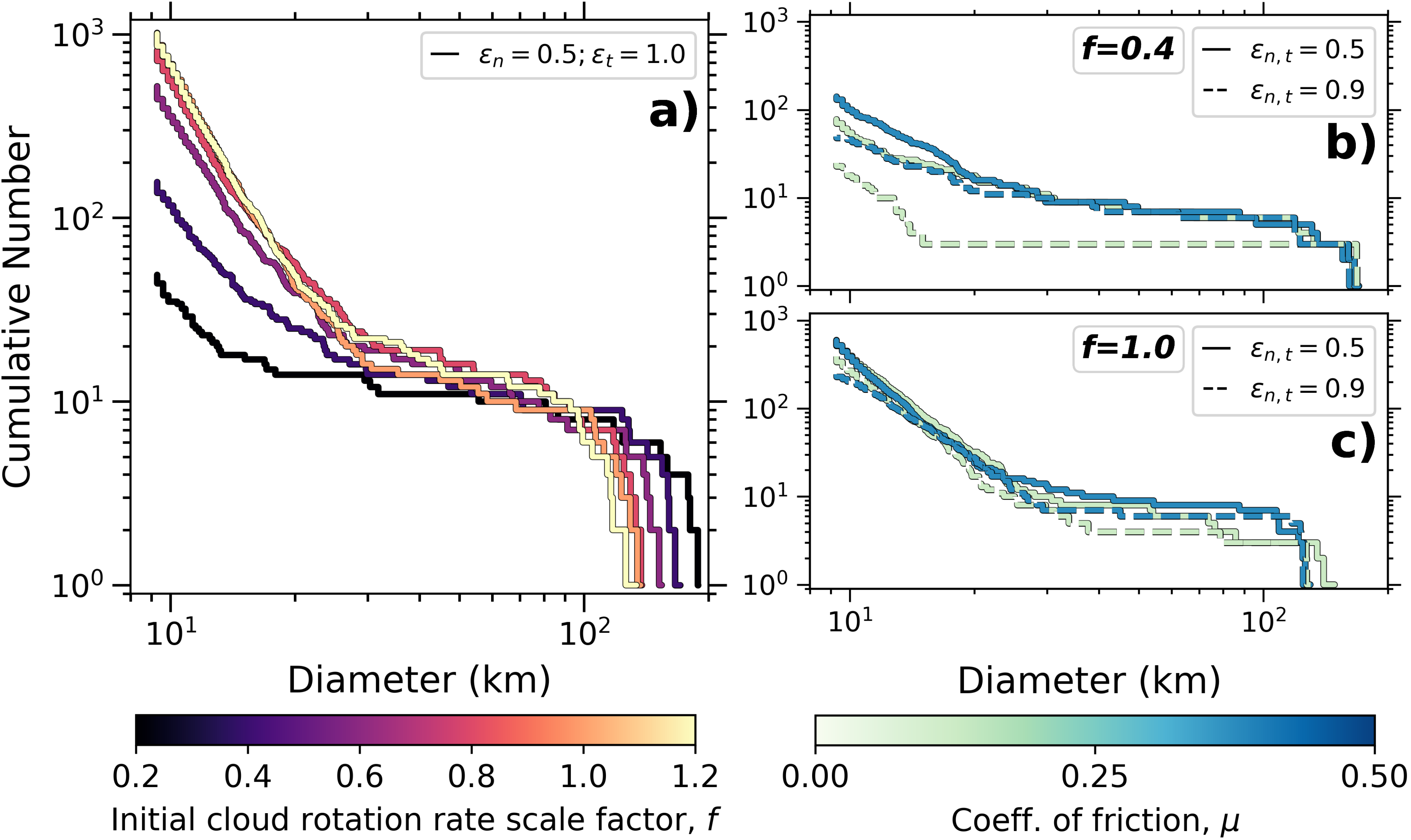}
    \caption{The cumulative number of planetesimals with a final planetesimal diameter greater than the diameter on the abscissa formed in \textit{Suite 1 --- Cloud Angular Velocity} (\textbf{a}) and \textit{Suite 2 --- Particle Contact Physics} (\textbf{b} and \textbf{c}) simulation suites.
    Each cumulative number distribution represents the sum over $5$~simulations in the case of \textit{Suite 1 --- Cloud Angular Velocity} (\textbf{a}) and $3$~simulations in the case of \textit{Suite 2 --- Particle Contact Physics} (\textbf{b} and \textbf{c}).
    These totals are colored according to either the initial cloud rotation rate scale factor~$f$ (\textbf{a}) or the particle coefficient of friction~$\mu$ (\textbf{b} and \textbf{c}).
    For \textit{Suite 2 --- Particle Contact Physics}, the subpanels are organized based on the initial cloud rotation rate scale factor~$f$, such that panels \textbf{b} and \textbf{c} include data associated with $f=0.4$ and $f=1.0$, respectively.}
    \label{fig:planetesimal_mass_SD}
\end{figure*}

Gravitational collapse always creates multiple planetesimals from each pebble cloud, many of which are ultimately gravitationally unbound to one another.
If we define a planetesimal as an accreted body with at least $10$~constituent SSDEM super-particles, then, in total, $713$~and $423$~individual planetesimals were created in the suites of $30$~angular velocity simulations and $24$~contact physics simulations, respectively.
The average number of individual planetesimals created in each cloud vary with both the initial cloud rotation rate scale factor~$f$ and particle contact physics parameters (see Table~\ref{table:planetesimal_stats} and Figure~\ref{fig:planetesimal_mass_SD}).
In both simulation suites, the average number of planetesimals created from collapse increases strongly with the initial cloud rotation rate scale factor~$f$.
This is a direct consequence of clouds with higher initial angular momenta ejecting planetesimals from the collapsing cloud thereby creating more planetesimals per cloud, since ejected planetesimals depart away from the cloud and stop accreting.

Planetesimal formation within a gravitationally collapsing cloud creates numerous planetesimals across a wide variety of sizes.
Within each simulated collapsing cloud, there are typically a couple large ($D \geq 100$~km) planetesimals formed from $\gtrsim 15\%$ of the initial cloud mass.
They are often bound to one another, but sometimes they form a handful of separate systems near the bottom of the cloud's gravitational potential well.
There are often a similar number of intermediate-sized ($30$~km~$\lesssim D \lesssim 100$~km) planetesimals, which are either ejected late from the cloud or sometimes satellites of the larger planetesimals. 
The cumulative size distribution of these intermediate-sized bodies is relatively shallow and is comparable across all simulation suites, see  Figure~\ref{fig:planetesimal_mass_SD}.

The strongest effect of varying the initial cloud rotation rate scale factor~$f$ is the associated strong increase in the number of small ($D \lesssim 30$~km) planetesimals, see Figure~\ref{fig:planetesimal_mass_SD}. 
This effect is responsible for the overall observation that faster initially rotating clouds generate a lot more planetesimals, see Table~\ref{table:planetesimal_stats}.
In general, clouds with higher initial cloud rotation rate scale factors $f \geq 0.8$ exhibit similarly steep cumulative size distribution slopes at small sizes and irrespective of most particle contact physics parameters, see Figure~\ref{fig:planetesimal_mass_SD}\hyperref[fig:planetesimal_mass_SD]{(a,~c)}.
This population of smaller ($D \lesssim 30$~km) planetesimals are produced early in the collapse, but since they are ejected from the cloud, their growth stalls and they are not accreted onto larger bodies.
However, at lower initial cloud rotation rate scale factors $f \leq 0.6$, the cumulative size distribution slopes become increasingly shallow for decreasing initial cloud rotation rates.
This is because a significant fraction of the cloud mass is ultimately accreted into the largest planetesimals due to the much less significant angular momentum support, see Figure~\ref{fig:planetesimal_mass_SD}\hyperref[fig:planetesimal_mass_SD]{(a,~b)}.

Focusing on the results from \textit{Suite 2 --- Particle Contact Physics}, a higher coefficient of restitution $\varepsilon = 0.9$ decreases the total number of produced planetesimals because collisions are more elastic and so objects do not efficiently accrete, see Table~\ref{table:planetesimal_stats}. 
In particular, a high coefficient of restitution $\varepsilon = 0.9$ and a low coefficient of friction $\mu = 0.5$ combine to suppress the accretion of planetesimals, so most free-moving aggregates contain fewer particles than the 10 particle criteria to be considered a planetesimal.
This is especially noticeable at low initial cloud rotation rate scale factor $f = 0.4$, since fewer objects are ejected from the collapsing cloud at this initial rotation rate so accretion must be shut-off by the contact physics rather than the dynamics, see Figure~\ref{fig:planetesimal_mass_SD}\hyperref[fig:planetesimal_mass_SD]{(b)}.
In contrast, when collapsing clouds have a higher initial cloud rotation rate scale factor $f = 1.0$, i.e., more objects are ejected from the collapsing cloud, see Table~\ref{table:planetesimal_stats}, the cumulative size distribution of created planetesimals has a much more muted dependence on the coefficient of restitution~$\varepsilon$ and the coefficient of friction~$\mu$, see also Figure~\ref{fig:planetesimal_mass_SD}\hyperref[fig:planetesimal_mass_SD]{(b,~c)}.

\subsection{Planetesimal Multiplicity}\label{sec:multiplicity}
The soft-sphere discrete element method (SSDEM) simulations also reproduce the finding from \citet{Nesvorny2010} and \citet{Robinson2020} that many collapsed pebble clouds create planetesimal systems with multiple bound members.
These bound systems include both satellites bound gravitationally to the largest planetesimal in each collapsing cloud as well as planetesimals with satellites that are not gravitationally bound to the largest planetesimal.
In total, $122$~binary planetesimal systems formed from the $30$~simulations in the angular velocity simulation suite (averaging to $4.3$~binaries per collapsing cloud), and $87$~from the $24$~simulations in the contact physics simulation suite (averaging to $3.6$~binaries per collapsing cloud).
As above, we have only considered particle aggregates that consist of more than $10$~SSDEM particles to be planetesimals.
This threshold applies equally to the satellites that comprise binary or higher multiplicity systems, so that each planetesimal member must meet this same mass-threshold requirement in order to be considered the component of a binary or higher multiplicity system.

\begin{figure*}[t!]
    \centering
    \includegraphics[width=\textwidth]{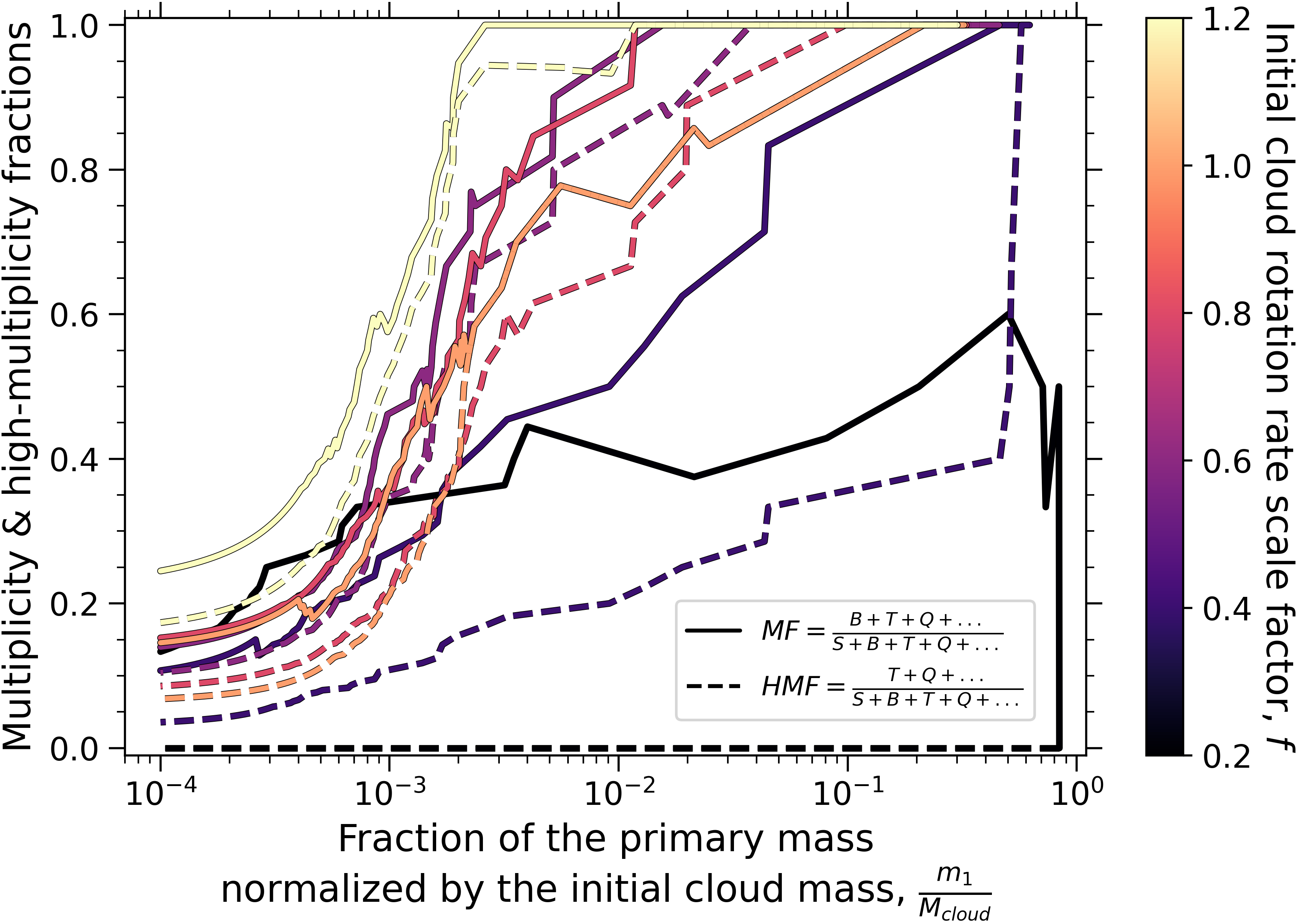}
    \caption{The cumulative multiplicity (solid) and high multiplicity (dashed) fractions of planetesimal systems produced by gravitational collapse as a function of the primary mass normalized by the initial cloud mass.
    Each cumulative fraction includes all planetary systems with normalized primary masses greater than the value on the abscissa, and each line represents the average of 5 simulations with the same initial cloud angular velocity.
    Each average set of runs samples initial cloud angular velocity scale factors $f$ from $0.2$ to $1.2$, in steps of $0.2$, which dictates their colors.}
    \label{fig:mult_higho_indiv_3}
\end{figure*}

To explore planetesimal multiplicity further than previous analyses, we borrow metrics from the star formation community to better quantify how multiplicity varies within each collapsing cloud simulation \citep[e.g.,][]{Offner2023}.
We define the multiplicity fraction of each collapsed cloud as $MF = (B+T+Q+...)/(S+B+T+Q+...)$ and the high multiplicity fraction as $HMF = (T+Q+...)/(S+B+T+Q+...)$, where the letters represent the number of planetesimal systems produced across a given suite of simulations with 1 ($S$, singles), 2 ($B$, binaries), 3 ($T$, triples), 4 ($Q$, quadruples), or more ($...$,~quintuples or higher) gravitationally bound planetesimal members.
These functions ($MF$ and $HMF$) are shown as cumulative distribution functions in Figure~\ref{fig:mult_higho_indiv_3} accumulating from the largest to the smallest systems, i.e., from right-to-left.

Almost all of the largest planetesimals formed from the cloud's gravitational collapse are part of planetesimal systems with binary or higher multiplicity, whereas most of the smallest planetesimals do not have bound companions, which is a consequence of their early ejection from the collapsing cloud. 
More broadly, planetesimal multiplicity and high multiplicity fractions increase with planetesimal mass.
This is similar to trends observed from star formation for stellar multiples across spectral types \citep[e.g.,][]{Offner2023}.

Clouds with a greater initial rotation rate scale factor~$f$ create higher multiplicity planetesimal systems.
For instance, in Figure~\ref{fig:mult_higho_indiv_3}, we compare the difference in multiplicity and high multiplicity fractions for simulations with initial rotation rate scale factors~$f=0.2$--$1.2$.
For the higher initial rotation rate case~$f=1.2$, the multiplicity and high multiplicity fractions are near unity for all collapsed planetesimals that contain more than 1\% of the initial cloud mass.
However, even for collapsing clouds with lower initial rotation rates, there is still a multiplicity fraction~$MF$ greater than 50\% and a high multiplicity fraction~$HMF$ of more than 20\%.
For simulations across nearly all initial rotation rate scale factors~$f$, the multiplicity and high multiplicity fractions do not trend down to zero but instead to some non-zero fraction.
It is clear that planetesimal systems with more than one component form even out of the material that is dynamically ejected and unbound from the cloud during its collapse.
The only exceptions are the clouds simulated with very low initial rotation rate scale factors~$f=0.2$ which do not produce ternary or higher member systems.

\begin{figure*}[t!]
    \centering
    \includegraphics[width=\textwidth]{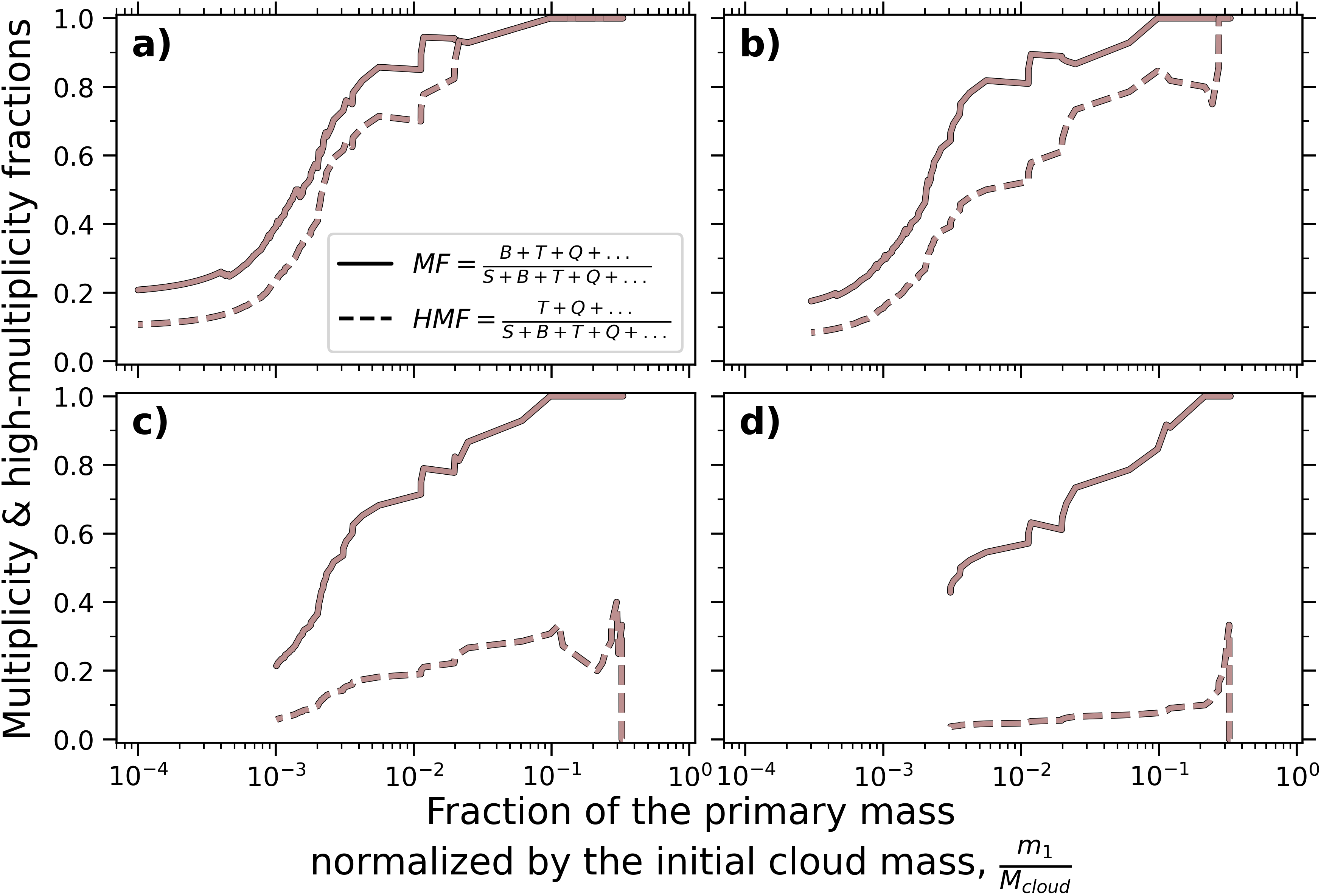}
    \caption{Each panel shows the cumulative  multiplicity fractions (solid) and high multiplicity fractions (dashed) of planetesimal systems as a function of the primary mass normalized by the initial cloud mass.
    Each cumulative fraction includes all planetary systems with normalized primary masses greater than the value on the abscissa.
    The panels have different minimum cut-off planetesimal masses from $10$ (\textbf{a}), $30$ (\textbf{b}), $100$ (\textbf{c}), to $300$ (\textbf{d}) particles to show how multiplicity varies with planetesimal size.
    Each line represents the average of 10 simulations from two similar initial cloud angular velocity scale factors, $f = 0.8$ and $1.0$.}
    \label{fig:mult_higho_cumul_10-300}
\end{figure*}

To assess the robustness of these planetesimal multiplicity trends with respect to our definition of an accreted planetesimal, we re-did our analysis modifying the minimum number of SSDEM particles within an aggregate to be considered a planetesimal.
We considered four different mass cutoffs ($10$, $30$, $100$, and $300$~particles), as shown in Figure~\ref{fig:mult_higho_cumul_10-300}.
All of the previously identified trends of multiplicity and high multiplicity fractions with planetesimal mass remain.
In detail, when restricting to the largest planetesimals, i.e., Figures~\ref{fig:mult_higho_cumul_10-300}\hyperref[fig:mult_higho_cumul_10-300]{(c,~d)}, where each member of the planetesimal system needs to have at least $100$ or $300$~particles, respectively, the high multiplicity fraction is a much more limited fraction overall.
This is a direct consequence of the particle number restriction, since many of the tertiary or higher bound members of collapsed planetesimal systems do not reach these higher particle counts.
This also indicates that these high multiplicity systems are hierarchical in nature---the two largest components of the high multiplicity system are much larger than the other components.

Focusing on binary systems, defined as the two largest components with a planetesimal system of two or more components, increasing the initial cloud rotation rate~$f$ directly increases the total number of binary systems created by each collapsing cloud, as shown in Table~\ref{table:binary_stats}.
For instance, clouds with higher initial cloud rotation rate scale factors $f = 1.0$ and $1.2$ create the most binary systems with an average of $6.4$ binaries per cloud, whereas an average of $1.1$ binaries form per cloud with lower initial cloud rotation rate scale factors $f=0.2$ and $0.4$.
Clearly, planetesimal formation via gravitational collapse efficiently creates a central planetesimal but if the parent cloud has any substantial rotation, then this mechanism is a very potent binary formation mechanism capable of creating many binary systems per collapsing cloud, reproducing a finding from~\citet{Nesvorny2010}.

\begin{table}[t!]
\noindent
\begin{center}
\begin{tabularx}{0.36\columnwidth}{YXYp{0.34\columnwidth}}
\multicolumn{2}{@{}M{0.36}}{\textit{\textbf{\shortstack{Suite 1 --- Cloud \\ Angular Velocity}}}} \\
\toprule
\textit{f} & \textit{N\textsubscript{b}} \\
\midrule
0.2 & $0.8\pm1.0$    \\   
0.4 & $1.4\pm0.7$    \\   
0.6 & $4.2\pm1.6$    \\   
0.8 & $5.2\pm3.7$    \\   
1.0 & $6.6\pm3.0$    \\   
1.2 & $6.2\pm1.0$    \\   
\bottomrule
\\
\\
\end{tabularx}
\hspace{-0.25\columnwidth}
\begin{tabularx}{0.47\columnwidth}{YXYXYXYp{0.23\columnwidth}}
\multicolumn{4}{@{}M{0.47}}{\textit{\textbf{\shortstack{Suite 2 --- Particle \\ Contact Physics}}}} \\
\toprule
\textit{f} & $\varepsilon$ & $\mu$ & \textit{N\textsubscript{b}} \\
\midrule
0.4 & 0.5 & 0.0 & $1.7 \pm 1.6$     \\           
0.4 & 0.5 & 0.5 & $2.0 \pm 1.4$     \\           
0.4 & 0.9 & 0.0 & $0.3 \pm 0.8$     \\           
0.4 & 0.9 & 0.5 & $1.0 \pm 0.0$     \\           
1.0 & 0.5 & 0.0 & $8.3 \pm 5.0$     \\           
1.0 & 0.5 & 0.5 & $5.7 \pm 2.9$     \\           
1.0 & 0.9 & 0.0 & $4.3 \pm 1.6$     \\           
1.0 & 0.9 & 0.5 & $5.7 \pm 0.8$     \\           
\bottomrule
\end{tabularx}
\end{center}
\caption{The average number of binary planetesimal systems $N_\text{b}$ created from each collapsing cloud for each group of simulations in their associated suite.
Averages shown with $3$-$\sigma$ standard errors.
Each simulation suite was defined by their initial cloud rotation rate scale factor ($f$) and  particle coefficients of restitution ($\varepsilon$) and friction ($\mu$), see Table~\ref{table:suite_params} for details. 
For \textit{Suite 1 -- Cloud Angular Velocity}, values are averaged over 5 simulations.
For \textit{Suite 2 -- Particle Contact Physics}, values are averaged over 3 simulations.
Here we define a binary as a system that contains at least 2 planetesimals that have each accreted 10 or more particles.}
\vspace{-1\baselineskip}
\label{table:binary_stats}
\end{table}

In contrast to the role of the initial cloud angular velocity, changes in the coefficient of friction have smaller effects on the binary system creation rate, as shown in Table~\ref{table:binary_stats}.
As in the angular velocity suite, an increase in the initial cloud rotation rate factor~$f$ creates a significant increase in the number of formed binary planetesimals, and this is again the dominant trend across the contact physics suite of simulations.
The consequences of changing the coefficient of restitution~$\varepsilon$ and friction~$\mu$ are much smaller and not always monotonic suggesting either a role for stochasticity that was not suppressed by averaging the relatively small number of simulations with a given set of parameters and/or a complicated relationship between contact physics and binarity.
Examining Table~\ref{table:binary_stats}, independent of their initial rotation rate scale factor~$f$ and the coefficient of friction~$\mu$, clouds with lower coefficients of restitution $\varepsilon=$~0.5 create more or, at least, the same number of binary planetesimal systems than simulations with higher coefficients of restitution $\varepsilon=$~0.9.

Simulations with very high elasticity of collisions (coefficient of restitution~$\varepsilon=$~0.9) can lead to peculiar systems with bouncing components at the end of the simulation.
Planetesimal systems are considered to be bouncing if their mutual periapses are less than their combined radii but they do not come to rest on each other after repeated collisions but instead bounce.
We did not consider these systems to be binaries for the purpose of our analysis since their orbits are not stable even on orbit timescales, and bouncing systems are expected to eventually accrete into a single planetesimal, just on much longer timescales than the simulation runtime.
In the specific circumstance of a slower rotating cloud with a large coefficient of restitution  and a low coefficient of friction, binary formation can be inhibited entirely (e.g., two out of three simulations with initial rotation rate scale factor~$f=0.4$, coefficient of restitution~$\varepsilon=0.9$, and coefficient of friction~$\mu=0.0$ created no gravitationally bound binary systems).
Critically, such contact physics parameters also impede effective planetesimal accretion and growth, and this prevents many particle aggregates from growing beyond the 10-particle threshold to form planetesimals according to that definition (see Table~\ref{table:planetesimal_stats}).
Given this unusual behavior, very likely this choice of contact physics parameters (coefficient of restitution~$\varepsilon=0.9$) is unphysical with respect to planetesimal formation by gravitational collapse, and these bouncing systems are only a computational curiosity.

\begin{figure*}[t!]
    \centering
    \includegraphics[width=\textwidth]{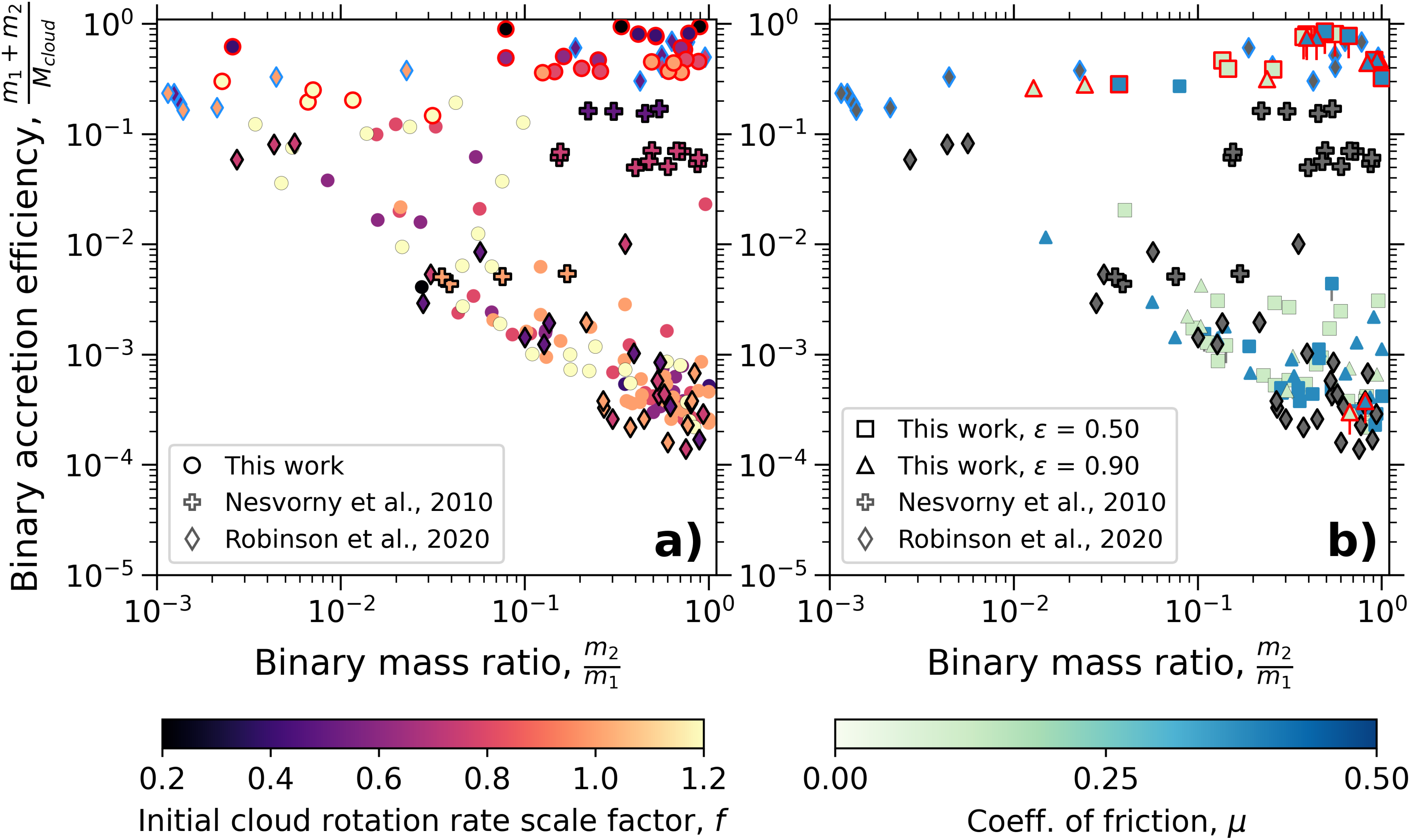}
    \caption{Binary planetesimal system accretion efficiency $(m_1 + m_2)/M_{\text{cloud}}$ as a function of the binary components' secondary-to-primary mass ratio $m_2/m_1$ from the suites of 30 simulations (\textbf{a}) with varying initial cloud angular velocities ($\circ$) and   24 simulations (\textbf{b}) with varying coefficients of friction and restitution, $\epsilon = 0.5$ ($\square$) and $\epsilon = 0.9$ ($\triangle$).
    Results are compared directly to results from \citet{Nesvorny2010} ($\Plus$) and \citet{Robinson2020} ($\blacklozenge$).
    The interior symbol coloring is according to the initial cloud rotation rate scale factor $f$ (\textbf{a}) and coefficient of friction (\textbf{b}). 
    Perfect merging models do not have a coefficient of friction so these points are colored gray in \textbf{b}.
    Larger points with red outlines indicate the most massive binary systems from each of the simulations.
    Similarly, blue outlines indicate systems that have achieved $(m_1 + m_2)/M_{\text{cloud}}\geq$~0.1 from~\citet{Robinson2020} simulations.
    In \textbf{b}, symbols with lines extending directly below them represent binaries formed in the simulations with more slowly initially rotating clouds, $f=0.4$.
    Those without this mark formed in more rapidly initially rotating clouds, $f=1.0$.
    Outside the frame of \textbf{b}, one binary system from the $f=0.4$, $\varepsilon=$~0.5, and $\mu=$~0.5 suite plots at \{$2.1\times10^{-4}$, 0.62\}.
    The lower left of each panel is empty of data because of both the monodispersed nature of the particle masses and an arbitrary cut-off mass (10 particles) to consider an aggregate as an accreted planetesimal.}
    \label{fig:mr_fig}
\end{figure*}

\subsection{Binary Planetesimal Masses and Radii} \label{sec:masses}
The SSDEM simulations produce binary systems with a wide variety of mass ratios that are generally similar to those produced in \citet{Nesvorny2010} and \citet{Robinson2020}.
Figure~\ref{fig:mr_fig} shows the accretion efficiency $(m_1 + m_2)/M_{\text{cloud}}$ of each binary (i.e., how much of the collapsing cloud was incorporated into the binary components) as a function of the binary mass ratio $m_2/m_1$.
The most massive planetesimal systems (larger circles with red outlines in Figure~\ref{fig:mr_fig}) formed at the center of the gravitational potential of the collapsing cloud, whereas less massive planetesimal systems (smaller circles with fainter outlines in Figure~\ref{fig:mr_fig}) accreted very little of the cloud mass and were dynamically ejected from the collapsing cloud.

\begin{figure*}[t!]
    \centering    
    \includegraphics[width=\textwidth]{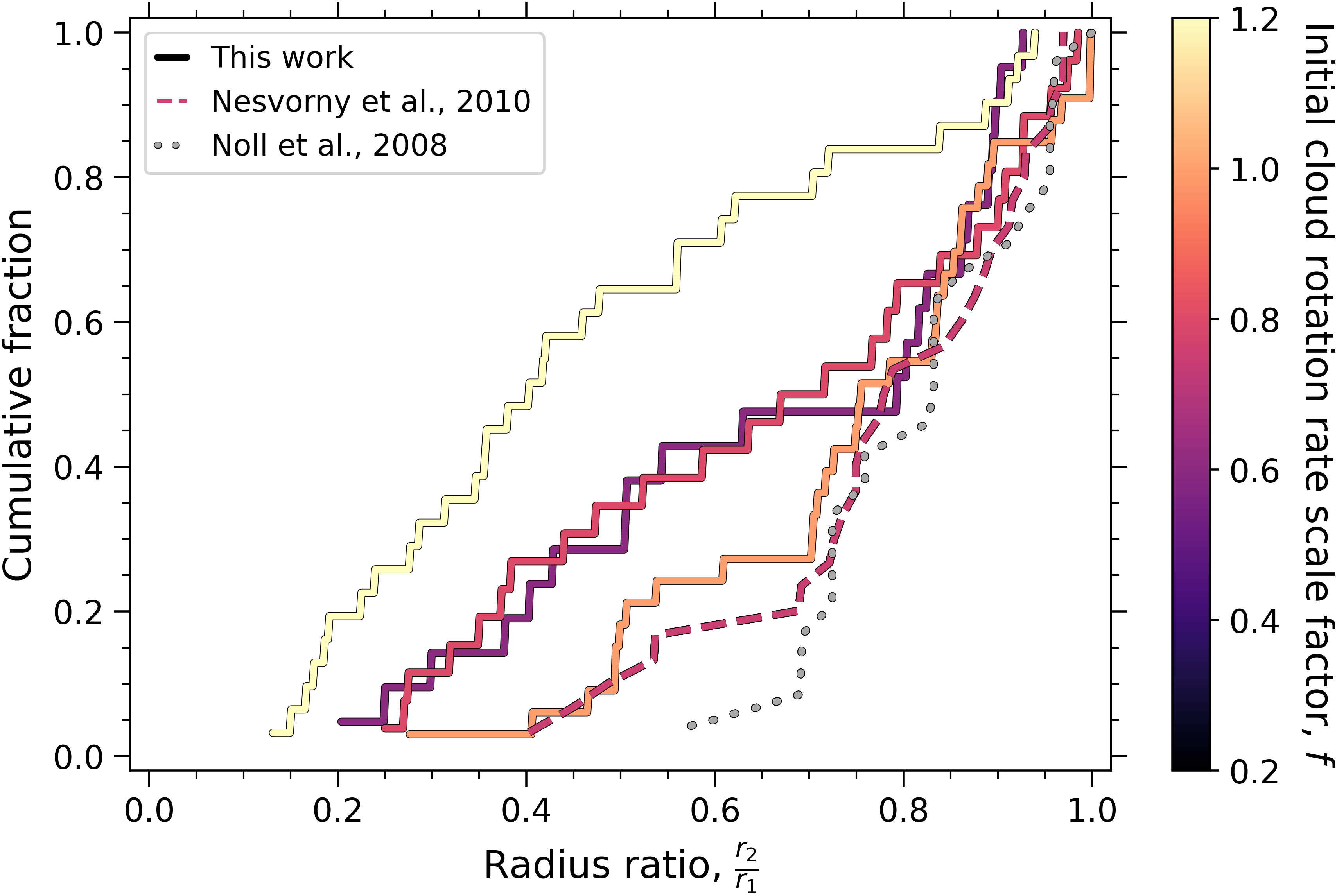}
    \caption{Cumulative distribution function of radius ratios for the all formed binary systems shown as solid lines for each simulation suite of the initial cloud angular velocity scale factors $f=0.6$--$1.2$, in steps of 0.2, which dictates their colors.
    Lower values $f = 0.2, 0.4$ are not included because very few binaries are created from their associated clouds. 
    For comparison, the same functions are shown but calculated from simulated binaries \citep[][$f = 0.75$, magenta dashed line]{Nesvorny2010} and observed binary systems in the Kuiper Belt \citep[][gray dotted line]{Noll2008}.
}
    \label{fig:cumul_radius_ratios_dist_all_planetesimals_comparison.png}
\end{figure*}


Similar to \citet{Nesvorny2010} and \citet{Robinson2020}, we find in the SSDEM simulations that increasing the angular velocity results in reduced binary planetesimal accretion efficiencies for the most massive binary in each simulation.
Across all of the simulations, as the initial cloud rotation rate scale factor~$f$ increases, the binary accretion efficiency decreases, i.e., less of the initial cloud mass ends up in the largest accreted planetesimal binary.
This is most evident in clouds with the highest angular velocity, i.e., initial cloud rotation scale factors $f=1.2$.
Only about 10--30\% of the initial cloud mass ends up in the largest binary systems of these collapsing clouds, as shown in Figure~\ref{fig:mr_fig}.
This relationship between the initial cloud rotation rate and the mass of the most massive binary can be attributed to angular momentum support increasing the number of particles that were ejected from the collapsing cloud either as individual particles, aggregates, or less-massive binary systems.
In other words, during the formation of the largest binaries, material scattered from the collapsing cloud carried away excess angular momentum enabling further collapse.
However, this scattered material also carried away mass.

Among the most-massive binary systems, a majority of the near-equal mass systems are created from clouds with initial cloud rotations scale factors ranging from $f=0.2$--$1.0$ from the angular velocity suite of simulations.
Notably, the clouds with an initial rotation rate scale factor $f = 1.2$ have $m_2/m_1< 0.4$ and accretion efficiencies $(m_1 + m_2)/M_{\text{cloud}}< 0.2$ (shown as pale yellow circles with red outlines in Figure~\ref{fig:mr_fig}\hyperref[fig:mr_fig]{(a)}).
Contrastingly, clouds with initial rotation rate factors ranging from $f=0.2$--$1.0$ produce massive systems with binary mass ratios $m_2/m_1 > 0.8$ and accretion efficiencies $(m_1 + m_2)/M_{\text{cloud}}> 0.2$ (shown as the other circles with red outlines in Figure~\ref{fig:mr_fig}\hyperref[fig:mr_fig]{(a)}).
For the second suite of simulations exploring contact physics, near-equal mass and equal-mass binary systems are created from simulations associated with initial cloud rotation rate factors of $f=0.4$ and~$1.0$, coefficients of restitution $\varepsilon = 0.5$ and~$0.9$, and high coefficients of friction $\mu = 0.5$, as seen in the right half of Figure~\ref{fig:mr_fig}.

One notable difference between these results and those of \citet{Nesvorny2010} is that the most efficiently accreting systems are distinctly isolated from the most massive binaries, effectively reproducing a difference that \citet{Robinson2020} also discovered when comparing their results with \citet{Nesvorny2010}.  
\citet{Nesvorny2010} used a large numerical timestep of $0.3$~days, and so particles may entirely pass through other particles during a step effectively skipping the collision.
Thus, accreting planetesimals may have missed some collisions, resulting in reduced accretion efficiencies and final binary masses.
We direct the reader to \citet{Robinson2020} $\S$4.2.4 for an in-depth assessment on the correlation between collision detection timestep and binary accretion efficiencies.
As discussed above, the SSDEM requires a very small timestep of about $6 \times 10^{-5}$~days---collisions must not just be identified but contact physics needs to be resolved---missing collisions is extremely improbable.

\begin{figure*}[t]
    \centering
    \includegraphics[width=\textwidth]{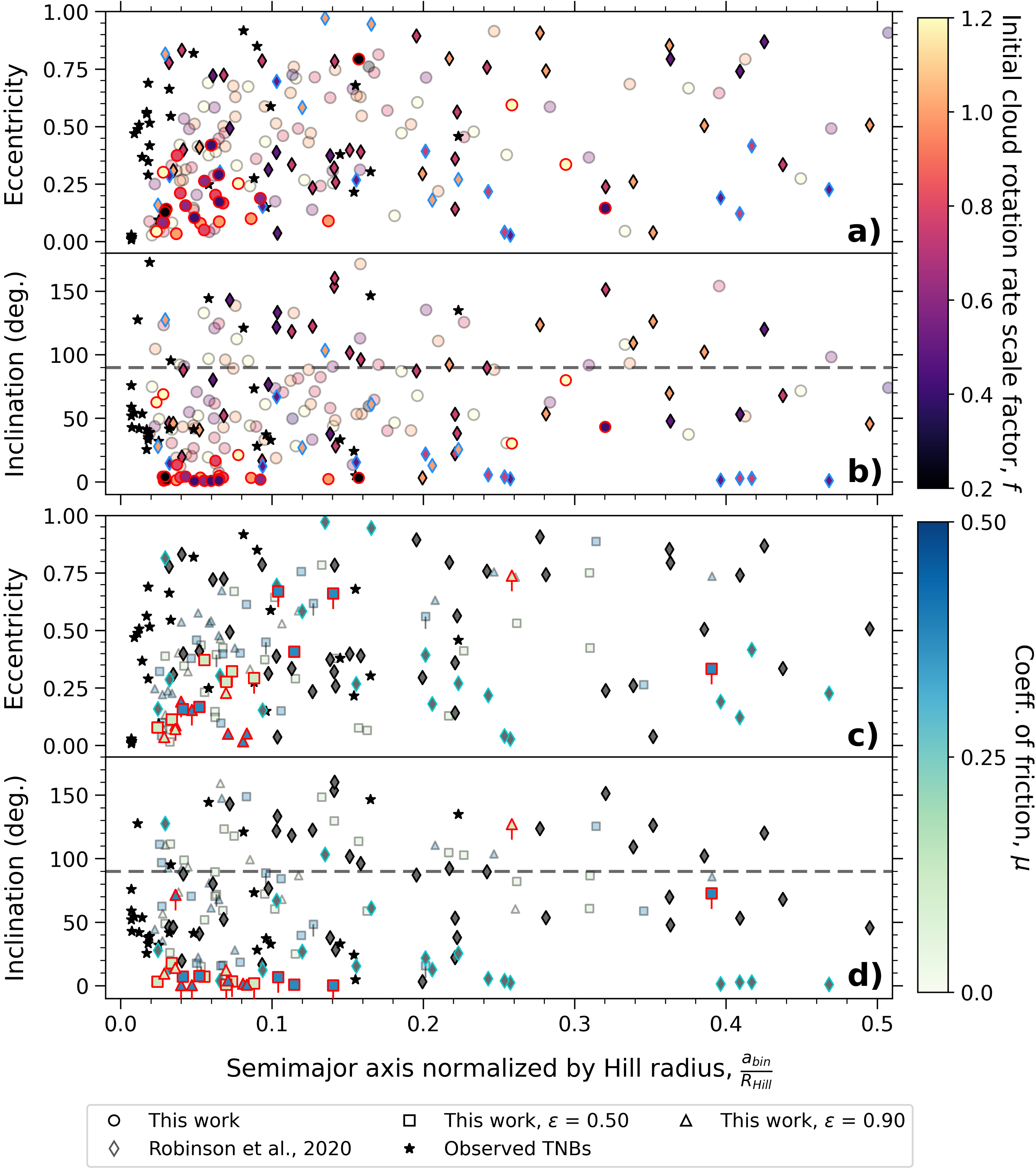}
    \hypertarget{fig:aRhill_all}{}
\end{figure*}

\renewcommand{\thefigure}{8 \emph{(previous page)}}
\begin{figure*}
    \hrulefill
    \caption{The eccentricity and inclination of binary planetesimals shown as a function of their semi-major axis normalized by their mutual Hill radius.
    Panels \textbf{a} and \textbf{b} show the suite of angular velocity simulations and panels \textbf{c} and \textbf{d} show the suite of contact physics simulations.
    For panels \textbf{a} and \textbf{b}, filling color indicates the initial cloud angular velocity.
    For panels \textbf{c} and \textbf{d}, different coefficients of restitution are indicated by shape: large squares ($\varepsilon=0.5$) and triangles ($\varepsilon=0.9$), while the filling color indicates the coefficient of friction $\mu$. 
    Since only the largest Kuiper Belt binaries are observable, we highlight the most massive simulated binary created in each SSDEM simulation with a red outline, and all other binary systems are shown as small squares and triangles with black outlines. 
    Trans-Neptunian binaries with full orbit solutions are shown as $\star$ \citep{Grundy2023}.
    The numerical results from \citet{Robinson2020} are shown as $\blacklozenge$, and to better compare with \citet{Robinson2020}, we outline in blue those binaries with mass accretion efficiencies $(m_1+m_2)/M_{\text{cloud}}> 0.1$ from their simulations.
    The numerical results from \citet{Robinson2020} are also colored according to their initial cloud angular velocity for panels \textbf{a} and \textbf{b}.
    In panels \textbf{c} and \textbf{d}, the \citet{Robinson2020} simulated binaries are still shown for reference but since these simulations assumed perfect merging, they have no defined coefficient of friction and so are colored gray.
    \label{fig:aRhill_all_caption} }
\end{figure*}
\setcounter{figure}{8}
\renewcommand{\thefigure}{\arabic{figure}}

The binary systems from SSDEM simulations match the overall size ratio distribution of observed relict binary planetesimals as well as those from prior simulations \citep{Noll2008,Nesvorny2010}.
In Figure~\ref{fig:cumul_radius_ratios_dist_all_planetesimals_comparison.png}, we show the cumulative distribution function of binary planetesimal radius ratios, from smallest to largest, compared to observed cold classical Kuiper Belt objects \citep{Noll2008} and other simulations \citep{Nesvorny2010}.
When the initial cloud angular rotation rate factor is $0.6 \leq f \leq 1.2$, the radius ratio cumulative distribution functions match both of the comparison distributions at higher radius ratios, $r_2/r_1 \gtrsim 0.75$.
Clouds with initial angular rotation rate factor $f = 1.2$ result in a greater fraction of unequal-sized binary systems.
At lower radius ratios, the observed classical Kuiper Belt object radius ratio cumulative distribution is likely underestimated due to the challenges of observing very faint secondaries orbiting about their much brighter primaries \citep[e.g.,][]{Showalter2021}.

\subsection{Binary Planetesimal Orbits}\label{sec:orbits}
Binary planetesimal systems created by gravitational collapse appear to possess a wide variety of mutual orbits, but some specific patterns are identifiable, see Figure~\hyperlink{fig:aRhill_all}{8}.
The mutual orbits of binary systems created within the SSDEM simulations either in the angular velocity suites (see Figure~\hyperlink{fig:aRhill_all}{8(a,\,b)}) or the contact physics suites (see Figure~\hyperlink{fig:aRhill_all}{8(c,\,d)}) are generally very similar to one another, but key differences appear when comparing to simulated binaries from previously published perfect-merger gravitational collapse models \citep{Robinson2020}.
Similarly, when we compare the simulated binary planetesimal systems to 41 trans-Neptunian binaries, which have been observed enough for the determination of full orbit solutions \citep{Grundy2023}\footnote{Data taken from the personal webpage of Dr. William Grundy, ``Mutual Orbits of Transneptunian Binaries" on January 10, 2025, \url{http://www2.lowell.edu/users/grundy/tnbs/orbits.html}, which is an updated dataset as described in \citet{Grundy2019}.}, we find a lot of similarity but also a few specific differences.

The most massive binary systems from each SSDEM simulation, i.e., those that form in the center of the gravitational well of the collapsing cloud, have noticeably tighter orbits than the most massive binary systems from the perfect merging simulations---in particular, compare the red outlined symbols from the SSDEM simulations to the blue outlined symbols from~\cite{Robinson2020} in Figure~\hyperlink{fig:aRhill_all}{8}.
The most massive binaries that formed in the perfect merging simulations have scaled semi-major axes between $0.02 \lesssim a/R_{\text{Hill}} \lesssim 0.5$ \citep{Robinson2020}, which stands in strong contrast with the most massive binaries formed in each of the SSDEM simulations which have $0.02 \lesssim a/R_{\text{Hill}} \lesssim 0.15$ with only a few massive binaries between $0.15 \lesssim a/R_{\text{Hill}} \lesssim 0.4$.
Notably, the 41 observed Kuiper Belt binaries with full orbits solutions also have $a/R_{\text{Hill}} \leq 0.16$ with one exception (2001 QW$_{322}$ at $a/R_{\text{Hill}} \simeq 0.22$), and many are much tighter $a/R_{\text{Hill}} < 0.02$ than any produced by numerical simulations \citep{Grundy2023}.
Even the smaller binaries produced from the SSDEM simulations are not as tight as the observed binaries.
This is surprising because binary softening due to flyby interactions with other Kuiper Belt objects has been hypothesized to be a dominant process that on average expands the primordial orbits of Kuiper Belt objects \citep{StoneKaib2021}.

The observed Kuiper Belt binaries show a wide distribution of mutual eccentricities from circular $e \sim 0.0$ to very elliptical $e \sim 0.9$ \citep{Grundy2023} similar to the wide distribution observed across all binaries in the perfect merging \citep{Robinson2020} and SSDEM simulations, as shown in Figure~\hyperlink{fig:aRhill_all}{8}.
However, the most massive binary systems from each SSDEM simulation were generally more circular $e \lesssim 0.5$ just like the most massive binaries from the perfect merging simulations.
Unlike the size of the orbits, the shape of the orbits is consistent with future evolution due to flyby interactions with other Kuiper Belt objects, which are expected to increase the eccentricity of binary systems \citep{StoneKaib2021}.

Simulated binaries formed within the perfect merging \citep{Robinson2020} and SSDEM simulations display a wide variety of inclinations just like the observed Kuiper Belt binaries \citep{Grundy2023} including many in retrograde mutual orbits, as shown in Figure~\hyperlink{fig:aRhill_all}{8}.
However, comparing the orientation of the orbit planes of the simulated and observed binaries is fraught because the Kuiper Belt inclinations are measured relative to the invariable plane of the solar system whereas the inclination of the simulated binaries are relative to the plane orthogonal to the angular momentum vector of the gravitationally collapsing cloud. 
It is likely that these two planes are nearly parallel, but unlikely that they are identically so.

\begin{figure}[t!]
    \centering
    \includegraphics[width=\columnwidth]{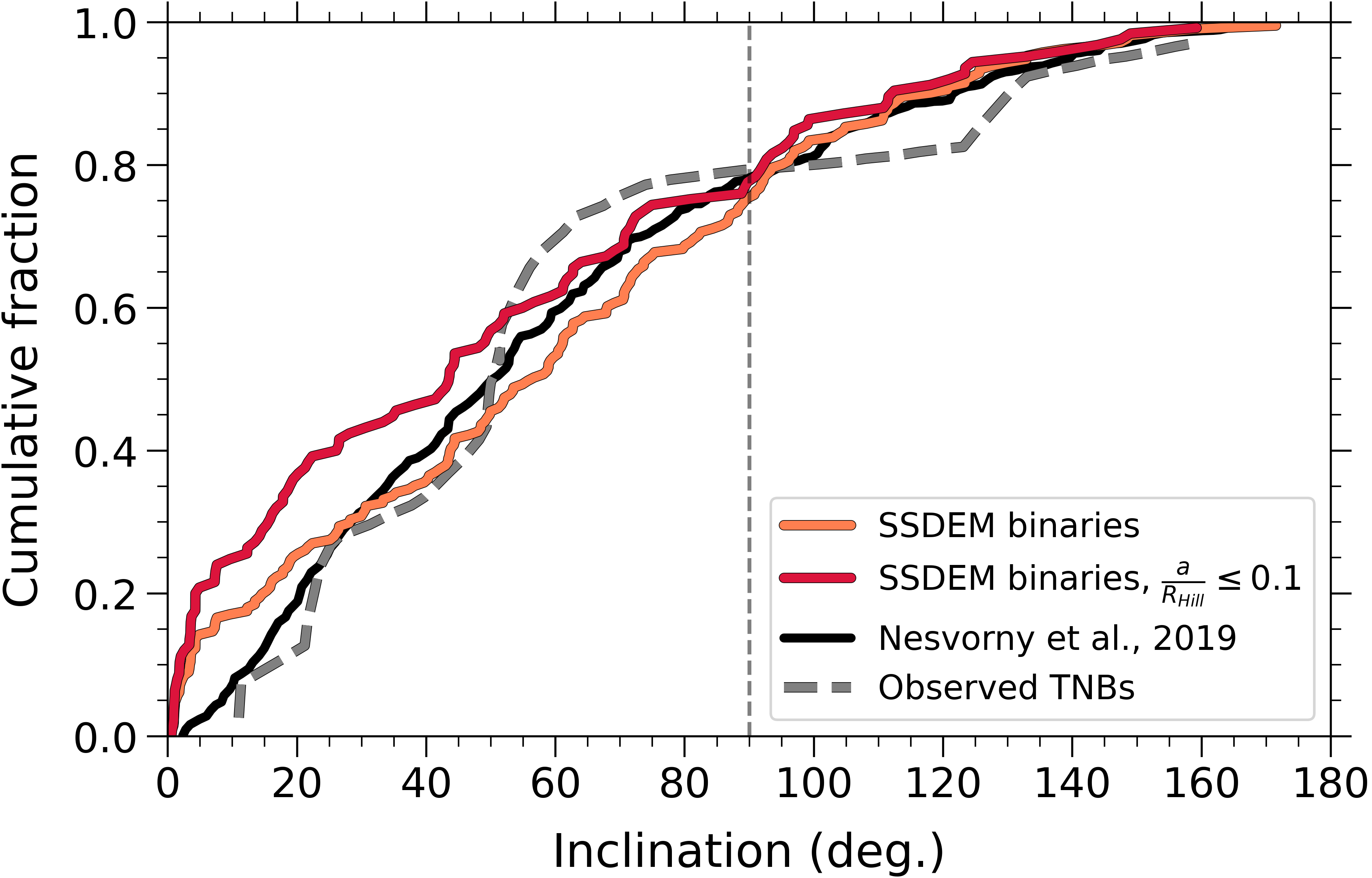}
    \caption{The cumulative fraction of binary planetesimals with an  inclination greater than the value on the abscissa.
    Results from the SSDEM simulations are colored red and orange, and they are compared to the streaming instability model from \citet{Nesvorny2019a} shown in black and the observed trans-Neptunian binaries from \citet{Grundy2019} shown as a gray dashed line.
    The orange line represents the entire SSDEM binary planetesimal dataset, whereas the red line shows the SSDEM binaries with a/$R_{Hill}\leq$~0.1, which is similar to the semi-major axis cutoff employed in \citet{Nesvorny2019a} for their binary systems and for the observed trans-Neptunian binaries from \citet{Grundy2023}.
    The vertical dashed line at 90$^{\circ}$ indicates the separation between prograde and retrograde orbits.}
    \label{fig:inclination_cdf}
\end{figure}

With that warning in mind, the most-massive binary system from each SSDEM simulation typically exhibits inclinations $i\leq$~20$^{\circ}$ (angular velocity suite) and $i\leq$~15$^{\circ}$ (contact physics suite). 
This is very similar to the largest systems of \citet{Robinson2020}, which are also characterized by inclinations very close to the orbit plane, i~$\lesssim$~20$^{\circ}$. 
However, there are a few binaries from SSDEM simulations exhibiting inclinations ranging from $i=20^{\circ}$ to approximately $i=80^{\circ}$.
Many of the most inclined binaries are from gravitational collapse simulations with the highest initial cloud rotation rate scale factor $f=1.2$.
In general, though, the most massive binaries formed from gravitational collapse generally conserve the orientation of the collapsing cloud's angular momentum vector.

All of the most-massive binaries in the angular velocity simulations remain on prograde orbits, and only one of the most-massive binaries in the contact-physics simulations is on a retrograde orbit.
When considering all simulated binaries, the fraction of systems with retrograde increases to approximately 25\%, see Figure~\ref{fig:inclination_cdf}.
The initial angular momentum of the cloud is preserved in the largest binaries but smaller binary systems do not preserve this bias.
Observed trans-Neptunian binary planetesimal systems also possess predominately prograde orbits with only about $\sim$20\% of trans-Neptunian binaries on retrograde orbits \citep{Grundy2023}.

While there is an overall good match of the distribution of binary inclinations between simulation and observation, specific binary sub-populations have distinct characteristics.
The most tightly ($a/R_{Hill}\leq$~0.1) bound binaries from the SSDEM simulations show an appreciable excess of very low inclination orbits relative to both the perfect merging simulations and observed trans-Neptunian binaries, see Figure~\ref{fig:inclination_cdf}.
Following a similar convention in categorizing planetesimals as in \citet{Robinson2020}, planetesimals are sorted into (1) \textit{observable binaries} with high mass ratios ($m_{1}+m_{2}/M_{\text{cloud}}\geq$ 0.075) and accretion efficiencies ($m_{2}/m_{1}\geq$ 0.075), (2) \textit{satellite systems} with significantly less-massive companions (10$^{-2}<m_{2}/m_{1}<$10$^{-3}$), and (3) \textit{atomistic binaries} with very little accretion $m_{1}+m_{2}/M_{\text{cloud}}\geq$10$^{-2.5}$.
The observable, satellite, and atomistic binaries exhibit 96\%, 78\%, and 68\% prograde orbits, respectively.
This again identifies a connection between the pebble cloud's initial angular momentum and the final binary angular momentum mediated by the mass of the binary components.

\begin{figure}[t!]
    \includegraphics[width=\columnwidth]{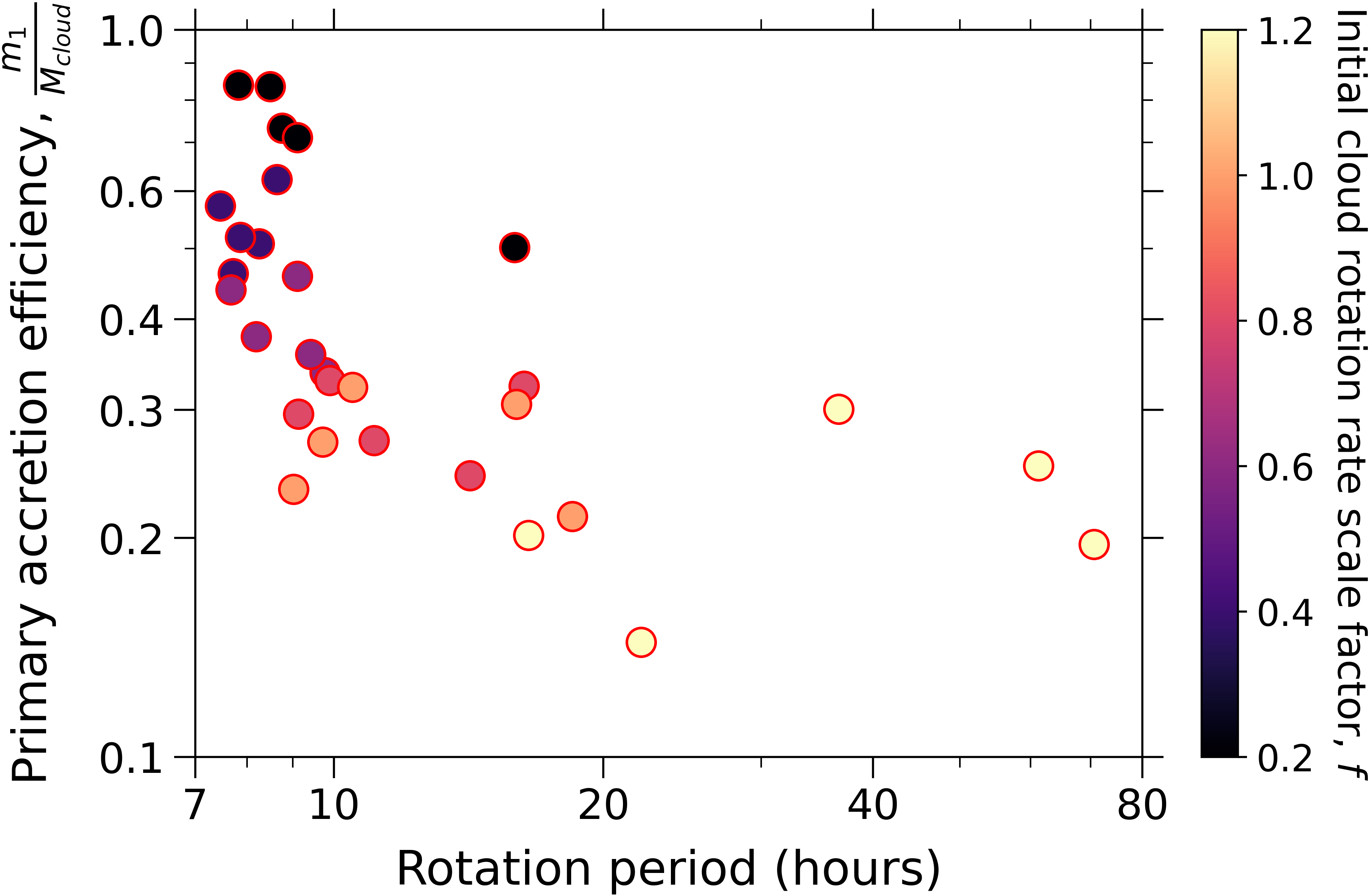}
    \caption{The most massive planetesimals from each simulation in the angular velocity suite with their rotation periods (in hours) plotted against their accretion efficiency.
    Planetesimals are colored according to their initial cloud rotation rates, as in Figs.~\ref{fig:planetesimalgrowth}~and~\ref{fig:mr_fig}.}
    \label{fig:spin_accr_eff_v_spin_largest_primary}
\end{figure}

\subsection{Planetesimal Spins}\label{sec:spins}
\begin{figure*}[t!]
    \includegraphics[width=\textwidth]{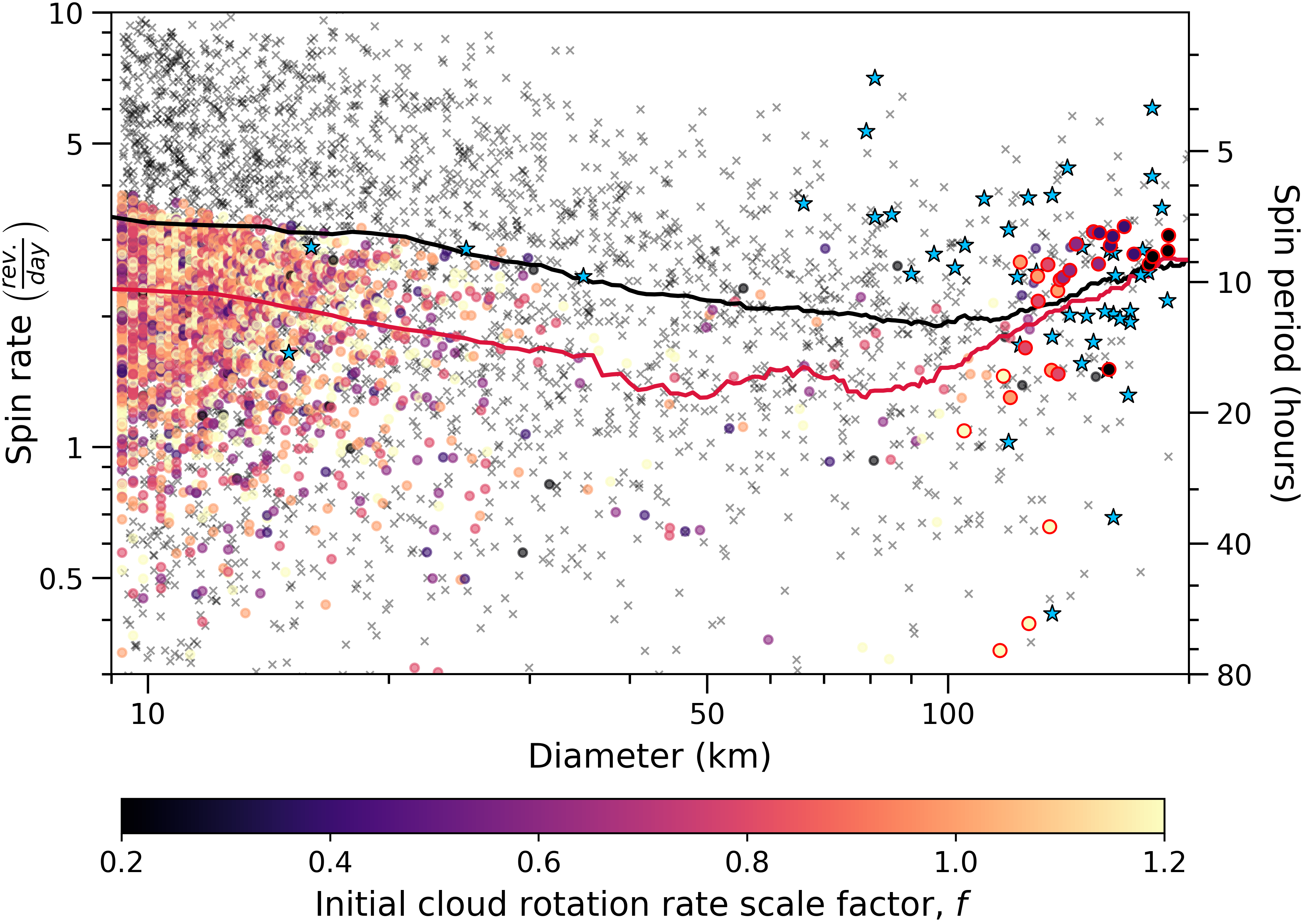}
    \caption{The planetesimal spin rate (revolutions per day) and diameter (km) for SSDEM simulated planetesimals ($\circ$) as well as for each asteroid in the MinorPlanet.info asteroid lightcurve database \citep[$\times$, ][]{Warner2009} and trans-Neptunian objects \citep[blue $\star$, ][]{Thirouin2010, Thirouin2014, Thirouin2018, Thirouin2019a, Thirouin2024, Vilenius2014}.
    The SSDEM simulated planetesimals are colored as in Figure~\ref{fig:mr_fig}, with the most massive planetesimal from each simulation outlined in red.
    A cut-off diameter of D~$\sim$~9~km has been implemented for both SSDEM planetesimals and asteroids, as this diameter is approximately equal to the 10-constituent-particle threshold for each planetesimal.
    A moving average is shown as a black line for the asteroid population and a red line for the SSDEM simulated planetesimal population.
    The moving averages for both asteroids and SSDEM planetesimals have associated moving window thresholds D$\sigma\leq$ D $\leq$ D/$\sigma$, where $\sigma = 0.75$.}
    \label{fig:diameter_v_spin}
\end{figure*}

Unlike perfect merging simulations, the SSDEM simulations produced planetesimals with resolved spin states, and so these spin states can be examined in detail and compared to observed relict planetesimal populations.
All of the most massive planetesimals from each simulation and the vast majority of the remaining planetesimals across all suites of simulations precess about their maximum moment of inertia axis, i.e., in short-axis mode (SAM) rotation.
Few are found to be in an excited state precessing about their minimum moment of inertia axis, i.e., in long-axis mode (LAM) rotation.
While collisions can be a source of rotational excitation, post-impact motion of internal material creates dissipation that removes excess spin energy leaving bodies in SAM and close to a fully relaxed spin state.
The angular velocity and contact physics simulation suites had nearly indistinguishable differences.
The mean period for the suite of angular velocity simulations is 12.9~hr ($\sigma=8.8$~hr) seen in Figure~\ref{fig:diameter_v_spin}, and the contact physics experiments had a slightly longer mean rotation period of 13.3~hr ($\sigma=8.5$~hr).
Moreover, there is little to no variation in rotation periods when comparing planetesimals created from the contact physics suite simulations with different coefficients of restitution and friction.

The most massive planetesimals from each SSDEM simulation show a clear direct trend in spin rate with their size and an inverse trend with the initial rotation rate of the collapsing pebble cloud, i.e., the slowest rotating clouds make the largest and fastest spinning planetesimals, as shown in Figure~\ref{fig:spin_accr_eff_v_spin_largest_primary}.
At lower initial cloud angular velocities ($f=0.2$--$0.6$) and higher primary accretion efficiencies ($m_1/M_{\text{cloud}}\geq0.35$), planetesimals exhibit the most rapid rotation periods, ranging between $7$--$10$~hr, on average.
A stark increase in primary rotation periods occurs when the initial cloud rotation rate scale factor $f>0.8$, and the average spin period increases beyond the 9~hr average of for initial cloud rotation rate scale factors $f= 0.2$--$0.6$, to a maximum of $\sim70$~hr for $f = 1.2$.
A similar trend is observed for planetesimals created in the contact physics suite when comparing clouds with $f=0.4$ and $f=1.0$.
That a high initial cloud angular velocity produces the slowest rotating most-massive planetesimals seems counter-intuitive, but more rapidly rotating initial clouds lead to less efficient growth of the most massive planetesimal, recall from Figure~\ref{fig:planetesimalgrowth}.
For the most massive planetesimals at the center of the collapsing cloud, the more mass that is accreted, then the more angular momentum is accreted as well, spinning them up. 
In other words, the initial cloud angular velocity determines the rotation periods for objects with the highest accretion efficiencies and which reside at the bottom of the local gravitational potential, as shown in Figure~\ref{fig:spin_accr_eff_v_spin_largest_primary}.

The spin rates of the largest simulated SSDEM planetesimals (diameters greater than 100~km) as a function of their size show patterns with remarkable similarities and differences to the trans-Neptunian object and Main Belt asteroid populations, as shown in Figure~\ref{fig:diameter_v_spin}.
First and foremost, the largest simulated SSDEM planetesimals spin over roughly the same range of periods $7$ to $70$ hours and with a very similar mean spin period as observed trans-Neptunian objects and Main Belt asteroids, which change with diameter but average approximately $10$ hours.
The SSDEM simulations do not reproduce the most rapidly rotating large relict planetesimals with spin periods $<7$ hours.
These largest trans-Neptunian objects and Main Belt asteroids are the relict planetesimals that best preserved their primordial spin states since there is only a narrow range of projectile sizes that could have significantly added or removed spin angular momentum while also not disrupting the body \citep{McAdoo1973,Harris1979}.
Even so, the spin states of these objects have certainly undergone some evolution over the last 4.5~Gyr \citep{Pravec2000,Pravec2002}, and this subsequent evolution may explain why the relict planetesimal population has rapid rotation periods ($<7$ hours) that are missing from the simulated SSDEM population.
On the other hand, this lack of rapid rotators could be revealing an unknown factor in the understanding or modeling of gravitational collapse, so it should be a focus of future work with SSDEM-like tools that can resolve rotation.
No previous modeling effort could test this aspect of the gravitational collapse planetesimal formation hypothesis, and, in general, it is another significant demonstration of the consistency between the hypothesis and observation.

The spin states of simulated SSDEM planetesimals do not consider the consequences of 4.5 Gyr of collisional evolution, so direct comparisons of smaller relict planetesimals (diameters less than 100~km) can be made only from the cold classical region of the Kuiper Belt, where relict planetesimals have not undergone significant collisional evolution.
When the New Horizons spacecraft flew by 486958 Arrokoth, a cold classical object with a diameter of about 20 km, it confirmed an object with limited collisional evolution \citep{Stern2019,Grundy2020,Lisse2021}.
The spin periods of the simulated SSDEM planetesimals at the size of Arrokoth range from about 6.5 to 60~hr, encompassing the current spin period of Arrokoth, which is about 16~hr \citep{Keane2022}.
Observed spin periods for other trans-Neptunian objects with diameters less than 100 km, which are from more collisionally evolved sub-populations of the trans-Neptunian region, range between 3 and 10~hr \citep{Thirouin2010, Thirouin2014, Thirouin2018, Thirouin2019a, Thirouin2024, Vilenius2014}.
These spin rates are generally higher than the simulated SSDEM planetesimal population, as shown in Figure \ref{fig:diameter_v_spin}.
This difference likely has its origin in two parts.
First, slow rotators require a lot more observing time and so establishing slow rotation is harder \citep{Binzel1989}, and, second, subsequent collisional evolution may have accelerated the rotation rates of these more collisionally evolved trans-Neptunian objects.

Comparing the spin period distribution of the simulated SSDEM planetesimals and relict planetesimals from the Main Belt is complicated by the likelihood that significant rotational evolution has occurred amongst the observed asteroids.
The spin rate of smaller (diameters less than 100~km) simulated SSDEM planetesimals generally increase with decreasing diameter.
Similarly, so do the spin periods of relict planetesimals in the Main Belt, however the simulated SSDEM planetesimals spin more slowly, on average by about 5 hours, than the modern asteroid population \citep{Pravec2002,Warner2009}, as shown in Fig~\ref{fig:diameter_v_spin}. 
Despite this difference, the moving averages of the SSDEM and asteroid population spin periods decrease at a similar rate as a function of diameter between 10 to 100~km.
At diameters larger than 100~km, the mean spin periods of the SSDEM planetesimals and the modern asteroids are very similar.
At all sizes, there are more rapidly rotating large asteroids than large SSDEM planetesimals, however.
At smaller sizes, this deficit is more extreme with asteroids as small as 10~km in diameter exhibiting spin periods up to 2.4~hr, but the fastest SSDEM planetesimals rotate at greater than 6~hr.
The distribution of spin rates of these asteroids have relaxed to a Maxwellian, which is expected for a collisionally evolved population \citep{Harris1979}, and, likely, many, if not most, are the fragments of disrupted larger asteroids, whether from identified families or not \citep{Bottke2015}.
While the properties of the equilibrium Maxwellian spin distribution are determined by the velocities and geometries of the collisions \citep{Salo1987}, it is noteworthy that the SSDEM simulations of gravitational collapse generally produce much slower rotating objects, so these models suggest that the history of collisional evolution of Main Belt asteroids includes a general acceleration of their spin rates, which has implications for maintaining shapes and surface structures.

\begin{figure}
    \centering
    \includegraphics[width=\columnwidth]{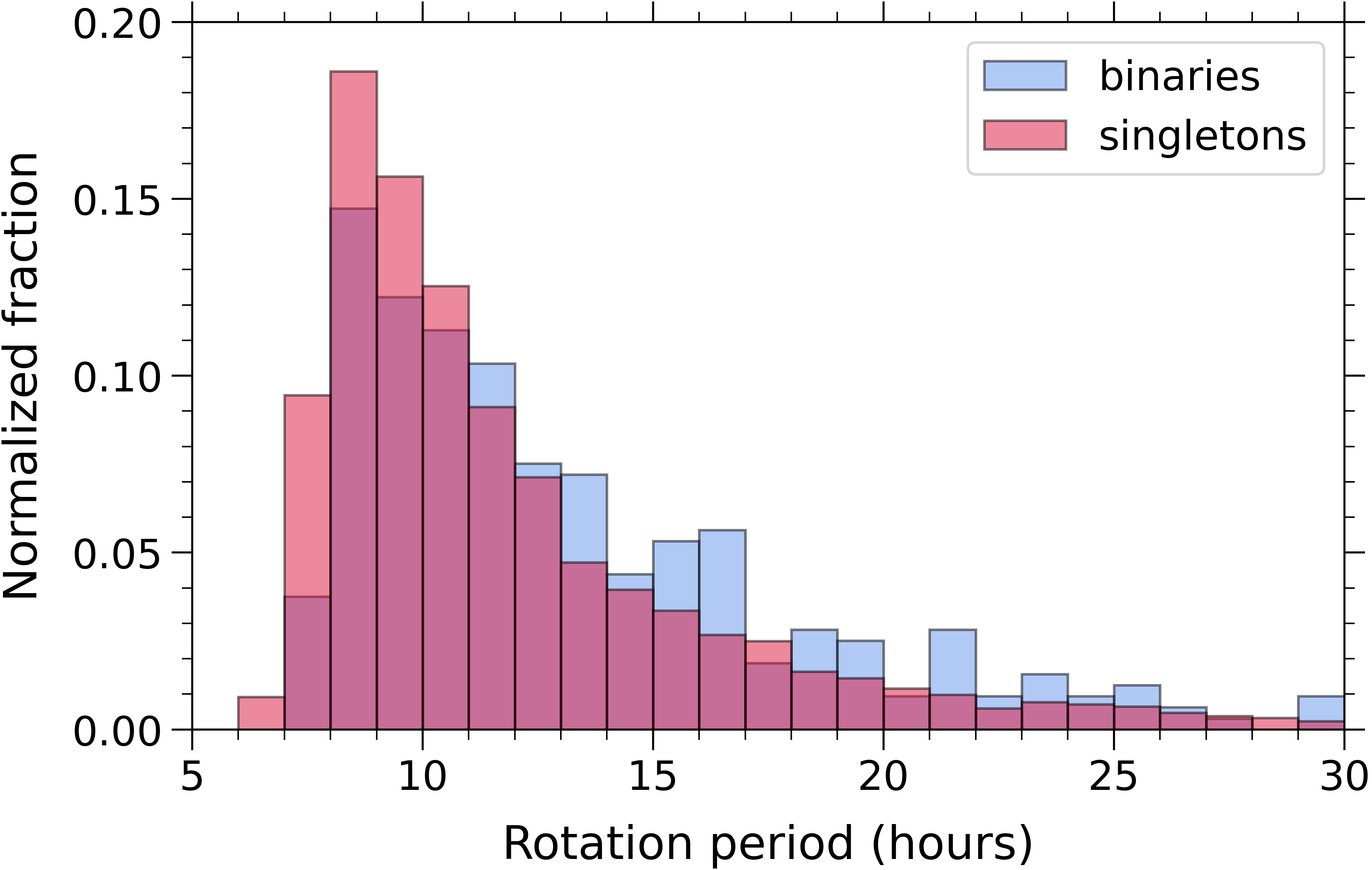}
    \caption{The rotation periods of binary planetesimal systems (blue) or singleton planetesimals (red) relative to their normalized fraction for the angular velocity and contact physics simulation suites.
    Only rotation periods less than 30 hours are examined above, as the vast majority of the SSDEM planetesimal rotations are below this threshold.}
    \label{fig:Maxwellian_spins}
\end{figure}
    
Additionally, there is a small but noticeable difference between the distributions of rotation periods for binary and non-binary trans-Neptunian objects.
Observed planetesimals in binary systems rotate markedly slower (10.1~hour mean) than non-binaries (8.4~hour mean) for an average difference of approximately 1.7~hours \citep{Thirouin2014}.
Planetesimals created directly from gravitational collapse with the SSDEM exhibit a similar preference for slow and more-rapid rotations for binary (15.8~hour mean) and non-binary (12.7~hour mean) planetesimal systems, respectively, with a slightly greater but comparable difference in means of 3.1~hours (Figure~\ref{fig:Maxwellian_spins}).
The distributions of modeled spins from both the angular velocity and contact physics suites may indicate a primordial preference for spins among binary and solitary planetesimals that are enhanced over Myr to Gyr timescales via collisions, tidal de-spinning, or some other mechanism \citep[e.g.,][]{Nesvorny2020a}.

\subsection{Planetesimal Shapes}\label{sec:shapes}
These SSDEM simulations are the first that can measure the shape of planetesimals formed from gravitational collapse, and the formed planetesimals are created with a wide variety of shapes.
To assess the physical shapes of the SSDEM planetesimals, we examined the created planetesimals by eye from the angular velocity and contact physics simulation suites.
We instituted a size cut-off of $50$ particles for the planetesimals that we would classify by shape, as it was difficult to discern the shapes of planetesimals made with fewer particles.
The following results are based on the $793$ planetesimals from both simulation suites above this cut-off.

\begin{figure*}[t!]
    \centering
    \includegraphics[width=\textwidth]{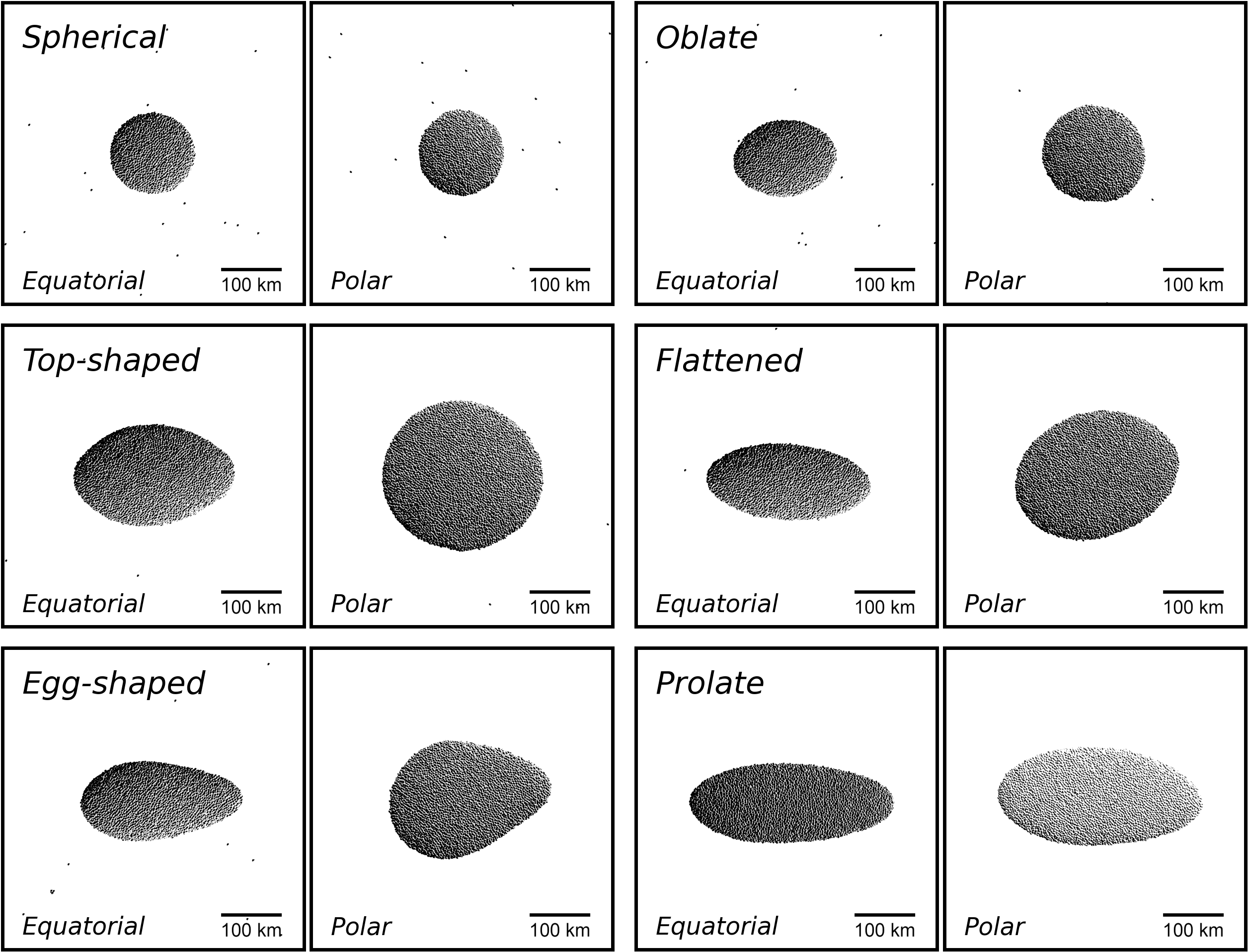}
    \caption{Each example planetesimal is shown with views along the equator (left panels) as well as from above the pole (right panels).
    The example planetesimal shapes include spherical (first panel), oblate (second panel), top-shaped (third panel), flattened (fourth panel), egg-shaped (fifth panel), and prolate (sixth panel).
    These planetesimals are only examples of their denoted shapes, and there is variation about these representative shapes.}
    \label{fig:shapes_composite_1}
\end{figure*}

We identified by eye six types of planetesimal shapes: spherical, oblate, top-shaped, flattened, egg-shaped, and prolate, with examples shown in Figure~\ref{fig:shapes_composite_1}.
Spherical, oblate, and flattened planetesimals form a continuum of objects with increasingly short semi-axes parallel to their spin axes but similar length equatorial semi-axes.
Flattened planetesimals often show some limited prolateness whereas prolate objects are truly tri-axial planetesimals.
Top-shaped planetesimals have a more prominent equatorial ridge than spheres or oblate-spheres, and egg-shaped planetesimals have an effectively out-of-balance appearance such that one half of the planetesimal appears oblate but the other half bulges out and terminates with a point similar to prolate planetesimals.
All planetesimals were sorted by eye into one of these six broad categories, but it is imaginable that many more specialized categories could be developed with more quantitative criteria that go beyond the tri-axial analysis introduced later, but this is left for future work.

\begin{table}[t!]
\noindent
\begin{center}
\begin{tabularx}{0.45\columnwidth}{cc}
\multicolumn{2}{@{}M{0.45}}{\textit{\textbf{\shortstack{Planetesimal \\ shape frequency}}}} \\
\toprule
\textit{spherical}   &  $41\pm3\%$    \\   %
\textit{oblate}      &  $33\pm3\%$    \\   %
\textit{egg-shaped}  &  $9\pm2\%$     \\   %
\textit{prolate}     &  $8\pm2\%$     \\   %
\textit{top-shaped}  &  $6\pm1\%$     \\   %
\textit{flattened}   &  $3\pm1\%$     \\   %
\bottomrule
\end{tabularx}
\end{center}
\caption{The frequency of planetesimal shapes created with the SSDEM across the angular velocity and contact physics simulation suites, in decreasing order.
Averages are shown with 1~$\sigma$ standard errors calculated using the score confidence interval \citep{Agresti1998}.}
\label{table:shape_stats}
\end{table}

The shape-by-eye statistics across the entire population are provided in Table~\ref{table:shape_stats}, and the series of shapes with similar length equatorial semi-axes (i.e., spherical, oblate, top-shaped, and flattened) are the vast majority $\sim$83\% of the population.
Indeed, spherical and oblate planetesimals by themselves constitute $\sim$74\% of all planetesimals.
Moreover, despite the diversity of shapes across the entire population of simulated planetesimals, the most massive planetesimals in each simulation are typically characterized as either near-spheres or oblate spheroids.

\begin{figure}[t!]
    \includegraphics[width=\columnwidth]{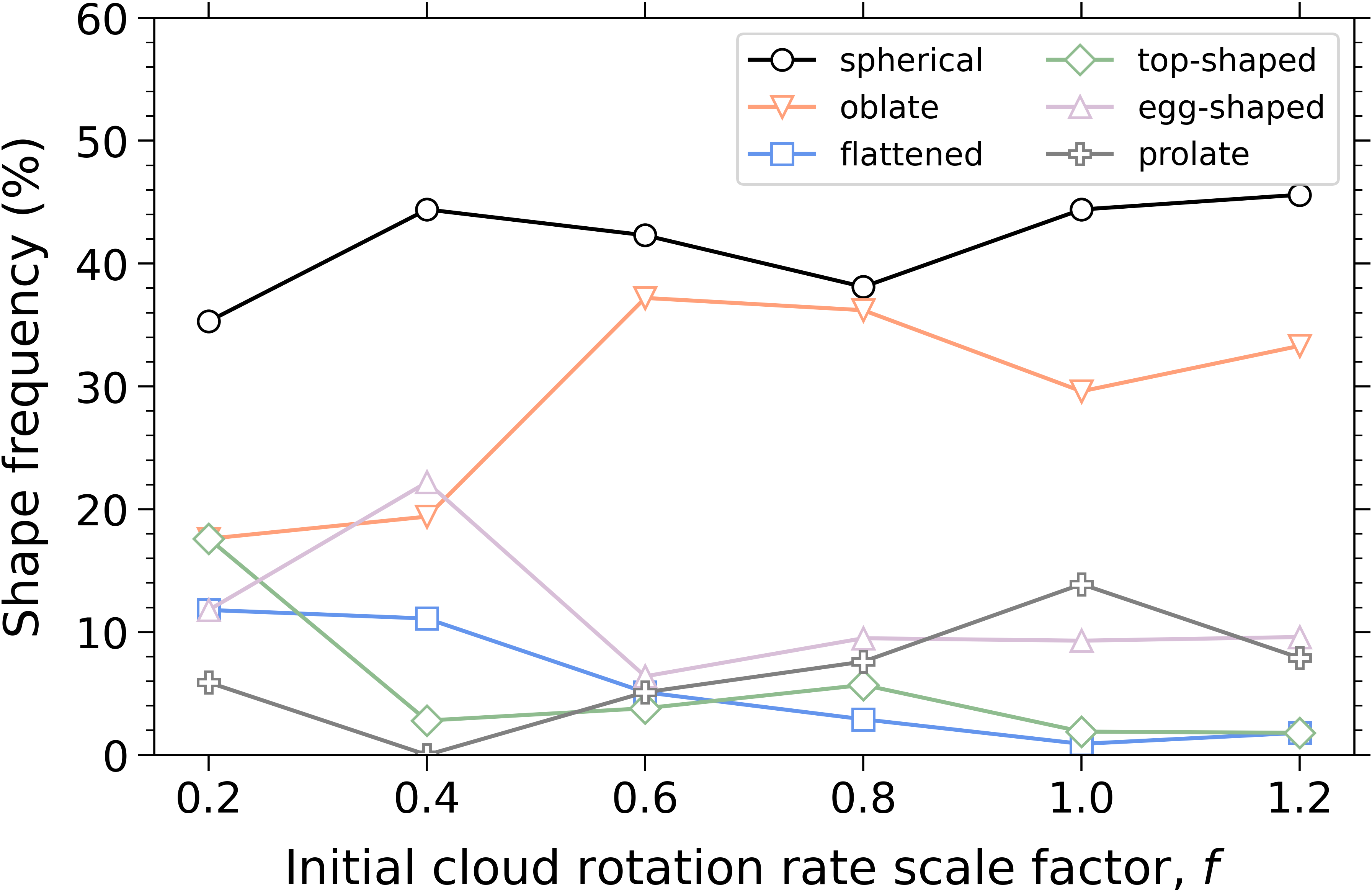}
    \caption{The frequency of planetesimal shapes across all simulated initial cloud rotation states.
    Shapes are denoted in the following manner: spherical (black $\circ$), oblate (red $\nabla$), flattened (blue $\square$), top-shaped (green $\diamond$), egg-shaped (purple $\triangle$), prolate (gray $\Plus$).}
    \label{fig:shape_stats_ang_vel_by_sim}
\end{figure}

The distribution of planetesimal shapes is linked to the rotation of the overall collapsing gravitational cloud, since the frequency of specific planetesimal shapes vary as a function of the initial cloud rotation rate scale factor $f$ within the angular velocity suite, as shown in Figure~\ref{fig:shape_stats_ang_vel_by_sim}.
Recall that there is an inverse relationship between the spin rate of the largest planetesimals and the initial cloud rotation rate scale factor, see Section~\ref{sec:spins} and Figure~\ref{fig:spin_accr_eff_v_spin_largest_primary}.
For instance, flattened, top-shaped, and egg-shaped planetesimals form at highest rates from collapsing clouds with initially slower rotating rates, which are associated with more rapidly rotating planetesimals.
Oblate and prolate planetesimals are created infrequently in clouds with slow initial rotation rate scale factors, and their rate of formation increases as the initial cloud rotation rate scale factor increases associating these shapes with relatively slower rotating planetesimals.
Lastly, spherical planetesimals form with the highest, and nearly constant, frequency across all simulations.

\begin{figure}[t]
    \includegraphics[width=\columnwidth]{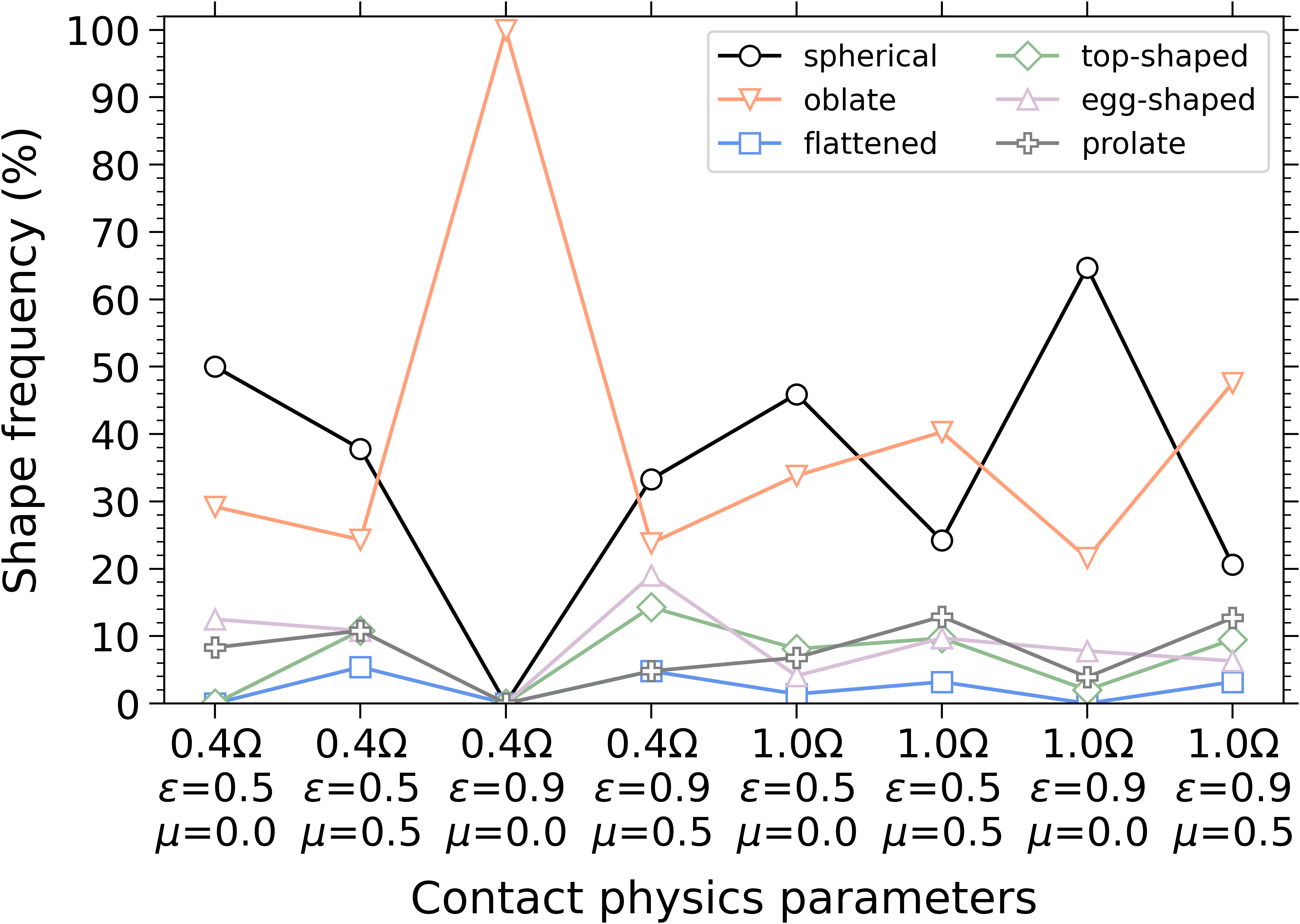}
    \caption{The frequency of planetesimal shapes across all simulated initial cloud contact physics.
    Shapes are denoted in the following manner: spherical (black $\circ$), oblate (red $\nabla$), flattened (blue $\square$), top-shaped (green $\diamond$), egg-shaped (purple $\triangle$), prolate (gray $\Plus$).}
    \label{fig:shape_stats_contact_by_sim}
\end{figure}

When considering the shapes of planetesimals created from the contact physics suite of simulations, similar patterns in planetesimal shape frequency were observed between clouds with $f=0.4$ and $1.0$, as shown in Figure~\ref{fig:shape_stats_contact_by_sim}. 
As in the angular velocity suite of simulations, spherical and oblate planetesimals form with the greatest frequency for all cloud rotations and contact physics parameters.
Looking more closely, prolate planetesimals preferentially form from simulations with initially rapidly rotating clouds and larger coefficients of friction, and top-shaped planetesimals form most readily from clouds with slow rotations and large coefficients of friction.

For the most part, the other contact physics suite simulations shared a similar planetesimal shape frequency as those in the angular velocity suite, but there is a striking difference.
Notably, collapsing clouds with an initial rotation rate factor $f=0.4$ and contact physics parameters characterized by a large coefficient of restitution $\varepsilon=0.9$ and no friction $\mu=0.0$ appear to only make oblate planetesimals.
However, this is because each of the three simulations with these parameters produced only one planetesimal with a particle number greater than the 50-particle cut-off for assessing shapes.
As noted in Section \ref{sec:multiplicity}, only one massive planetesimal formed from each collapsing cloud because a large coefficient of restitution with a low coefficient of friction often results in bouncing binary systems that require long timescales to finish bouncing and accrete into a single object.
The planetesimals that form from these clouds contain a substantial fraction of the initial cloud mass, and the lack of resistance to motion and rapid rotation result in a less spherical and more oblate shape.

Many observed relict planetesimals only have published tri-axial shapes, but these simple shapes combined with spin can reveal information about internal structure.
In order to quantitatively assign a tri-axial shape to each SSDEM planetesimal, we use a moment of inertia analysis.
The moment of inertia tensor was calculated for each planetesimal from its constituent particles.
We assumed a constant bulk density of 0.66 g cm$^{-3}$ to approximately account for the packing efficiency of the 1 g cm$^{-3}$ spherical constituent particles.
Then, we calculated the corresponding semi-axis lengths $a$, $b$, and $c$ of a tri-axial ellipsoid from the inertial tensor such that $a \geq b \geq c$.
The principal moment semi-axes of each planetesimal then define the semi-axes of a tri-axial ellipsoid describing an idealized shape for the body.
These tri-axial ellipsoid shapes can then be characterized according to their oblateness $\gamma = 1 - c/a$ and their prolateness $\beta = 1 - b/a$.
Since we report all results in terms of these scaled shape factors, the density choice above is unimportant.

\begin{figure*}[t!]
    \centering
    \includegraphics[width=\textwidth]{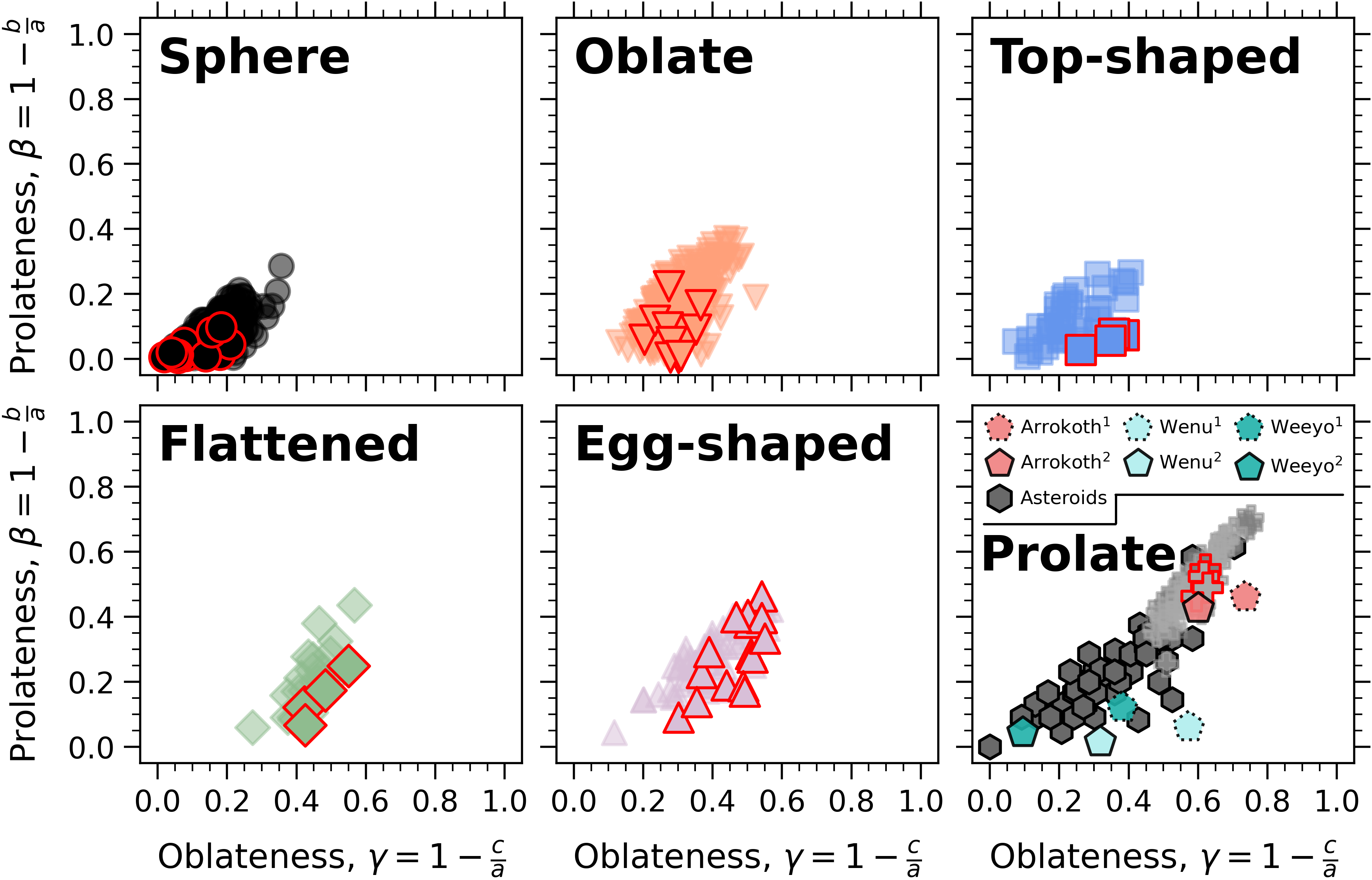}
    \caption{Each panel shows the prolateness $\beta = 1 - b/a$ and oblateness $\gamma = 1 - c/a$ of the $793$ planetesimals from the SSDEM simulations, organized by shape as in Figs.~\ref{fig:shapes_composite_1}, \ref{fig:shape_stats_ang_vel_by_sim}, and \ref{fig:shape_stats_contact_by_sim}.
    Each planetesimal's shape is described as a tri-axial ellipsoid that corresponds to the same moment of inertia tensor as the planetesimal assuming a density of 1 g cm$^{-3}$ and semi-axis lengths $a$, $b$, and $c$ such that $a \geq b \geq c$.
    The most massive planetesimals from each simulation are indicated with larger markers and red outlines.
    For comparison, the shapes of asteroids are shown as gray hexagons \citep{Kryszczynska2007} and the shape of 286958 Arrokoth including lobes Wenu and Weeyo ($^{1}$: \citet{Keane2022}; $^{2}$: \citet{Porter2024b}) are shown as colored pentagons.
    All of these relict planetesimal shapes are shown in the prolate panel, because that is where they overlap with the simulated planetesimals the least.
    These relict planetesimals overlap substantially with the other shapes.}
    \label{fig:axes_combined}
\end{figure*}

Corroborating the shape identification by eye, different shapes have  distinct inertial tri-axial ellipsoid shapes, although there is particular overlap between the sphere, oblate, and top-shaped planetesimals, as shown in Figure~\ref{fig:axes_combined}.
Of course, the sphere planetesimals have about $\beta \sim \gamma \sim 0$ but the prolateness of sphere planetesimals can be as high as $\beta \lesssim 0.2$ and the oblateness $\gamma \lesssim 0.2$.
Oblate and top-shaped planetesimals trend toward larger values of oblateness on average, with oblate planetesimals ranging from $0.2 \lesssim \gamma \lesssim 0.4$ and top-shaped planetesimals ranging from $0.2 \lesssim \gamma \lesssim 0.4$---what distinguishes these two by-eye shapes from each other is an equatorial ridge feature that is not captured by the tri-axial moment of inertia approach.
Flattened planetesimals have even more oblateness $0.4 \lesssim \gamma \lesssim 0.6$ but have a similar limited prolateness $\beta \lesssim 0.4$ to the sphere, oblate, and top-shaped planetesimals.
Again, the egg-shaped planetesimals' most distinguishing features are not well-captured by a tri-axial ellipsoid moment of inertia model, but they can be more prolate $0.1 \lesssim \beta \lesssim 0.5$ than the flattened shaped planetesimals with a similar range of oblateness $0.3 \lesssim \gamma \lesssim 0.6$.
Naturally, the prolate shaped planetesimals are the most prolate $0.4 \lesssim \beta \lesssim 0.7$ and possess even more extreme levels of oblateness $0.5 \lesssim \gamma \lesssim 0.8$.

Notably, the shapes of the most massive planetesimals from each of the gravitational collapse simulations span all six identified shapes and a wide variety of prolateness $\beta$ and oblateness $\gamma$ values.
The most massive planetesimals are generally symmetric about the longest $a$ and intermediate $b$ axes as indicated by their corresponding low prolateness ratios.
Interestingly, the most massive top-shaped, flattened, and egg-shaped planetesimals from each simulation possess the most oblateness relative to other less-massive planetesimals within the same shape category, as shown in Figure~\ref{fig:axes_combined}.
The most massive prolate planetesimals, however, do not exhibit either the least or the most extreme prolate shapes.

The simulated planetesimals have shapes that are comparable to the observed shape distributions among large objects in the asteroid belt \citep{Kryszczynska2007} and to Arrokoth, including its individual lobes Wenu and Weeyo \citep{Keane2022,Porter2024b}, as shown in the final subfigure of Figure~\ref{fig:axes_combined}.
The most extreme prolate planetesimals formed in the SSDEM simulations are rare in the asteroid belt, and most asteroid shapes are consistent with the other shape groups.

The shapes observed among the simulated planetesimals reflect both their assembly history and the shape-limits imposed by the material properties of the assembled bodies.
When considering gravitational and centrifugal accelerations, as the rotation of the body increases, the minimum energy shape of a fluid body follows the long-established Maclaurin spheroid and Jacobi ellipsoid sequence.
These shapes are equilibrium solutions for materials with fluid properties, i.e., cannot support shear stresses and so have an internal angle of friction $\phi=0^{\circ}$.
Alternatively, if the internal structure of the planetesimal can support shear stresses, i.e, an internal angle of friction $\phi>0^{\circ}$, then the simulated planetesimals can instead take a wider range of shapes \citep{Holsapple2001,Harris2009}.

The Mohr–Coloumb yield criteria is a condition for plastic failure often used when considering granular media including rubble pile asteroids, and it can be used to define a limiting set of shapes for a given internal angle of friction $\phi$ \citep{Holsapple2001}.
Here, we assert that this criteria is a reasonable starting place for the analysis of planetesimal shapes assembled via the aggregation of pebbles from a gravitationally collapsing cloud.
The Mohr–Coloumb yield criteria can be written:
\begin{equation}
    \left( \sigma_1 - \sigma_3 \right) \sqrt{1 + \tan^2 \phi} + \left( \sigma_1 + \sigma_3 \right) \tan \phi \leq 2 Y = 0
\end{equation}
where $\sigma_1 \geq \sigma_2 \geq \sigma_3$ are the sorted principal stresses, and $Y$ is the cohesion \citep{Holsapple2001}.
Note that $\sigma_3$ is the compressive stress with the highest absolute magnitude and $\sigma_1$ has the least absolute magnitude.
For this analysis, these bodies are presumed to possess only compressive strength and to not possess tensile strength, so no significant cohesion $Y=0$ and all principal stresses must be in compression $\sigma < 0$.
The Mohr-Coloumb criterion can be re-arranged to determine the minimum internal friction angle $\phi$ required to support the tri-axial shape of the body:
\begin{equation}
    \tan \phi \geq \frac{\frac{\sigma_1}{\sigma_3} - 1}{2 \sqrt{\frac{\sigma_1}{\sigma_3}}}
\end{equation}
We note that in the above condition only the ratios of principal stresses appear, which means that the common stress factor $\sigma_0$, defined below and which contains the common dependence on location within the tri-axial body, is always eliminated, even though which principal stress is the largest and smallest is not always the same \citep{Holsapple2001}.
In the case of a non-rotating tri-axial body, $\sigma_c \geq \sigma_b \geq \sigma_a$, so the largest $\sigma_1 = \sigma_c$ and smallest $\sigma_3 = \sigma_a$ principal stresses are always the same for non-rotating bodies.
On the other hand, when considering rotating bodies, which principal stress is largest and smallest depends strongly on the specific rotation rate.

\begin{figure*}[t]
    \centering
    \includegraphics[width=\textwidth]{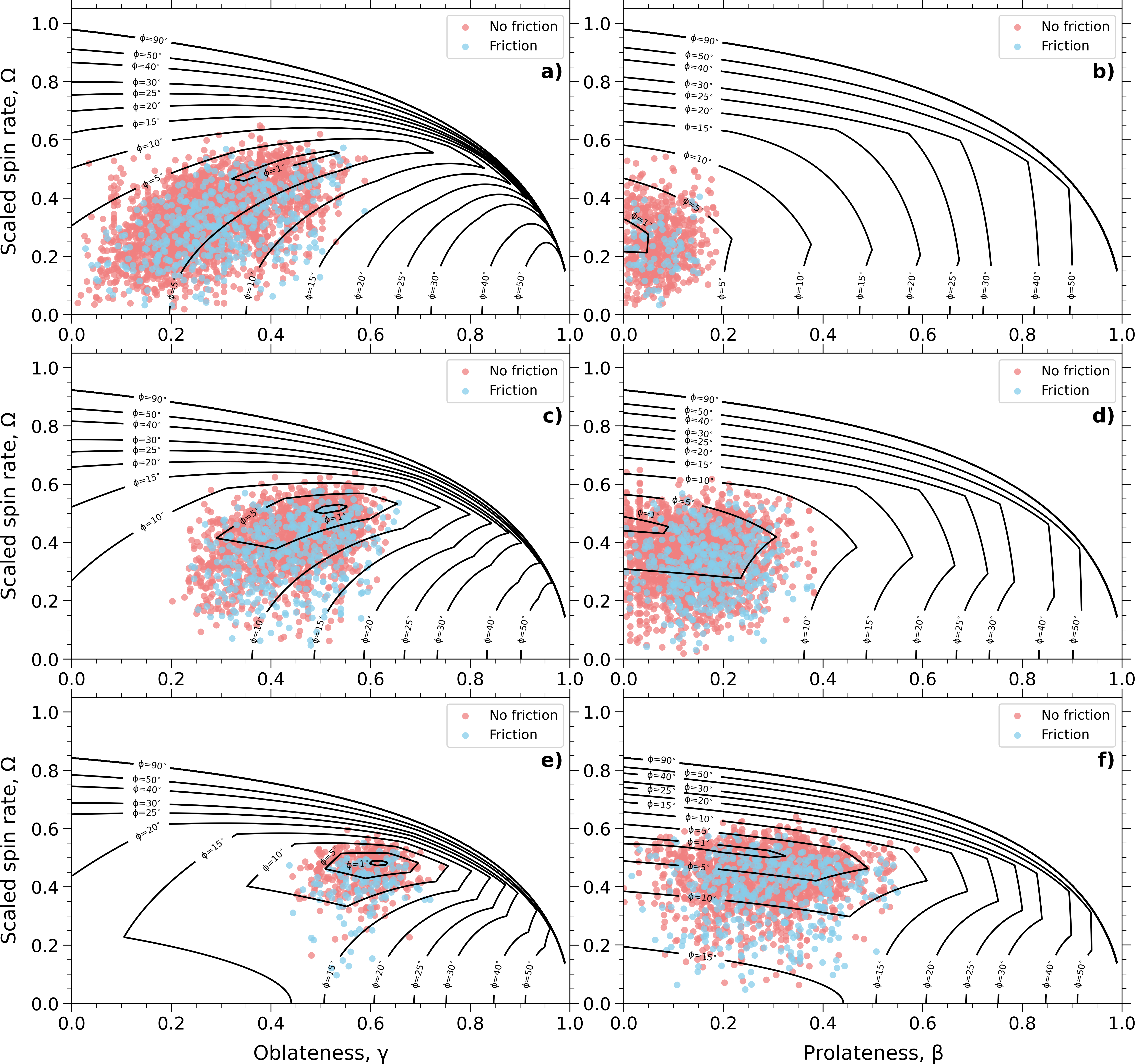}
    \caption{The scaled spin rates $\Omega = \omega/\omega{_d}$, the oblateness $\gamma = 1 - c/a$ (left column), and prolateness $\beta = 1 - b/a$ (right column) of simulated planetesimals created in the angular velocity or the contact physics simulation suites.
    Planetesimals created in simulations that did not include friction ($\mu=0$) are colored red, and those that did include friction ($\mu>0$) are colored blue.
    Contours show constant internal angles of friction $\phi$ indicating the minimum angle required to support that shape at that spin rate \citep{Holsapple2001}.
    The sub-panels are organized to show the consequence of varying oblateness while prolateness is held approximately constant (left column) and varying prolateness while oblateness is held approximately constant (right column).
    Each sub-panel shows a subset of the simulated planetesimals with specific tri-axial ellipsoid shapes: (\textbf{a}) $0.0<\beta<0.2$, (\textbf{b}) $0.0<\gamma<0.2$, (\textbf{c}) $0.2<\beta<0.4$, (\textbf{d}) $0.2<\gamma<0.4$, (\textbf{e}) $0.4<\beta<0.6$, and (\textbf{f}) $0.4<\gamma<0.6$.
    Correspondingly, in each sub-panel, the contours of minimum internal angles of friction are calculated asserting that: (\textbf{a}) $\beta=0.1$, (\textbf{b}) $\gamma=0.1$, (\textbf{c}) $\beta=0.3$, (\textbf{d}) $\gamma=0.3$, (\textbf{e}) $\beta=0.5$, and (\textbf{f}) $\gamma=0.5$.}
    \label{fig:contour_combined}
\end{figure*}

From an tri-axial inertial shape and a spin state, we calculate the principal normal stresses $\sigma_a$, $\sigma_b$, and $\sigma_c$ at location $(x,y,z)$ within the body in the direction of each of the orthogonal semi-axes of length $a$, $b$, and $c$, respectively:
\begin{align}
    \sigma_a & = - \sigma_0 \Bigl( E_a - \Omega^2  \Bigr)   \\
    \sigma_b & = - \sigma_0 \left( E_b - \Omega^2\left(1 - \beta \right)^{-2}  \right) \\
    \sigma_c & = - \sigma_0 E_c \\
    \sigma_0 & = \frac{2 a^2 \rho^2 G}{3} \left( 1 - \left( \frac{x}{a} \right)^2 - \left( \frac{y}{b} \right)^2 - \left( \frac{z}{c} \right)^2 \right) \\
    \Omega & =  \frac{\omega}{\omega_d} \leq 1 \\
    \omega_d & = \sqrt{ \frac{4 \pi \rho G}{3} } \\
    E_a & = \frac{3}{2}  \int_0^\infty \frac{\left(1 - \beta \right) \left(1 - \gamma \right)  }{\Delta \left(s + 1 \right) } ds \leq 1 \\
    E_b & = \frac{3}{2}  \int_0^\infty \frac{\left(1 - \beta \right)^3 \left(1 - \gamma \right)  }{\Delta (s + \left(1 - \beta \right)^2 ) }ds \leq 1 \\
    E_c & =  \frac{3}{2} \int_0^\infty \frac{\left(1 - \beta \right) \left(1 - \gamma \right)^3 }{\Delta (s + \left(1 - \gamma \right)^2 )}ds \leq 1 \\
    \Delta & = \sqrt{ \Bigl(s  + 1 \Bigr)\left(s + \left(1 - \beta \right)^2 \right)\left(s + \left(1 - \gamma \right)^2 \right)}.
\end{align}
where $\sigma_0$ is a common stress factor between the principal normal stresses, which also contains the common dependence on location within the tri-axial body, $\rho$ is the body density, $G$ is Newton's gravitational constant, $\Omega$ is the dimensionless scaled spin rate, $\omega$ is the spin rate of the planetesimal, $\omega_d$ is the catastrophic disruption spin rate limit for a sphere of density $\rho$, $E_a$, $E_b$, and $E_c$ are dimensionless elliptic integrals over the integration variable $s$ that depend only on the shape of the body and are encountered when solving for the gravitational potential within a tri-axial ellipsoid, and $\delta$ is a shared factor within the integrand of the elliptic integrals \citep[for derivation details, see][]{Holsapple2001}.
The scaled spin rate $\Omega$ cannot exceed $1$, since then even for a spherical shape, the body is no longer gravitationally bound to itself, i.e., the normal stresses are greater than zero.
More extended tri-axial shapes have lower spin rate limits.
The shared maximum value for $E_a$, $E_b$, and $E_c$ is $1$, and they are always ordered in value $E_a \geq E_b \geq E_c$ due to the construction of the tri-axial model so that the lengths of the semi-axes obey $a \geq b \geq c$.

Most simulated planetesimals have tri-axial ellipsoid shapes consistent with low internal angles of friction.
Figure~\ref{fig:contour_combined} shows simulated planetesimals collected by similar shapes, and contours of minimum required angles of friction as a function of spin rate and shape parameter, oblateness or prolateness.
For each of the shapes examined, the typical required minimum internal angles of friction are $\phi \lesssim 10^\circ$ with many consistent with $\phi \approx 0^\circ$.
Planetesimals requiring the largest minimum internal angles of friction $\phi \approx 20^\circ$ are rare, but these values are still low compared to the internal angles of friction $\phi \approx 32^\circ$ found for rubble pile asteroids \citep{Robin2024}.
Note that we only calculate the minimum internal angle of friction so these results are consistent; however, they indicate that the gravitational collapse formation process assembles planetesimals into relaxed shapes that are far from the stability limit.

Small impacts, slope failure, and mass shedding for specific planetesimal shapes contributed to the evolution of the shapes and rotation periods of simulated planetesimals.
As the planetesimals accommodated any extra angular momentum from accretion or significant impacts, their moments of inertia were modified accordingly.
This ultimately resulted in final periods that are noticeably slower than the theoretical spin limit.

\section{Discussion}\label{Discussion}
\subsection{How relict binary planetesimal orbits relate to gravitational collapse models}
One of the primary goals of this assessment was to test the validity of the soft-sphere discrete element method (SSDEM) for modeling the gravitational collapse of a cloud of super-particles by comparing binary planetesimal systems created with the SSDEM to systems created from similar numerical experiments, which used a perfect merging method combined with super-particles possessing inflated radii.
Although binary planetesimal accretion efficiencies and binary mass ratios match well with the results of both \citet{Nesvorny2010} and \citet{Robinson2020}, there are clear differences between the final binary orbits produced.
Notable is the presence of a large population of massive and tight binaries in the SSDEM simulations in comparison to \citet{Robinson2020}.
We classify each of the SSDEM binary systems by mutual semi-major axis $a_p$ into ultra-tight, tight, wide, and ultra-wide systems, which are similar to classifications discussed in \citet{Campbell2023}, see Table~\ref{table:orbit_stats}.
Of all binary systems created, the SSDEM simulations produce 60\% with $a/R_{\text{Hill}} \leq 0.1$, which is likely not possible in perfect merging simulations because of the inflated particle assumption, where potential tight binary systems merge into a single planetesimal.
However, it remains a mystery as to why even the SSDEM simulations cannot achieve orbits as tight as some of the observed population of trans-Neptunian objects \citep[see Figure~\hyperlink{fig:aRhill_all}{8};][]{Grundy2019, Grundy2023, Porter2024a}.
Possible post-formation orbital evolutionary processes, such as collisions and planetary encounters \citep{Nesvorny2019b}, gas-drag interactions with the nebular gas \citep{McKinnon2020}, or Kozai-Lidov combined with tides and drag \citep{Lyra2021} could lead to binary hardening and an explanation for this population.
        
\begin{table}[t!]
\centering
\begin{tabularx}{\columnwidth}
{lYp{0.482\columnwidth}Yp{0.09\columnwidth}Yp{0.09\columnwidth}}
\multicolumn{4}{@{}M{1}}{\textit{\textbf{Binary Orbit Classification}}} \\
\toprule
\textit{Orbit type} & \textit{Definition} & $f_{mm}$ & $f_{total}$ \\
\midrule
Ultra-tight  & $\frac{a_p}{R_{Hill}}<0.03$                  & 0.16 & 0.11 \\
Tight       & $0.03<\frac{a_p}{R_{Hill}}<0.05$   & 0.22 & 0.16 \\
Wide        & $0.05<\frac{a_p}{R_{Hill}}<0.07$   & 0.26 & 0.21 \\
Ultra-wide   & $\frac{a_p}{R_{Hill}}>0.07$                  & 0.36 & 0.52 \\
\bottomrule
\end{tabularx}
\caption{Classification of binary planetesimals by mutual semi-major axis $a_p$ into ultra-tight, tight, wide, and ultra-wide systems, which are similar to classifications discussed in \citet{Campbell2023}. 
$R_{Hill}$ is the Hill radius of the system. 
The fraction of the most massive systems $f_{mm}$ and all formed systems $f_{total}$ that match these orbital classifications are provided as well.}
\label{table:orbit_stats}
\end{table}

Like the perfect merging method, an abundance of wide binary planetesimals are also created using the SSDEM and match observed ultra-wide populations \citep{Grundy2019, Grundy2023}.
However, it is not clear if this primordial ultra-wide population is the ultra-wide population observed today in the Kuiper Belt.
Modeling shows that flyby events between planetesimal systems result in binary softening and increasing eccentricities \citep{StoneKaib2021,Campbell2023}, matching expectations from the Heggie Hills Law since flyby interaction times are much shorter than the mutual orbital periods.
Primordial ultra-wide binaries are shown to have a very low chance of survival in the dynamically hot population of the Kuiper Belt \citep{Nesvorny2019b} and even in the cold population \citep{Campbell2025}.
In other words, the observed ultra-wide population are evolved from primordial tighter systems.
However, if this were so, simulations seem to suggest that the observed population should be more eccentric and rarer \citep{Parker2012, Nesvorny2019b}.
Thus, the long-term evolution of the planetesimal systems generated by SSDEM is left as important future work.

\subsection{Comparing the spin states of simulated planetesimals and relict asteroids and trans-Neptunian objects}\label{sec:dichotomy}
The differences in the averaged spin rates between planetesimals created with the SSDEM and the population of observed asteroids (see Figure~\ref{fig:diameter_v_spin}) is likely the result of a prolonged period of collisional evolution following asteroid formation.
This difference is most evident for asteroids with D~$\lesssim$~40~km.
Below this 40-km threshold, the averaged spin rates diverge and asteroids exhibit noticeably greater spin rates.
This suggests that these asteroids, as the collisionally evolved remnants of a primordial sub-population \citep[e.g.,][]{Vokrouhlicky2003,Bottke2006,Steinberg2015}, had their spin rates driven upwards by an extensive collision history in the Main Belt.
The study of this collisionally-driven rotational acceleration is left for future work, but we note that the current study clearly establishes that the average long-term effect of collisions is to drive the spin rate upwards.

However, not all asteroid rotation states have been significantly altered over the age of the solar system.
The spins of large asteroids (D~$\gtrsim$~100 km) have long been theorized to be born with rotation periods on the order of their currently observed periods \citep{Alfven1964,Harris1979} because the observed collisional environment is not expected to change their angular momentum substantially without disrupting them, and so these spins reflect their primordial rates \citep{Pravec2002,Bottke2005}.
It has also been noted that large Main Belt asteroids with diameters greater than 40~km preserve a non-Maxwellian distribution, which is inconsistent with a collisional origin \citep{Pravec2002,Salo1987} and likely also reflects an individualized formation process rather than a process operating across the population like collisions.
Thus, the clear increase in the spin rate as a function of diameter observed amongst the large asteroids in the asteroid belt is a real reflection of the same trend observed in the SSDEM planetesimals.
There is also evidence of a slight decrease in the spin rate as a function of diameter up to a diameter of about 100~km, which matches a similar trend seen in the asteroid population \citep{Warner2009} and examined by \citet{Steinberg2015}.
Furthermore, despite limited observations of some of the most massive planetesimals in the Kuiper Belt, it is likely that they have undergone minimal collisional evolution and their spins are primordial \citep{Steinberg2015}.

\subsection{Comparing the shapes of simulated planetesimals and relict asteroids and trans-Neptunian objects}
Asteroids and trans-Neptunian objects typically exhibit elongated shapes, which may indicate a significant inheritance of angular momentum from their formation process \citep{Cibulkova2016,Sheppard2002}.
The shapes of these small bodies appear to be consistent with a rubble-pile-like internal structure with low cohesion, and the shapes of cohesionless rubble piles have been investigated at length \citep[e.g.,][]{Richardson2000,Richardson2005,Holsapple2006,Sanchez2016,Zhang2018,Zhang2021}.
For example, rubble pile asteroids spun up by some external perturbation (e.g., YORP) may result in landsliding effects to create top-like shapes \citep[e.g.,][]{Harris2009}, mass-shedding \citep[e.g.,][]{Walsh2008,Walsh2012,Hirabayashi2015a,Hirabayashi2015b,Hirabayashi2019,Barnouin2019}, or rotational failure \citep[e.g.,][]{Hirabayashi2015b,Zhang2018}.
Other experiments indicate a possible origin for top-shaped rubble piles through the catastrophic disruption of some parent body and the gravitational reaccumulation of its fragments \citep{Michel2001,Michel2020}.
Alternatively, these rubble piles can also maintain oblate or prolate shapes that are far-removed from fluid equilibrium \citep{Richardson2000,Holsapple2001,Richardson2005,Harris2009,Sanchez2016}.

Small bodies in fluid equilibrium reside within a phase-space of allowable shapes and spins, and a vast number of asteroids exhibit attributes consistent with a granular body with a porous interior \citep{Richardson2005,Holsapple2006}.
Experiments conducted by \citet{Richardson2005} investigated the permissible region of asteroid shapes and spins when following the Mohr--Coulomb yield criterion (i.e., a natural transition for particles locked together as in an elastic material to a more plastic and granular material).
Prolate rubble piles were discovered to be able to maintain a diverse range of shapes even in the absence of rotation.
With friction angles ranging between approximately 30$^{\circ}$--40$^{\circ}$, prolate bodies are able to rotate more rapidly than a fluid object while still maintaining their shape \citep{Holsapple2001,Richardson2005}.
Later work refined this range and found that, when employing a Drucker-Parger yield criterion \citep[i.e., an effective smooth outer envelope in regards to the Mohr--Colulomb criterion;][]{Sharma2009}, asteroids are likely to exhibit an internal friction angle of 31$^{\circ}$.
Planetesimals created with the SSDEM exhibit friction angles less than the quoted 31$^{\circ}$ of \citet{Sharma2009} or the approximate 30$^{\circ}$--40$^{\circ}$ range from \citet{Richardson2005}, and they instead exhibit friction angles ranging from approximately 0$^{\circ}$--20$^{\circ}$ (see Figure~\ref{fig:contour_combined}). 
It is likely that these planetesimals have already experienced a period of slope failure or mass shedding prior to achieving their final shapes but still maintain strong particle interlocking. 
Perhaps periods of catastrophic disruption and reformation into second- or third-generation rubble piles or perhaps future models that do not solely rely on a monodisperse distribution of spherical constituent particles may result in the creation of asteroids that more closely exhibit quoted friction angles $\phi\gtrsim$~30$^{\circ}$.

Lastly, simulations from \citet{Lorek2024} modeled the creation of planetesimals from rotating clumps of pebbles using a direct-simulation Monte Carlo (DSMC) method, and they discovered that  slower rotating clouds often produced planetesimals as flattened ellipsoids. 
For example, the interstellar object `Oumuamua may have either an extremely flattened or prolate shape \citet{Mashchenko2019}.
This is comparable to the sizeable fraction of flattened planetesimals that are the product of slowly rotating clouds ($f\sim$~0.2--0.6) that have experienced gravitational collapse with the SSDEM.
However, the investigation conducted by \citet{Lorek2024} was limited to the final aggregation of pebble cloud sub-clumps instead of direct planetesimal formation from gravitational collapse, and the final resolution of their model was limited due to the utilization of a fixed grid. 
Consequently, collisions essential to final planetesimal assembly may have been missed early in their collapse model, and the rotation states of their formed planetesimals were not adequately resolved.
As such, other diverse planetesimal shapes may have been missed (e.g., egg-like and prolate shapes; see Figures~\ref{fig:shape_stats_ang_vel_by_sim} and \ref{fig:shape_stats_contact_by_sim}).

\section{Conclusions}
We have demonstrated the effectiveness of the PKDGRAV SSDEM in modeling binary planetesimal formation via the gravitational collapse of a cloud of super-particles.
Binary accretion efficiencies and mass ratios for systems formed from the SSDEM simulations match well with previously published perfect-merger models of gravitational collapse as in \citet{Nesvorny2010} and \citet{Robinson2020}.
This means that for the first time, one can estimate the rotation periods and shapes of planetesimals resulting directly from gravitational collapse.
We also found that from the experiment of different coefficients of friction and restitution, that the choice of these parameters does not have a large effect on the resultant binary systems.

The SSDEM has proven effective at modeling the formation of planetesimal systems through gravitational collapse.
We have been able to characterize the rotation periods and shapes of newly formed planetesimals.
The SSDEM simulations also create binaries with tighter orbits as well as a wider distribution of eccentricities and inclinations than what are currently observed in either the cold classical Kuiper Belt or general trans-Neptunian populations.

The key results of the numerical experiments reported are summarized here:
\begin{enumerate}
    \item Gravitational collapse is a rapid and efficient planetesimal formation process.
    It often completes within less than an orbital period and enables the accretion of $>$~20\% of a cloud's mass by the largest planetesimal or binary planetesimal system.
    \item Gravitational collapse is capable of producing binary planetesimal systems such that $>$~10\% of all planetesimals are born as constituents of binary or higher multiplicity systems, and the most massive planetesimals from all collapsing clouds with initial rotation rate scale factors $f~\geq~0.6$ are components of such systems.
    Some planetesimals born from clouds with lower angular velocities or a combination of low angular velocities and high coefficients of restitution and friction instead create solitary planetesimals.
    \item Gravitational collapse is capable of producing approximately N~$>$~5 binary systems from each cloud with $f~\sim~0.8$--$1.2$ and N~$<$~5 systems for clouds with $f~\sim~0.2$--$0.6$.
    \item Binary systems can maintain tight orbits with the most-massive systems primarily characterized by low-inclination (i~$\lesssim$~15$^{\circ}$), moderate eccentricity (e~$\lesssim$~0.40), prograde orbits with $a/R_{\text{Hill}}$ predominately ranging from 0.02 to 0.15.
    \item Less-massive planetesimal systems exhibit a wide range of eccentricities and inclinations and are characterized by a wider range of $a/R_{\text{Hill}}$ than the most massive planetesimals, from $a/R_{\text{Hill}}~\sim~0.02$ to $a/R_{\text{Hill}}~\geq~0.5$.
    \item Simulated SSDEM planetesimals are created with a wide variety of rotation periods.
    The largest simulated planetesimals (D~$\geq$~100~km) that are comparable to the large asteroid and trans-Neptunian object populations show an ascending spin rate with size, suggesting a primordial origin for these spin states.
    Smaller simulated planetesimals D~$\leq$~100~km spin on average 5 hours slower than comparable asteroids today, indicating that subsequent collisional evolution has rotationally accelerated this population.
    \item Simulated SSDEM planetesimals can exhibit a wide range of shapes, with the most-massive planetesimals commonly forming as spheres or oblate-spheroids, but some also exhibit flattened-, egg-, top-like, or prolate shapes. These shapes are generally consistent with relaxation towards shapes consistent with a cohesionless fluid equilibrium. Indeed, many SSDEM planetesimal shapes can be characterized by low friction angles $\phi~\lesssim~10^{\circ}$, assuming a Mohr--Coulomb yield criterion.
\end{enumerate}

Due to the fact that the suites of experiments were meant to compare and expand upon prior investigations of binary systems formed from gravitational collapse, this analysis too is focused on binary systems and away from the dynamics of triple or higher-order systems.
These systems deserve their own separate analysis.
Looking forward, the marked increase in the current population of trans-Neptunian objects provided by the Vera Rubin Observatory will undoubtedly help constrain the shapes and rotation rates for cold classical Kuiper Belt and trans-Neptunian objects, which will provide a more robust sampling of primordial planetesimal characteristics and thus a stronger understanding as to their formation histories and evolutions.

\section{Data availability}
The processed output data from our PKDGRAV simulations and data used to directly generate the figures will be available on the Zenodo page affiliated with this article. 

\section*{Acknowledgments}
We are grateful to Dr. Audrey Thirouin for maintaining and providing access to an up-to-date table of trans-Neptunian object rotation periods from \citet{Thirouin2019b}, which was utilized in the analysis and presented in Figure~\ref{fig:diameter_v_spin}, and Dr. Jamie Robinson for access to the raw data used to compile Figure~9 in \citet{Robinson2020}, which was used for direct comparisons with the data in Figure~\hyperlink{fig:aRhill_all}{8}.
SAJ and JTB acknowledge direct support from the National Science Foundation (NSF) award number (AST-2406891).
SAJ and JTB used the Extreme Science and Engineering Discovery Environment (XSEDE), which was supported by National Science Foundation (NSF) award numbers (ACI-1053575 and ACI-1548562). 
JTB acknowledges the Michigan Space Grant Consortium (award number: 80NSSC20M0124) for their graduate fellowship support.

\bibliography{biblio}{}

\begin{thebibliography}{}
\expandafter\ifx\csname natexlab\endcsname\relax\def\natexlab#1{#1}\fi
\providecommand{\url}[1]{\href{#1}{#1}}
\providecommand{\dodoi}[1]{doi:~\href{http://doi.org/#1}{\nolinkurl{#1}}}
\providecommand{\doeprint}[1]{\href{http://ascl.net/#1}{\nolinkurl{http://ascl.net/#1}}}
\providecommand{\doarXiv}[1]{\href{https://arxiv.org/abs/#1}{\nolinkurl{https://arxiv.org/abs/#1}}}

\bibitem[{Agresti \& Coull(1998)}]{Agresti1998}
Agresti, A., \& Coull, B.~A. 1998, The American Statistician, 52, 119.
\newblock \url{http://www.jstor.org/stable/2685469}

\bibitem[{{Agrusa} {et~al.}(2024){Agrusa}, {Zhang}, {Richardson}, {Pravec},
  {{\'C}uk}, {Michel}, {Ballouz}, {Jacobson}, {Scheeres}, {Walsh}, {Barnouin},
  {Daly}, {Palmer}, {Pajola}, {Lucchetti}, {Tusberti}, {DeMartini}, {Ferrari},
  {Meyer}, {Raducan}, \& {S{\'a}nchez}}]{Agrusa2024}
{Agrusa}, H.~F., {Zhang}, Y., {Richardson}, D.~C., {et~al.} 2024, The Planetary
  Science Journal, 5, 54, \dodoi{10.3847/PSJ/ad206b}

\bibitem[{{Alfv{\'e}n}(1964)}]{Alfven1964}
{Alfv{\'e}n}, H. 1964, \icarus, 3, 52, \dodoi{10.1016/0019-1035(64)90030-2}

\bibitem[{{Ballouz} {et~al.}(2014){Ballouz}, {Richardson}, {Michel}, \&
  {Schwartz}}]{Ballouz2014}
{Ballouz}, R.-L., {Richardson}, D.~C., {Michel}, P., \& {Schwartz}, S.~R. 2014,
  The Astrophysical Journal, 789, 158, \dodoi{10.1088/0004-637X/789/2/158}

\bibitem[{{Ballouz} {et~al.}(2015){Ballouz}, {Richardson}, {Michel},
  {Schwartz}, \& {Yu}}]{Ballouz2015}
{Ballouz}, R.~L., {Richardson}, D.~C., {Michel}, P., {Schwartz}, S.~R., \&
  {Yu}, Y. 2015, Planetary and Space Science, 107, 29,
  \dodoi{10.1016/j.pss.2014.06.003}

\bibitem[{{Barnouin} {et~al.}(2019){Barnouin}, {Daly}, {Palmer}, {Gaskell},
  {Weirich}, {Johnson}, {Al Asad}, {Roberts}, {Perry}, {Susorney}, {Daly},
  {Bierhaus}, {Seabrook}, {Espiritu}, {Nair}, {Nguyen}, {Neumann}, {Ernst},
  {Boynton}, {Nolan}, {Adam}, {Moreau}, {Rizk}, {Drouet D'Aubigny}, {Jawin},
  {Walsh}, {Michel}, {Schwartz}, {Ballouz}, {Mazarico}, {Scheeres}, {McMahon},
  {Bottke}, {Sugita}, {Hirata}, {Hirata}, {Watanabe}, {Burke}, {Dellagiustina},
  {Bennett}, {Lauretta}, {The Osiris-Rex Team}, {Highsmith}, {Small},
  {Vokrouhlick{\'y}}, {Bowles}, {Brown}, {Donaldson Hanna}, {Warren}, {Brunet},
  {Chicoine}, {Desjardins}, {Gaudreau}, {Haltigin}, {Millington-Veloza},
  {Rubi}, {Aponte}, {Gorius}, {Lunsford}, {Allen}, {Grindlay}, {Guevel},
  {Hoak}, {Hong}, {Schrader}, {Bayron}, {Golubov}, {S{\'a}nchez}, {Stromberg},
  {Hirabayashi}, {Hartzell}, {Oliver}, {Rascon}, {Harch}, {Joseph}, {Squyres},
  {Richardson}, {Emery}, {McGraw}, {Ghent}, {Binzel}, {Asad}, {Johnson},
  {Philpott}, {Susorney}, {Cloutis}, {Hanna}, {Connolly}, {Ciceri},
  {Hildebrand}, {Ibrahim}, {Breitenfeld}, {Glotch}, {Rogers}, {Clark},
  {Ferrone}, {Thomas}, {Campins}, {Fernandez}, {Chang}, {Cheuvront}, {Trang},
  {Tachibana}, {Yurimoto}, {Brucato}, {Poggiali}, {Pajola}, {Dotto}, {Epifani},
  {Crombie}, {Lantz}, {Izawa}, {de Leon}, {Licandro}, {Garcia}, {Clemett},
  {Thomas-Keprta}, {van Wal}, {Yoshikawa}, {Bellerose}, {Bhaskaran}, {Boyles},
  {Chesley}, {Elder}, {Farnocchia}, {Harbison}, {Kennedy}, {Knight},
  {Martinez-Vlasoff}, {Mastrodemos}, {McElrath}, {Owen}, {Park}, {Rush},
  {Swanson}, {Takahashi}, {Velez}, {Yetter}, {Thayer}, {Adam}, {Antreasian},
  {Bauman}, {Bryan}, {Carcich}, {Corvin}, {Geeraert}, {Hoffman}, {Leonard},
  {Lessac-Chenen}, {Levine}, {McAdams}, {McCarthy}, {Nelson}, {Page},
  {Pelgrift}, {Sahr}, {Stakkestad}, {Stanbridge}, {Wibben}, {Williams},
  {Williams}, {Wolff}, {Hayne}, {Kubitschek}, {Barucci}, {Deshapriya},
  {Fornasier}, {Fulchignoni}, {Hasselmann}, {Merlin}, {Praet}, {Bierhaus},
  {Billett}, {Boggs}, {Buck}, {Carlson-Kelly}, {Cerna}, {Chaffin}, {Church},
  {Coltrin}, {Daly}, {Deguzman}, {Dubisher}, {Eckart}, {Ellis}, {Falkenstern},
  {Fisher}, {Fisher}, {Fleming}, {Fortney}, {Francis}, {Freund}, {Gonzales},
  {Haas}, {Hasten}, {Hauf}, {Hilbert}, {Howell}, {Jaen}, \&
  {Jayakody}}]{Barnouin2019}
{Barnouin}, O.~S., {Daly}, M.~G., {Palmer}, E.~E., {et~al.} 2019, Nature
  Geoscience, 12, 247, \dodoi{10.1038/s41561-019-0330-x}

\bibitem[{{Benecchi} {et~al.}(2009){Benecchi}, {Noll}, {Grundy}, {Buie},
  {Stephens}, \& {Levison}}]{Benecchi2009}
{Benecchi}, S.~D., {Noll}, K.~S., {Grundy}, W.~M., {et~al.} 2009, Icarus, 200,
  292, \dodoi{10.1016/j.icarus.2008.10.025}

\bibitem[{{Binney} \& {Tremaine}(2008)}]{Binney2008}
{Binney}, J., \& {Tremaine}, S. 2008, Galactic Dynamics: Second Edition
  (Princeton University Press)

\bibitem[{Binzel {et~al.}(1989)Binzel, Farinella, Zappala, \&
  Cellino}]{Binzel1989}
Binzel, R.~P., Farinella, P., Zappala, V., \& Cellino, A. 1989, Asteroids II,
  416–441

\bibitem[{{Bottke} {et~al.}(2006){Bottke}, {Vokrouhlick{\'y}}, {Rubincam}, \&
  {Nesvorn{\'y}}}]{Bottke2006}
{Bottke}, William~F., J., {Vokrouhlick{\'y}}, D., {Rubincam}, D.~P., \&
  {Nesvorn{\'y}}, D. 2006, Annual Review of Earth and Planetary Sciences, 34,
  157, \dodoi{10.1146/annurev.earth.34.031405.125154}

\bibitem[{{Bottke} {et~al.}(2015){Bottke}, {Bro{\v{z}}}, {O'Brien}, {Campo
  Bagatin}, {Morbidelli}, \& {Marchi}}]{Bottke2015}
{Bottke}, W.~F., {Bro{\v{z}}}, M., {O'Brien}, D.~P., {et~al.} 2015, in
  Asteroids IV, ed. P.~{Michel}, F.~E. {DeMeo}, \& W.~F. {Bottke}, 701--724,
  \dodoi{10.2458/azu_uapress_9780816532131-ch036}

\bibitem[{{Bottke} {et~al.}(2005){Bottke}, {Durda}, {Nesvorn{\'y}}, {Jedicke},
  {Morbidelli}, {Vokrouhlick{\'y}}, \& {Levison}}]{Bottke2005}
{Bottke}, W.~F., {Durda}, D.~D., {Nesvorn{\'y}}, D., {et~al.} 2005, \icarus,
  175, 111, \dodoi{10.1016/j.icarus.2004.10.026}

\bibitem[{{Buie} {et~al.}(2018){Buie}, {Zangari}, {Marchi}, {Levison}, \&
  {Mottola}}]{Buie2018}
{Buie}, M.~W., {Zangari}, A.~M., {Marchi}, S., {Levison}, H.~F., \& {Mottola},
  S. 2018, \aj, 155, 245, \dodoi{10.3847/1538-3881/aabd81}

\bibitem[{{Campbell} {et~al.}(2025){Campbell}, {Anderson}, \&
  {Kaib}}]{Campbell2025}
{Campbell}, H.~M., {Anderson}, K.~E., \& {Kaib}, N.~A. 2025, Nature Astronomy,
  9, 75, \dodoi{10.1038/s41550-024-02388-4}

\bibitem[{{Campbell} {et~al.}(2023){Campbell}, {Stone}, \&
  {Kaib}}]{Campbell2023}
{Campbell}, H.~M., {Stone}, L.~R., \& {Kaib}, N.~A. 2023, \aj, 165, 19,
  \dodoi{10.3847/1538-3881/aca08e}

\bibitem[{{Cibulkov{\'a}} {et~al.}(2016){Cibulkov{\'a}}, {{\v{D}}urech},
  {Vokrouhlick{\'y}}, {Kaasalainen}, \& {Oszkiewicz}}]{Cibulkova2016}
{Cibulkov{\'a}}, H., {{\v{D}}urech}, J., {Vokrouhlick{\'y}}, D., {Kaasalainen},
  M., \& {Oszkiewicz}, D.~A. 2016, \aap, 596, A57,
  \dodoi{10.1051/0004-6361/201629192}

\bibitem[{Cundall {et~al.}(1978)Cundall, Marti, Beresford, Last, \&
  Asgian}]{Cundall1978}
Cundall, P., Marti, J., Beresford, P., Last, N., \& Asgian, M. 1978, Computer
  modelling of jointed rock masses (Dames and Moore)

\bibitem[{Dikaiakos \& Stadel(1996)}]{Dikaiakos1996}
Dikaiakos, M.~D., \& Stadel, J. 1996, in Proceedings of the 10th International
  Conference on Supercomputing, ICS '96 (New York, NY, USA: Association for
  Computing Machinery), 94--101, \dodoi{10.1145/237578.237590}

\bibitem[{{Drazkowska} {et~al.}(2016){Drazkowska}, {Alibert}, \&
  {Moore}}]{Drazkowska2016}
{Drazkowska}, J., {Alibert}, Y., \& {Moore}, B. 2016, Astronomy and
  Astrophysics, 594, A105, \dodoi{10.1051/0004-6361/201628983}

\bibitem[{{Durda} {et~al.}(2004){Durda}, {Bottke}, {Enke}, {Merline},
  {Asphaug}, {Richardson}, \& {Leinhardt}}]{Durda2004}
{Durda}, D.~D., {Bottke}, W.~F., {Enke}, B.~L., {et~al.} 2004, Icarus, 167,
  382, \dodoi{10.1016/j.icarus.2003.09.017}

\bibitem[{{Fraser} {et~al.}(2017){Fraser}, {Bannister}, {Pike}, {Marsset},
  {Schwamb}, {Kavelaars}, {Lacerda}, {Nesvorn{\'y}}, {Volk}, {Delsanti},
  {Benecchi}, {Lehner}, {Noll}, {Gladman}, {Petit}, {Gwyn}, {Chen}, {Wang},
  {Alexandersen}, {Burdullis}, {Sheppard}, \& {Trujillo}}]{Fraser2017}
{Fraser}, W.~C., {Bannister}, M.~T., {Pike}, R.~E., {et~al.} 2017, Nature
  Astronomy, 1, 0088, \dodoi{10.1038/s41550-017-0088}

\bibitem[{{Fraser} {et~al.}(2021){Fraser}, {Benecchi}, {Kavelaars}, {Marsset},
  {Pike}, {Bannister}, {Schwamb}, {Volk}, {Nesvorny}, {Alexandersen}, {Chen},
  {Gwyn}, {Lehner}, \& {Wang}}]{Fraser2021}
{Fraser}, W.~C., {Benecchi}, S.~D., {Kavelaars}, J.~J., {et~al.} 2021, The
  Planetary Science Journal, 2, 90, \dodoi{10.3847/PSJ/abf04a}

\bibitem[{{Grundy}(2023)}]{Grundy2023}
{Grundy}, W. 2023, Flagstaff, AZ: Lowell Observatory.
\newblock \url{http://www2.lowell.edu/users/grundy/tnbs/orbits.html}

\bibitem[{{Grundy} {et~al.}(2019){Grundy}, {Noll}, {Roe}, {Buie}, {Porter},
  {Parker}, {Nesvorn{\'y}}, {Levison}, {Benecchi}, {Stephens}, \&
  {Trujillo}}]{Grundy2019}
{Grundy}, W.~M., {Noll}, K.~S., {Roe}, H.~G., {et~al.} 2019, Icarus, 334, 62,
  \dodoi{10.1016/j.icarus.2019.03.035}

\bibitem[{{Grundy} {et~al.}(2020){Grundy}, {Bird}, {Britt}, {Cook},
  {Cruikshank}, {Howett}, {Krijt}, {Linscott}, {Olkin}, {Parker}, {Protopapa},
  {Ruaud}, {Umurhan}, {Young}, {Dalle Ore}, {Kavelaars}, {Keane}, {Pendleton},
  {Porter}, {Scipioni}, {Spencer}, {Stern}, {Verbiscer}, {Weaver}, {Binzel},
  {Buie}, {Buratti}, {Cheng}, {Earle}, {Elliott}, {Gabasova}, {Gladstone},
  {Hill}, {Horanyi}, {Jennings}, {Lunsford}, {McComas}, {McKinnon}, {McNutt},
  {Moore}, {Parker}, {Quirico}, {Reuter}, {Schenk}, {Schmitt}, {Showalter},
  {Singer}, {Weigle}, \& {Zangari}}]{Grundy2020}
{Grundy}, W.~M., {Bird}, M.~K., {Britt}, D.~T., {et~al.} 2020, Science, 367,
  aay3705, \dodoi{10.1126/science.aay3705}

\bibitem[{{G{\"u}ttler} {et~al.}(2010){G{\"u}ttler}, {Blum}, {Zsom}, {Ormel},
  \& {Dullemond}}]{Guttler2010}
{G{\"u}ttler}, C., {Blum}, J., {Zsom}, A., {Ormel}, C.~W., \& {Dullemond},
  C.~P. 2010, Astronomy and Astrophysics, 513, A56,
  \dodoi{10.1051/0004-6361/200912852}

\bibitem[{{Harris}(1979)}]{Harris1979}
{Harris}, A.~W. 1979, \icarus, 40, 145, \dodoi{10.1016/0019-1035(79)90059-9}

\bibitem[{{Harris} {et~al.}(2009){Harris}, {Fahnestock}, \&
  {Pravec}}]{Harris2009}
{Harris}, A.~W., {Fahnestock}, E.~G., \& {Pravec}, P. 2009, \icarus, 199, 310,
  \dodoi{10.1016/j.icarus.2008.09.012}

\bibitem[{{Hayashi}(1981)}]{Hayashi1981}
{Hayashi}, C. 1981, Progress of Theoretical Physics Supplement, 70, 35,
  \dodoi{10.1143/PTPS.70.35}

\bibitem[{{Hirabayashi}(2015)}]{Hirabayashi2015b}
{Hirabayashi}, M. 2015, Monthly Notices of the Royal Astronomical Society, 454,
  2249, \dodoi{10.1093/mnras/stv2017}

\bibitem[{{Hirabayashi} {et~al.}(2015){Hirabayashi}, {S{\'a}nchez}, \&
  {Scheeres}}]{Hirabayashi2015a}
{Hirabayashi}, M., {S{\'a}nchez}, D.~P., \& {Scheeres}, D.~J. 2015, The
  Astrophysical Journal, 808, 63, \dodoi{10.1088/0004-637X/808/1/63}

\bibitem[{{Hirabayashi} \& {Scheeres}(2019)}]{Hirabayashi2019}
{Hirabayashi}, M., \& {Scheeres}, D.~J. 2019, Icarus, 317, 354,
  \dodoi{10.1016/j.icarus.2018.08.003}

\bibitem[{{Holsapple}(2001)}]{Holsapple2001}
{Holsapple}, K.~A. 2001, \icarus, 154, 432, \dodoi{10.1006/icar.2001.6683}

\bibitem[{{Holsapple} \& {Michel}(2006)}]{Holsapple2006}
{Holsapple}, K.~A., \& {Michel}, P. 2006, \icarus, 183, 331,
  \dodoi{10.1016/j.icarus.2006.03.013}

\bibitem[{{Ivezi{\'c}} {et~al.}(2019){Ivezi{\'c}}, {Kahn}, {Tyson}, {Abel},
  {Acosta}, {Allsman}, {Alonso}, {AlSayyad}, {Anderson}, {Andrew}, \&
  et~al.}]{Ivezic2019}
{Ivezi{\'c}}, {\v Z}., {Kahn}, S.~M., {Tyson}, J.~A., {et~al.} 2019, \apj, 873,
  111, \dodoi{10.3847/1538-4357/ab042c}

\bibitem[{{Jacobson} \& {Margot}(2007)}]{Jacobson2007}
{Jacobson}, S., \& {Margot}, J.~L. 2007, in AAS/Division for Planetary Sciences
  Meeting Abstracts, Vol.~39, AAS/Division for Planetary Sciences Meeting
  Abstracts \#39, 52.11

\bibitem[{{Johansen} {et~al.}(2015){Johansen}, {Mac Low}, {Lacerda}, \&
  {Bizzarro}}]{Johansen2015}
{Johansen}, A., {Mac Low}, M.-M., {Lacerda}, P., \& {Bizzarro}, M. 2015,
  Science Advances, 1, 1500109, \dodoi{10.1126/sciadv.1500109}

\bibitem[{{Johansen} {et~al.}(2007){Johansen}, {Oishi}, {Mac Low}, {Klahr},
  {Henning}, \& {Youdin}}]{Johansen2007a}
{Johansen}, A., {Oishi}, J.~S., {Mac Low}, M.-M., {et~al.} 2007, Nature, 448,
  1022, \dodoi{10.1038/nature06086}

\bibitem[{{Johansen} \& {Youdin}(2007)}]{Johansen2007b}
{Johansen}, A., \& {Youdin}, A. 2007, The Astrophysical Journal, 662, 627,
  \dodoi{10.1086/516730}

\bibitem[{{Keane} {et~al.}(2022){Keane}, {Porter}, {Beyer}, {Umurhan},
  {McKinnon}, {Moore}, {Spencer}, {Stern}, {Bierson}, {Binzel}, {Hamilton},
  {Lisse}, {Mao}, {Protopapa}, {Schenk}, {Showalter}, {Stansberry}, {White},
  {Verbiscer}, {Parker}, {Olkin}, {Weaver}, \& {Singer}}]{Keane2022}
{Keane}, J.~T., {Porter}, S.~B., {Beyer}, R.~A., {et~al.} 2022, Journal of
  Geophysical Research (Planets), 127, e07068, \dodoi{10.1029/2021JE007068}

\bibitem[{{Kryszczy{\'n}ska} {et~al.}(2007){Kryszczy{\'n}ska}, {La Spina},
  {Paolicchi}, {Harris}, {Breiter}, \& {Pravec}}]{Kryszczynska2007}
{Kryszczy{\'n}ska}, A., {La Spina}, A., {Paolicchi}, P., {et~al.} 2007,
  \icarus, 192, 223, \dodoi{10.1016/j.icarus.2007.06.008}

\bibitem[{{Leinhardt} \& {Richardson}(2005)}]{Leinhardt2005}
{Leinhardt}, Z.~M., \& {Richardson}, D.~C. 2005, The Astrophysical Journal,
  625, 427, \dodoi{10.1086/429402}

\bibitem[{{Leinhardt} {et~al.}(2009){Leinhardt}, {Richardson}, {Lufkin}, \&
  {Haseltine}}]{Leinhardt2009}
{Leinhardt}, Z.~M., {Richardson}, D.~C., {Lufkin}, G., \& {Haseltine}, J. 2009,
  Monthly Notices of the Royal Astronomical Society, 396, 718,
  \dodoi{10.1111/j.1365-2966.2009.14769.x}

\bibitem[{{Levison} {et~al.}(2021){Levison}, {Olkin}, {Noll}, {Marchi}, {Bell},
  {Bierhaus}, {Binzel}, {Bottke}, {Britt}, {Brown}, {Buie}, {Christensen},
  {Emery}, {Grundy}, {Hamilton}, {Howett}, {Mottola}, {P{\"a}tzold}, {Reuter},
  {Spencer}, {Statler}, {Stern}, {Sunshine}, {Weaver}, \& {Wong}}]{Levison2021}
{Levison}, H.~F., {Olkin}, C.~B., {Noll}, K.~S., {et~al.} 2021, \psj, 2, 171,
  \dodoi{10.3847/PSJ/abf840}

\bibitem[{{Li} {et~al.}(2019){Li}, {Youdin}, \& {Simon}}]{Li2019a}
{Li}, R., {Youdin}, A.~N., \& {Simon}, J.~B. 2019, The Astrophysical Journal,
  885, 69, \dodoi{10.3847/1538-4357/ab480d}

\bibitem[{{Lisse} {et~al.}(2021){Lisse}, {Young}, {Cruikshank}, {Sandford},
  {Schmitt}, {Stern}, {Weaver}, {Umurhan}, {Pendleton}, {Keane}, {Gladstone},
  {Parker}, {Binzel}, {Earle}, {Horanyi}, {El-Maarry}, {Cheng}, {Moore},
  {McKinnon}, {Grundy}, {Kavelaars}, {Linscott}, {Lyra}, {Lewis}, {Britt},
  {Spencer}, {Olkin}, {McNutt}, {Elliott}, {Dello-Russo}, {Steckloff}, {Neveu},
  \& {Mousis}}]{Lisse2021}
{Lisse}, C.~M., {Young}, L.~A., {Cruikshank}, D.~P., {et~al.} 2021, Icarus,
  356, 114072, \dodoi{10.1016/j.icarus.2020.114072}

\bibitem[{{Lorek} \& {Johansen}(2024)}]{Lorek2024}
{Lorek}, S., \& {Johansen}, A. 2024, \aap, 683, A38,
  \dodoi{10.1051/0004-6361/202347742}

\bibitem[{{Lyra} {et~al.}(2021){Lyra}, {Youdin}, \& {Johansen}}]{Lyra2021}
{Lyra}, W., {Youdin}, A.~N., \& {Johansen}, A. 2021, \icarus, 356, 113831,
  \dodoi{10.1016/j.icarus.2020.113831}

\bibitem[{{Mashchenko}(2019)}]{Mashchenko2019}
{Mashchenko}, S. 2019, \mnras, 489, 3003, \dodoi{10.1093/mnras/stz2380}

\bibitem[{{Matsumura} {et~al.}(2014){Matsumura}, {Richardson}, {Michel},
  {Schwartz}, \& {Ballouz}}]{Matsumura2014}
{Matsumura}, S., {Richardson}, D.~C., {Michel}, P., {Schwartz}, S.~R., \&
  {Ballouz}, R.-L. 2014, Monthly Notices of the Royal Astronomical Society,
  443, 3368, \dodoi{10.1093/mnras/stu1388}

\bibitem[{{Maurel} {et~al.}(2017){Maurel}, {Ballouz}, {Richardson}, {Michel},
  \& {Schwartz}}]{Maurel2017}
{Maurel}, C., {Ballouz}, R.-L., {Richardson}, D.~C., {Michel}, P., \&
  {Schwartz}, S.~R. 2017, Monthly Notices of the Royal Astronomical Society,
  464, 2866, \dodoi{10.1093/mnras/stw2641}

\bibitem[{{McAdoo} \& {Burns}(1973)}]{McAdoo1973}
{McAdoo}, D.~C., \& {Burns}, J.~A. 1973, Icarus, 18, 285,
  \dodoi{10.1016/0019-1035(73)90211-X}

\bibitem[{{McKinnon} {et~al.}(2020){McKinnon}, {Richardson}, {Marohnic},
  {Keane}, {Grundy}, {Hamilton}, {Nesvorn{\'y}}, {Umurhan}, {Lauer}, {Singer},
  {Stern}, {Weaver}, {Spencer}, {Buie}, {Moore}, {Kavelaars}, {Lisse}, {Mao},
  {Parker}, {Porter}, {Showalter}, {Olkin}, {Cruikshank}, {Elliott},
  {Gladstone}, {Parker}, {Verbiscer}, {Young}, \& {New Horizons Science
  Team}}]{McKinnon2020}
{McKinnon}, W.~B., {Richardson}, D.~C., {Marohnic}, J.~C., {et~al.} 2020,
  Science, 367, aay6620, \dodoi{10.1126/science.aay6620}

\bibitem[{{Michel} {et~al.}(2001){Michel}, {Benz}, {Tanga}, \&
  {Richardson}}]{Michel2001}
{Michel}, P., {Benz}, W., {Tanga}, P., \& {Richardson}, D.~C. 2001, Science,
  294, 1696, \dodoi{10.1126/science.1065189}

\bibitem[{{Michel} {et~al.}(2020){Michel}, {Ballouz}, {Barnouin}, {Jutzi},
  {Walsh}, {May}, {Manzoni}, {Richardson}, {Schwartz}, {Sugita}, {Watanabe},
  {Miyamoto}, {Hirabayashi}, {Bottke}, {Connolly}, {Yoshikawa}, \&
  {Lauretta}}]{Michel2020}
{Michel}, P., {Ballouz}, R.~L., {Barnouin}, O.~S., {et~al.} 2020, Nature
  Communications, 11, 2655, \dodoi{10.1038/s41467-020-16433-z}

\bibitem[{{Mottola} {et~al.}(2020){Mottola}, {Hellmich}, {Buie}, {Zangari},
  {Marchi}, {Brown}, \& {Levison}}]{Mottola2020}
{Mottola}, S., {Hellmich}, S., {Buie}, M.~W., {et~al.} 2020, \psj, 1, 73,
  \dodoi{10.3847/PSJ/abb942}

\bibitem[{{Mottola} {et~al.}(2023){Mottola}, {Hellmich}, {Buie}, {Zangari},
  {Stephens}, {Di Martino}, {Proffe}, {Marchi}, {Olkin}, \&
  {Levison}}]{Mottola2023}
---. 2023, \psj, 4, 18, \dodoi{10.3847/PSJ/acaf79}

\bibitem[{{Murray} \& {Dermott}(1999)}]{Murray1999}
{Murray}, C.~D., \& {Dermott}, S.~F. 1999, Solar System Dynamics (Cambridge
  University), \dodoi{10.1017/CBO9781139174817}

\bibitem[{{Nakagawa} {et~al.}(1986){Nakagawa}, {Sekiya}, \&
  {Hayashi}}]{Nakagawa1986}
{Nakagawa}, Y., {Sekiya}, M., \& {Hayashi}, C. 1986, Icarus, 67, 375,
  \dodoi{10.1016/0019-1035(86)90121-1}

\bibitem[{{Nesvorn{\'y}} {et~al.}(2021){Nesvorn{\'y}}, {Li}, {Simon}, {Youdin},
  {Richardson}, {Marschall}, \& {Grundy}}]{Nesvorny2021}
{Nesvorn{\'y}}, D., {Li}, R., {Simon}, J.~B., {et~al.} 2021, The Planetary
  Science Journal, 2, 27, \dodoi{10.3847/PSJ/abd858}

\bibitem[{{Nesvorn{\'y}} {et~al.}(2019){Nesvorn{\'y}}, {Li}, {Youdin}, {Simon},
  \& {Grundy}}]{Nesvorny2019a}
{Nesvorn{\'y}}, D., {Li}, R., {Youdin}, A.~N., {Simon}, J.~B., \& {Grundy},
  W.~M. 2019, Nature Astronomy, 3, 808, \dodoi{10.1038/s41550-019-0806-z}

\bibitem[{{Nesvorn{\'y}} \& {Vokrouhlick{\'y}}(2019)}]{Nesvorny2019b}
{Nesvorn{\'y}}, D., \& {Vokrouhlick{\'y}}, D. 2019, \icarus, 331, 49,
  \dodoi{10.1016/j.icarus.2019.04.030}

\bibitem[{{Nesvorn{\'y}} {et~al.}(2020{\natexlab{a}}){Nesvorn{\'y}},
  {Vokrouhlick{\'y}}, {Bottke}, {Levison}, \& {Grundy}}]{Nesvorny2020a}
{Nesvorn{\'y}}, D., {Vokrouhlick{\'y}}, D., {Bottke}, W.~F., {Levison}, H.~F.,
  \& {Grundy}, W.~M. 2020{\natexlab{a}}, The Astrophysical Journal, 893, L16,
  \dodoi{10.3847/2041-8213/ab8311}

\bibitem[{{Nesvorn{\'y}} {et~al.}(2022){Nesvorn{\'y}}, {Vokrouhlick{\'y}}, \&
  {Fraser}}]{Nesvorny2022}
{Nesvorn{\'y}}, D., {Vokrouhlick{\'y}}, D., \& {Fraser}, W.~C. 2022, \aj, 163,
  137, \dodoi{10.3847/1538-3881/ac4bc9}

\bibitem[{{Nesvorn{\'y}} {et~al.}(2020{\natexlab{b}}){Nesvorn{\'y}}, {Youdin},
  {Marschall}, \& {Richardson}}]{Nesvorny2020b}
{Nesvorn{\'y}}, D., {Youdin}, A.~N., {Marschall}, R., \& {Richardson}, D.~C.
  2020{\natexlab{b}}, The Astrophysical Journal, 895, 63,
  \dodoi{10.3847/1538-4357/ab89a1}

\bibitem[{{Nesvorn{\'y}} {et~al.}(2010){Nesvorn{\'y}}, {Youdin}, \&
  {Richardson}}]{Nesvorny2010}
{Nesvorn{\'y}}, D., {Youdin}, A.~N., \& {Richardson}, D.~C. 2010, The
  Astronomical Journal, 140, 785, \dodoi{10.1088/0004-6256/140/3/785}

\bibitem[{{Noll} {et~al.}(2020){Noll}, {Grundy}, {Nesvorn{\'y}}, \&
  {Thirouin}}]{Noll2020}
{Noll}, K., {Grundy}, W.~M., {Nesvorn{\'y}}, D., \& {Thirouin}, A. 2020, in The
  Trans-Neptunian Solar System, ed. D.~{Prialnik}, M.~A. {Barucci}, \&
  L.~{Young} (Elsevier), 201--224, \dodoi{10.1016/B978-0-12-816490-7.00009-6}

\bibitem[{{Noll} {et~al.}(2008){Noll}, {Grundy}, {Stephens}, {Levison}, \&
  {Kern}}]{Noll2008}
{Noll}, K.~S., {Grundy}, W.~M., {Stephens}, D.~C., {Levison}, H.~F., \& {Kern},
  S.~D. 2008, Icarus, 194, 758, \dodoi{10.1016/j.icarus.2007.10.022}

\bibitem[{{Offner} {et~al.}(2022){Offner}, {Moe}, {Kratter}, {Sadavoy},
  {Jensen}, \& {Tobin}}]{Offner2023}
{Offner}, S. S.~R., {Moe}, M., {Kratter}, K.~M., {et~al.} 2022, arXiv e-prints,
  arXiv:2203.10066, \dodoi{10.48550/arXiv.2203.10066}

\bibitem[{{Parker} \& {Kavelaars}(2012)}]{Parker2012}
{Parker}, A.~H., \& {Kavelaars}, J.~J. 2012, \apj, 744, 139,
  \dodoi{10.1088/0004-637X/744/2/139}

\bibitem[{{Parker} {et~al.}(2011){Parker}, {Kavelaars}, {Petit}, {Jones},
  {Gladman}, \& {Parker}}]{Parker2011}
{Parker}, A.~H., {Kavelaars}, J.~J., {Petit}, J.-M., {et~al.} 2011, \apj, 743,
  1, \dodoi{10.1088/0004-637X/743/1/1}

\bibitem[{{Petit} \& {Mousis}(2004)}]{Petit2004}
{Petit}, J.~M., \& {Mousis}, O. 2004, \icarus, 168, 409,
  \dodoi{10.1016/j.icarus.2003.12.013}

\bibitem[{{Porter} {et~al.}(2024){Porter}, {Benecchi}, {Verbiscer}, {Grundy},
  {Noll}, \& {Parker}}]{Porter2024a}
{Porter}, S.~B., {Benecchi}, S.~D., {Verbiscer}, A.~J., {et~al.} 2024, \psj, 5,
  143, \dodoi{10.3847/PSJ/ad3f19}

\bibitem[{{Porter} {et~al.}(in rev.){Porter}, {Singer}, {Schenk}, {Verbiscer},
  {Benecchi}, {Spencer}, {Parker}, {Brandt}, \& {Stern}}]{Porter2024b}
{Porter}, S.~B., {Singer}, K.~N., {Schenk}, P.~M., {et~al.} in rev., The
  Planetary Science Journal

\bibitem[{{Pravec} \& {Harris}(2000)}]{Pravec2000}
{Pravec}, P., \& {Harris}, A.~W. 2000, \icarus, 148, 12,
  \dodoi{10.1006/icar.2000.6482}

\bibitem[{{Pravec} {et~al.}(2002){Pravec}, {Harris}, \&
  {Michalowski}}]{Pravec2002}
{Pravec}, P., {Harris}, A.~W., \& {Michalowski}, T. 2002, in Asteroids III
  (University of Arizona Press), 113--122

\bibitem[{{Richardson} {et~al.}(2005){Richardson}, {Elankumaran}, \&
  {Sanderson}}]{Richardson2005}
{Richardson}, D.~C., {Elankumaran}, P., \& {Sanderson}, R.~E. 2005, \icarus,
  173, 349, \dodoi{10.1016/j.icarus.2004.09.007}

\bibitem[{Richardson {et~al.}(2000)Richardson, Quinn, Stadel, \&
  Lake}]{Richardson2000}
Richardson, D.~C., Quinn, T., Stadel, J., \& Lake, G. 2000, Icarus, 143, 45,
  \dodoi{https://doi.org/10.1006/icar.1999.6243}

\bibitem[{{Richardson} {et~al.}(2011){Richardson}, {Walsh}, {Murdoch}, \&
  {Michel}}]{Richardson2011}
{Richardson}, D.~C., {Walsh}, K.~J., {Murdoch}, N., \& {Michel}, P. 2011,
  \icarus, 212, 427, \dodoi{10.1016/j.icarus.2010.11.030}

\bibitem[{{Robin} {et~al.}(2024){Robin}, {Duchene}, {Murdoch}, {Vincent},
  {Lucchetti}, {Pajola}, {Ernst}, {Daly}, {Barnouin}, {Raducan}, {Michel},
  {Hirabayashi}, {Stott}, {Cuervo}, {Jawin}, {Trigo-Rodriguez}, {Parro},
  {Sunday}, {Vivet}, {Mimoun}, {Rivkin}, \& {Chabot}}]{Robin2024}
{Robin}, C.~Q., {Duchene}, A., {Murdoch}, N., {et~al.} 2024, Nature
  Communications, 15, 6203, \dodoi{10.1038/s41467-024-50147-w}

\bibitem[{{Robinson} {et~al.}(2020){Robinson}, {Fraser}, {Fitzsimmons}, \&
  {Lacerda}}]{Robinson2020}
{Robinson}, J.~E., {Fraser}, W.~C., {Fitzsimmons}, A., \& {Lacerda}, P. 2020,
  Astronomy and Astrophysics, 643, A55, \dodoi{10.1051/0004-6361/202037456}

\bibitem[{{Ros} {et~al.}(2019){Ros}, {Johansen}, {Riipinen}, \&
  {Schlesinger}}]{Ros2019}
{Ros}, K., {Johansen}, A., {Riipinen}, I., \& {Schlesinger}, D. 2019, Astronomy
  and Astrophysics, 629, A65, \dodoi{10.1051/0004-6361/201834331}

\bibitem[{{Salo}(1987)}]{Salo1987}
{Salo}, H. 1987, Icarus, 70, 37, \dodoi{10.1016/0019-1035(87)90073-X}

\bibitem[{{S{\'a}nchez} \& {Scheeres}(2016)}]{Sanchez2016}
{S{\'a}nchez}, P., \& {Scheeres}, D.~J. 2016, Icarus, 271, 453,
  \dodoi{10.1016/j.icarus.2016.01.016}

\bibitem[{{Schoonenberg} \& {Ormel}(2017)}]{Schoonenberg2017}
{Schoonenberg}, D., \& {Ormel}, C.~W. 2017, Astronomy and Astrophysics, 602,
  A21, \dodoi{10.1051/0004-6361/201630013}

\bibitem[{{Schwartz} {et~al.}(2013){Schwartz}, {Michel}, \&
  {Richardson}}]{Schwartz2013}
{Schwartz}, S.~R., {Michel}, P., \& {Richardson}, D.~C. 2013, Icarus, 226, 67,
  \dodoi{10.1016/j.icarus.2013.05.007}

\bibitem[{Schwartz {et~al.}(2012)Schwartz, Richardson, \&
  Michel}]{Schwartz2012}
Schwartz, S.~R., Richardson, D.~C., \& Michel, P. 2012, Granular Matter, 14,
  363

\bibitem[{{Sharma} {et~al.}(2009){Sharma}, {Jenkins}, \& {Burns}}]{Sharma2009}
{Sharma}, I., {Jenkins}, J.~T., \& {Burns}, J.~A. 2009, \icarus, 200, 304,
  \dodoi{10.1016/j.icarus.2008.11.003}

\bibitem[{{Sheppard} \& {Jewitt}(2004)}]{Sheppard2004}
{Sheppard}, S.~S., \& {Jewitt}, D. 2004, \aj, 127, 3023, \dodoi{10.1086/383558}

\bibitem[{{Sheppard} \& {Jewitt}(2002)}]{Sheppard2002}
{Sheppard}, S.~S., \& {Jewitt}, D.~C. 2002, \aj, 124, 1757,
  \dodoi{10.1086/341954}

\bibitem[{{Showalter} {et~al.}(2021){Showalter}, {Benecchi}, {Buie}, {Grundy},
  {Keane}, {Lisse}, {Olkin}, {Porter}, {Robbins}, {Singer}, {Verbiscer},
  {Weaver}, {Zangari}, {Hamilton}, {Kaufmann}, {Lauer}, {Mehoke}, {Mehoke},
  {Spencer}, {Throop}, {Parker}, {Stern}, {New Horizons Geology}, \&
  Team}]{Showalter2021}
{Showalter}, M.~R., {Benecchi}, S.~D., {Buie}, M.~W., {et~al.} 2021, \icarus,
  356, 114098, \dodoi{10.1016/j.icarus.2020.114098}

\bibitem[{{Simon} {et~al.}(2022){Simon}, {Blum}, {Birnstiel}, \&
  {Nesvorn{\'y}}}]{Simon2022}
{Simon}, J.~B., {Blum}, J., {Birnstiel}, T., \& {Nesvorn{\'y}}, D. 2022, arXiv
  e-prints, arXiv:2212.04509, \dodoi{10.48550/arXiv.2212.04509}

\bibitem[{{Spencer} {et~al.}(2020){Spencer}, {Stern}, {Moore}, {Weaver},
  {Singer}, {Olkin}, {Verbiscer}, {McKinnon}, {Parker}, {Beyer}, {Keane},
  {Lauer}, {Porter}, {White}, {Buratti}, {El-Maarry}, {Lisse}, {Parker},
  {Throop}, {Robbins}, {Umurhan}, {Binzel}, {Britt}, {Buie}, {Cheng},
  {Cruikshank}, {Elliott}, {Gladstone}, {Grundy}, {Hill}, {Horanyi},
  {Jennings}, {Kavelaars}, {Linscott}, {McComas}, {McNutt}, {Protopapa},
  {Reuter}, {Schenk}, {Showalter}, {Young}, {Zangari}, {Abedin},
  {Beddingfield}, {Benecchi}, {Bernardoni}, {Bierson}, {Borncamp}, {Bray},
  {Chaikin}, {Dhingra}, {Fuentes}, {Fuse}, {Gay}, {Gwyn}, {Hamilton},
  {Hofgartner}, {Holman}, {Howard}, {Howett}, {Karoji}, {Kaufmann}, {Kinczyk},
  {May}, {Mountain}, {P{\"a}tzold}, {Petit}, {Piquette}, {Reid}, {Reitsema},
  {Runyon}, {Sheppard}, {Stansberry}, {Stryk}, {Tanga}, {Tholen}, {Trilling},
  \& {Wasserman}}]{Spencer2020}
{Spencer}, J.~R., {Stern}, S.~A., {Moore}, J.~M., {et~al.} 2020, Science, 367,
  aay3999, \dodoi{10.1126/science.aay3999}

\bibitem[{{Squire} \& {Hopkins}(2020)}]{Squire2020}
{Squire}, J., \& {Hopkins}, P.~F. 2020, Monthly Notices of the Royal
  Astronomical Society, 498, 1239, \dodoi{10.1093/mnras/staa2311}

\bibitem[{{Stadel}(2001)}]{Stadel2001}
{Stadel}, J.~G. 2001, PhD thesis, University of Washington, Seattle

\bibitem[{{Steinberg} \& {Sari}(2015)}]{Steinberg2015}
{Steinberg}, E., \& {Sari}, R. 2015, \aj, 149, 124,
  \dodoi{10.1088/0004-6256/149/4/124}

\bibitem[{{Stern} {et~al.}(2019){Stern}, {Weaver}, {Spencer}, {Olkin},
  {Gladstone}, {Grundy}, {Moore}, {Cruikshank}, {Elliott}, {McKinnon},
  {Parker}, {Verbiscer}, {Young}, {Aguilar}, {Albers}, {Andert}, {Andrews},
  {Bagenal}, {Banks}, {Bauer}, {Bauman}, {Bechtold}, {Beddingfield}, {Behrooz},
  {Beisser}, {Benecchi}, {Bernardoni}, {Beyer}, {Bhaskaran}, {Bierson},
  {Binzel}, {Birath}, {Bird}, {Boone}, {Bowman}, {Bray}, {Britt}, {Brown},
  {Buckley}, {Buie}, {Buratti}, {Burke}, {Bushman}, {Carcich}, {Chaikin},
  {Chavez}, {Cheng}, {Colwell}, {Conard}, {Conner}, {Conrad}, {Cook}, {Cooper},
  {Custodio}, {Dalle Ore}, {Deboy}, {Dharmavaram}, {Dhingra}, {Dunn}, {Earle},
  {Egan}, {Eisig}, {El-Maarry}, {Engelbrecht}, {Enke}, {Ercol}, {Fattig},
  {Ferrell}, {Finley}, {Firer}, {Fischetti}, {Folkner}, {Fosbury}, {Fountain},
  {Freeze}, {Gabasova}, {Glaze}, {Green}, {Griffith}, {Guo}, {Hahn}, {Hals},
  {Hamilton}, {Hamilton}, {Hanley}, {Harch}, {Harmon}, {Hart}, {Hayes},
  {Hersman}, {Hill}, {Hill}, {Hofgartner}, {Holdridge}, {Hor{\'a}nyi},
  {Hosadurga}, {Howard}, {Howett}, {Jaskulek}, {Jennings}, {Jensen}, {Jones},
  {Kang}, {Katz}, {Kaufmann}, {Kavelaars}, {Keane}, {Keleher}, {Kinczyk},
  {Kochte}, {Kollmann}, {Krimigis}, {Kruizinga}, {Kusnierkiewicz}, {Lahr},
  {Lauer}, {Lawrence}, {Lee}, {Lessac-Chenen}, {Linscott}, {Lisse}, {Lunsford},
  {Mages}, {Mallder}, {Martin}, {May}, {McComas}, {McNutt}, {Mehoke}, {Mehoke},
  {Nelson}, {Nguyen}, {N{\'u}{\~n}ez}, {Ocampo}, {Owen}, {Oxton}, {Parker},
  {P{\"a}tzold}, {Pelgrift}, {Pelletier}, {Pineau}, {Piquette}, {Porter},
  {Protopapa}, {Quirico}, {Redfern}, {Regiec}, {Reitsema}, {Reuter},
  {Richardson}, {Riedel}, {Ritterbush}, {Robbins}, {Rodgers}, {Rogers}, {Rose},
  {Rosendall}, {Runyon}, {Ryschkewitsch}, {Saina}, {Salinas}, {Schenk},
  {Scherrer}, {Schlei}, {Schmitt}, {Schultz}, {Schurr}, {Scipioni}, {Sepan},
  {Shelton}, {Showalter}, {Simon}, {Singer}, {Stahlheber}, {Stanbridge},
  {Stansberry}, {Steffl}, {Strobel}, {Stothoff}, {Stryk}, {Stuart}, {Summers},
  {Tapley}, {Taylor}, {Taylor}, {Tedford}, {Throop}, {Turner}, {Umurhan}, {Van
  Eck}, {Velez}, {Versteeg}, {Vincent}, {Webbert}, {Weidner}, {Weigle},
  {Wendel}, {White}, {Whittenburg}, {Williams}, {Williams}, {Williams},
  {Winters}, {Zangari}, \& {Zurbuchen}}]{Stern2019}
{Stern}, S.~A., {Weaver}, H.~A., {Spencer}, J.~R., {et~al.} 2019, Science, 364,
  aaw9771, \dodoi{10.1126/science.aaw9771}

\bibitem[{Stone \& Kaib(2021)}]{StoneKaib2021}
Stone, L.~R., \& Kaib, N.~A. 2021, Monthly Notices of the Royal Astronomical
  Society: Letters, 505, L31, \dodoi{10.1093/mnrasl/slab044}

\bibitem[{{Testi} {et~al.}(2014){Testi}, {Birnstiel}, {Ricci}, {Andrews},
  {Blum}, {Carpenter}, {Dominik}, {Isella}, {Natta}, {Williams}, \&
  {Wilner}}]{Testi2014}
{Testi}, L., {Birnstiel}, T., {Ricci}, L., {et~al.} 2014, in Protostars and
  Planets VI, ed. H.~{Beuther}, R.~S. {Klessen}, C.~P. {Dullemond}, \&
  T.~{Henning}, 339--361, \dodoi{10.2458/azu_uapress_9780816531240-ch015}

\bibitem[{{Thirouin} {et~al.}(2014){Thirouin}, {Noll}, {Ortiz Moreno}, \&
  {Morales }}]{Thirouin2014}
{Thirouin}, A., {Noll}, K.~S., {Ortiz Moreno}, J.~L., \& {Morales }, N. 2014,
  in AAS/Division for Planetary Sciences Meeting Abstracts, Vol.~46,
  AAS/Division for Planetary Sciences Meeting Abstracts \#46, 421.09

\bibitem[{{Thirouin} {et~al.}(2010){Thirouin}, {Ortiz}, {Duffard},
  {Santos-Sanz}, {Aceituno}, \& {Morales}}]{Thirouin2010}
{Thirouin}, A., {Ortiz}, J.~L., {Duffard}, R., {et~al.} 2010, \aap, 522, A93,
  \dodoi{10.1051/0004-6361/200912340}

\bibitem[{{Thirouin} \& {Sheppard}(2018)}]{Thirouin2018}
{Thirouin}, A., \& {Sheppard}, S.~S. 2018, \aj, 155, 248,
  \dodoi{10.3847/1538-3881/aac0ff}

\bibitem[{{Thirouin} \& {Sheppard}(2019{\natexlab{a}})}]{Thirouin2019b}
---. 2019{\natexlab{a}}, \aj, 158, 53, \dodoi{10.3847/1538-3881/ab27bc}

\bibitem[{{Thirouin} \& {Sheppard}(2019{\natexlab{b}})}]{Thirouin2019a}
---. 2019{\natexlab{b}}, \aj, 157, 228, \dodoi{10.3847/1538-3881/ab18a9}

\bibitem[{{Thirouin} \& {Sheppard}(2024)}]{Thirouin2024}
---. 2024, \psj, 5, 84, \dodoi{10.3847/PSJ/ad2933}

\bibitem[{{Vilenius} {et~al.}(2014){Vilenius}, {Kiss}, {M{\"u}ller}, {Mommert},
  {Santos-Sanz}, {P{\'a}l}, {Stansberry}, {Mueller}, {Peixinho}, {Lellouch},
  {Fornasier}, {Delsanti}, {Thirouin}, {Ortiz}, {Duffard}, {Perna}, \&
  {Henry}}]{Vilenius2014}
{Vilenius}, E., {Kiss}, C., {M{\"u}ller}, T., {et~al.} 2014, \aap, 564, A35,
  \dodoi{10.1051/0004-6361/201322416}

\bibitem[{{Vokrouhlick{\'y}} {et~al.}(2003){Vokrouhlick{\'y}}, {Nesvorn{\'y}},
  \& {Bottke}}]{Vokrouhlicky2003}
{Vokrouhlick{\'y}}, D., {Nesvorn{\'y}}, D., \& {Bottke}, W.~F. 2003, \nat, 425,
  147, \dodoi{10.1038/nature01948}

\bibitem[{{Volk} \& {Malhotra}(2019)}]{Volk2019}
{Volk}, K., \& {Malhotra}, R. 2019, \aj, 158, 64,
  \dodoi{10.3847/1538-3881/ab2639}

\bibitem[{{Walsh} {et~al.}(2008){Walsh}, {Richardson}, \& {Michel}}]{Walsh2008}
{Walsh}, K.~J., {Richardson}, D.~C., \& {Michel}, P. 2008, Nature, 454, 188,
  \dodoi{10.1038/nature07078}

\bibitem[{{Walsh} {et~al.}(2012){Walsh}, {Richardson}, \& {Michel}}]{Walsh2012}
---. 2012, Icarus, 220, 514, \dodoi{10.1016/j.icarus.2012.04.029}

\bibitem[{Warner {et~al.}(2009)Warner, Harris, \& Pravec}]{Warner2009}
Warner, B., Harris, A., \& Pravec, P. 2009, Icarus, 202, 134.
\newblock \url{http://www.MinorPlanet.info/php/lcdb.php}

\bibitem[{{Weidenschilling}(1977)}]{Weidenschilling1977b}
{Weidenschilling}, S.~J. 1977, Monthly Notices of the Royal Astronomical
  Society, 180, 57, \dodoi{10.1093/mnras/180.2.57}

\bibitem[{{Windmark} {et~al.}(2012){Windmark}, {Birnstiel}, {Ormel}, \&
  {Dullemond}}]{Windmark2012a}
{Windmark}, F., {Birnstiel}, T., {Ormel}, C.~W., \& {Dullemond}, C.~P. 2012,
  Astronomy and Astrophysics, 544, L16, \dodoi{10.1051/0004-6361/201220004}

\bibitem[{{Youdin} \& {Johansen}(2007)}]{Youdin2007b}
{Youdin}, A., \& {Johansen}, A. 2007, The Astrophysical Journal, 662, 613,
  \dodoi{10.1086/516729}

\bibitem[{{Youdin} \& {Goodman}(2005)}]{Youdin2005b}
{Youdin}, A.~N., \& {Goodman}, J. 2005, The Astrophysical Journal, 620, 459,
  \dodoi{10.1086/426895}

\bibitem[{{Youdin} \& {Lithwick}(2007)}]{Youdin2007a}
{Youdin}, A.~N., \& {Lithwick}, Y. 2007, Icarus, 192, 588,
  \dodoi{10.1016/j.icarus.2007.07.012}

\bibitem[{{Yu} {et~al.}(2014){Yu}, {Richardson}, {Michel}, {Schwartz}, \&
  {Ballouz}}]{Yu2014}
{Yu}, Y., {Richardson}, D.~C., {Michel}, P., {Schwartz}, S.~R., \& {Ballouz},
  R.-L. 2014, Icarus, 242, 82, \dodoi{10.1016/j.icarus.2014.07.027}

\bibitem[{{Zhang} {et~al.}(2015){Zhang}, {Baoyin}, {Li}, {Richardson}, \&
  {Schwartz}}]{Zhang2015}
{Zhang}, Y., {Baoyin}, H., {Li}, J., {Richardson}, D.~C., \& {Schwartz}, S.~R.
  2015, Astrophysics and Space Science, 360, 30,
  \dodoi{10.1007/s10509-015-2536-8}

\bibitem[{{Zhang} \& {Michel}(2021)}]{Zhang2021}
{Zhang}, Y., \& {Michel}, P. 2021, Astrodynamics, 5, 293,
  \dodoi{10.1007/s42064-021-0128-7}

\bibitem[{{Zhang} {et~al.}(2018){Zhang}, {Richardson}, {Barnouin}, {Michel},
  {Schwartz}, \& {Ballouz}}]{Zhang2018}
{Zhang}, Y., {Richardson}, D.~C., {Barnouin}, O.~S., {et~al.} 2018, \apj, 857,
  15, \dodoi{10.3847/1538-4357/aab5b2}

\bibitem[{{Zhang} {et~al.}(2017){Zhang}, {Richardson}, {Barnouin}, {Maurel},
  {Michel}, {Schwartz}, {Ballouz}, {Benner}, {Naidu}, \& {Li}}]{Zhang2017}
---. 2017, \icarus, 294, 98, \dodoi{10.1016/j.icarus.2017.04.027}

\bibitem[{{Zhao} {et~al.}(2021){Zhao}, {Rezac}, {Skorov}, {Hu}, {Samarasinha},
  \& {Li}}]{Zhao2021}
{Zhao}, Y., {Rezac}, L., {Skorov}, Y., {et~al.} 2021, Nature Astronomy, 5, 139,
  \dodoi{10.1038/s41550-020-01218-7}

\bibitem[{{Zsom} {et~al.}(2010){Zsom}, {Ormel}, {G{\"u}ttler}, {Blum}, \&
  {Dullemond}}]{Zsom2010}
{Zsom}, A., {Ormel}, C.~W., {G{\"u}ttler}, C., {Blum}, J., \& {Dullemond},
  C.~P. 2010, Astronomy and Astrophysics, 513, A57,
  \dodoi{10.1051/0004-6361/200912976}

\end{thebibliography}
\bibliographystyle{aasjournal}

\end{document}